\documentclass[twocolumn]{jpsj3} 

\def\eg{e.g.}
\def\ie{i.e.}
\def\etal{{\it et\ al.}}

\newcommand{\lsim}
 {\ \raise.35ex\hbox{$<$}\kern-0.75em\lower.5ex\hbox{$\sim$}\ }
\newcommand{\gsim}
 {\ \raise.35ex\hbox{$>$}\kern-0.75em\lower.5ex\hbox{$\sim$}\ }
\def\journal #1#2#3#4{#1 {\bf #2} (#4) #3}
\def\AP{Adv.~Phys.}
\def\APNY{Ann.\ Phys.\ (New York)}

\def\IJMP{Int.\ J.\ Mod.\ Phys.}

\def\JLTP{J.~Low Temp.~Phys.}
\def\JMMM{J.~Mag.~Mag.~Mat.}

\def\JPParis{J.~Phys.~(Paris)}
\def\JPCM{J.\ Phys.\ Cond.\ Mat.}
\def\JPCS{J.\ Phys.\ Chem.\ Solids}

\def\JPSJ{J.\ Phys.\ Soc.\ Jpn.}

\def\NJP{New J.~Phys.}
\def\NP{Nat.~Phys.}

\def\PP{Phys.~Proc.}
\def\PRep{Phys.\ Rep.}
\def\PR{Phys.\ Rev.}

\def\PRB{Phys.\ Rev.\ B}
\def\PRL{Phys.\ Rev.\ Lett.}

\def\PTP{Prog.\ Theor.\ Phys.}
\def\RMP{Rev.\ Mod.\ Phys.}

\def\EPL{Europhys.\ Lett.}
\def\RPP{Rep.~Prog.~Phys.}
\usepackage{color}

\title{
Crossover between BCS Superconductor and Doped Mott Insulator \\
of $d$-wave Pairing State in Two-Dimensional Hubbard Model
}

\author{
Hisatoshi \textsc{Yokoyama}$^{1}$\thanks{E-mail address: 
yoko@cmpt.phys.tohoku.ac.jp},
Masao \textsc{Ogata}$^{2}$, 
Yukio \textsc{Tanaka}$^{3}$,
Kenji \textsc{Kobayashi}$^{4}$, \\ and 
Hiroki \textsc{Tsuchiura}$^{5}$ 
}

\inst{
$^{1}$Department of Physics, Tohoku University, Sendai 980-8578, Japan \\
$^{2}$Department of Physics, University of Tokyo, Bunkyo-ku, Tokyo 113-0033, 
Japan \\
$^{3}$Department of Applied Physics, Nagoya University, Nagoya 464-8603, 
Japan \\
$^{4}$Department of Natural Science, Chiba Institute of Technology, 
Narashino 275-0023, Japan \\
$^{5}$Department of Applied Physics, Tohoku University, Sendai 980-8579
Japan 
}

\abst{
With high-$T_{\rm c}$ cuprates in mind, the properties of correlated 
$d_{x^2-y^2}$-wave superconducting (SC) and antiferromagnetic (AF) states 
are studied for the Hubbard ($t$-$t'$-$U$) model on square lattices 
using a variational Monte Carlo method. 
We employ simple trial wave functions including only crucial parameters, 
such as a doublon-holon binding factor indispensable for describing correlated 
SC and normal states as doped Mott insulators. 
The $U/t$, $t'/t$, and $\delta$ (doping rate) dependences of relevant 
quantities 
are systematically calculated. 
As $U/t$ increases, a sharp crossover of SC properties occurs 
at $U_{\rm co}/t\sim 10$ from a conventional BCS type to a 
kinetic-energy-driven type for any $t'/t$. 
As $\delta$ decreases, $U_{\rm co}/t$ is smoothly connected to the Mott 
transition point at half filling. 
For $U/t\lsim 5$, steady superconductivity corresponding to the cuprates 
is not found, whereas the $d$-wave SC correlation function $P_d^\infty$ 
rapidly increases for $U/t\gsim 6$ and becomes maximum at $U=U_{\rm co}$. 
Comparing the $\delta$ dependence of $P_d^\infty$ with an experimentally 
observed dome-shaped $T_{\rm c}$ and condensation energy, we find that 
the effective value of $U$ for cuprates should be larger than 
the bandwidth, for which the $t$-$J$ model is valid. 
Analyzing the kinetic energy, we reveal that, for $U>U_{\rm co}$, only doped 
holes (electrons) become charge carriers, which will make a small Fermi 
surface (hole pocket), but for $U<U_{\rm co}$ all the electrons (holes) 
contribute to conduction and will make an ordinary large Fermi surface, 
which is contradictory to the feature of cuprates. 
By introducing an appropriate negative (positive) $t'/t$, the SC (AF) state is 
stabilized. 
In the underdoped regime, the strength of SC for $U>U_{\rm co}$ is 
determined by two factors, i.e., the AF spin correlation, which creates 
singlet pairs (pseudogap), and the charge mobility 
dominated by Mott physics. 
In this connection, we argue that the electrons near the antinodal points 
in the momentum space play a leading role in stabilizing the $d$-wave state, 
in contrast to the dichotomy of electronic roles in the momentum space 
proposed for the two-gap problem. 
We also show the instability of the hole-doped AF state against phase 
separation. 
}

\kword{superconductivity, cuprate, antiferromagnetism, pseudogap, 
doped Mott insulator, doublon-holon binding, Hubbard model, 
strong correlation, variational Monte Carlo method.}
\begin{document}
\maketitle

\section{Introduction\label{sec:intro}}
%
The essence of superconductivity (SC) in high-$T_{\rm c}$ cuprates 
is most probably captured using two-dimensional (2D) $t$-$J$ and Hubbard 
models;\cite{Anderson,Zhang-Rice} a variety of theories have been 
developed on the basis of these models.\cite{Theory,PALee,OF} 
In the ground state of the $t$-$J$-type models, which deal with strongly 
correlated regimes, different approaches concluded that superconducting 
(SC) states with $d_{x^2-y^2}$-wave symmetry are stabilized in the parameter 
range relevant to cuprates, as we will see further in \S\ref{sec:model}. 
Meanwhile, approaches starting from the weak-correlation 
limit\cite{Weak,NRG} of the Hubbard model provided complementary, but 
often qualitative, views of SC for cuprates, particularly in terms of 
thermal and dynamical properties. 
Nevertheless, quantum Monte Carlo (QMC) methods,\cite{QMC,Aimi,YanagisawaQMC} 
which are quantitatively reliable approaches for finite correlation strengths,
unanimously yielded a result that the SC correlation function never develops 
in weakly correlated regimes as compared with the noninteracting case, 
and therefore cast serious doubt on the validity of applying weak-correlation 
theories to realistic correlation strengths. 
Thus, it is important to clarify how the properties of a $d_{x^2-y^2}$-wave 
SC state evolve as the correlation strength is increased in the Hubbard 
model. 
\par

To this end, the variational Monte Carlo (VMC) 
method\cite{McMillan,Ceperley} is useful for its continuous applicability 
from weakly to strongly correlated regimes in a quantitative manner. 
This is by virtue of the correct treatment of local electron correlations. 
Since the early stage of research on cuprates, VMC 
studies\cite{YStJ2,Gros,YO} have mainly treated $t$-$J$-type models and 
yielded results basically consistent with the behavior of 
cuprates.\cite{OF} 
Later, the present authors studied a many-body SC state, applying it 
to the 2D Hubbard ($t$-$t'$-$U$) model [$U$: onsite Coulomb correlation, 
$t$, $t'$: hopping integrals to the nearest-neighbor (NN) and diagonal 
sites].\cite{YTOT,YOT,Oproc,Yproc} 
In refs.~\citen{YTOT} (for $t'=0$) and \citen{YOT}, we discussed 
superconductor (SC)-insulator (Mott) transitions at half filling 
arising at $U=U_{\rm c}\sim 7t$ without directly introducing an 
antiferromagnetic (AF) correlation. 
According to these and recent studies,\cite{Miyagawa,YMO} the binding effect 
between a doubly occupied site (doublon, D) and an empty site (holon, H) 
plays a leading role in nonmagnetic Mott transitions. 
Reference \citen{YTOT} also showed in doped cases that a sharp crossover 
exists at $U=U_{\rm co}\sim W$ [$W(=8t)$: bandwidth], where the nature 
of the $d_{x^2-y^2}$-wave SC state, $\Psi_Q^d$, changes qualitatively 
from a conventional Bardeen-Cooper-Schrieffer (BCS) type for $U<U_{\rm co}$ 
to an unconventional strongly 
correlated type for $U>U_{\rm co}$, in which $\Psi_Q^d$ is stabilized 
by the reduction in kinetic energy. 
This crossover for SC is smoothly connected to the Mott transition 
at half filling ($U_{\rm co}\rightarrow U_{\rm c}$) as the doping rate 
$\delta$ is decreased. 
Thus, the strongly correlated SC state for $U>U_{\rm co}$ corresponds to 
a doped Mott insulator,\cite{Imada} and is strongly affected 
by the effect of D-H binding correlation. 
\par

In this paper, as an extension of refs.~\citen{YTOT} and \citen{YOT},  
we study SC and antiferromagnetism (AF) independently for doped cases 
by systematically changing $U/t$, $t'/t$, and $\delta$, 
using D-H binding wave functions. 
The primary purpose of this paper is to establish that the nature of SC 
is qualitatively 
different between the regimes of $U<U_{\rm co}$ and $U>U_{\rm co}$ 
by studying various quantities, and that the behavior of cuprates is 
contradictory to that of $d$-wave BCS-type SC realized for $U<U_{\rm co}$ 
in various aspects, but consistent with the behavior of doped Mott 
insulators ($U>U_{\rm co}$). 
\par

The second purpose of this study is to consider the differentiation or 
dichotomy of electronic roles in momentum space,\cite{twogap,JPSJ-ARPES} 
in connection with the pseudogap problem. 
Early experimental studies on the so-called two-gap 
problem\cite{Raman,ARPES,AIPES,STM} challenged the familiar view 
that the pseudogap stems from incoherent pairs of 
electrons\cite{Ong,Kanigel,Valla,ARPESanti} for $T_{\rm c}<T<T^*$. 
Two characteristic energy scales (gaps) were discovered whose 
$\delta$ dependences are mutually opposite in the underdoped regime. 
They seemed to stem from electrons with wave numbers 
near the nodal ($\pi/2,\pi/2$) and antinodal ($\pi,0$) points, 
and correspond to the SC gap and pseudogap, respectively, indicating 
that the origin of the pseudogap is not directly connected to pairing. 
Recent experiments on the pseudogap have taken on more complicated 
aspects,\cite{Ma,Yoshida,Nakayama,Yu,Dubroka,Daou,liquid-crystal,Gomes} 
and are still actively developing.\cite{Kohsaka,JPSJ-STS} 
In this study, we argue that the two energy scales can be interpreted 
as a pairing gap and its correction by the charge fluctuation released 
from a Mott insulating state,
and that the electronic states near the antinodal point $(\pi,0)$ 
[not in the nodal region $\sim(\pi/2,\pi/2)$] are crucial to the 
$d$-wave SC.\cite{Young,ARPESanti} 
\par

Some of the results were reported previously.\cite{YOK,ISS2011}
\par

The rest of this paper is organized as follows: 
In \S\ref{sec:formalism}, we explain the model and method used 
in this paper. 
In \S\ref{sec:square}, we present the results for $t'=0$ to grasp basic 
features of the $U/t$ and $\delta$ dependences of relevant quantities, and 
argue the existence of a crossover and its relevance to cuprates. 
In \S\ref{sec:diag}, we discuss the effects of the diagonal transfer $t'$, 
which provides a clue to understanding the stability of the SC and AF states. 
In \S\ref{sec:discussions}, we argue that the electrons near the 
antinodal point play a leading role in realizing the $d$-wave SC. 
In \S\ref{sec:summary}, the main results are summarized.
In Appendix~\ref{sec:derivation}, we give supplementary explanations 
of the model. 
In Appendix~\ref{sec:dh}, we compare some forms of doublon-holon binding 
factors as a supplement to \S\ref{sec:wf}. 
In Appendix~\ref{sec:pairfunc}, we describe in detail the definition and 
justification of the long-distance value of the $d$-wave pair correlation 
function in the present cases. 
\par

\section{Formalism\label{sec:formalism}}
In \S\ref{sec:model}, we explain the Hubbard model and the background 
of this study. 
In \S\ref{sec:wf}, we introduce the many-body trial wave functions used 
in this paper. 
In \S\ref{sec:VMCtech}, we briefly summarize the setting of VMC 
calculations. 

\subsection{Hubbard model for cuprates\label{sec:model}}
As a model of the CuO$_2$ planes, we consider the Hubbard model 
on a square lattice: 
\begin{eqnarray}
{\cal H}&=&{\cal H}_{\rm kin}+{\cal H}_U \nonumber \\
        &=&\sum_{{\bf k}\sigma} \varepsilon({\bf k})
        c^\dagger_{{\bf k}\sigma}c_{{\bf k}\sigma} 
         +U\sum_jd_j, 
\label{eq:Hamil}
\end{eqnarray} 
with $d_j=n_{j\uparrow}n_{j\downarrow}$ (doublon projector) and 
$n_{j\sigma}=c^\dag_{j\sigma}c_{j\sigma}$. 
Although the band structure $\varepsilon({\bf k})$ is qualitatively 
important,\cite{Fukuyama,Tohyama,Gooding,Raimondi,Pavarini,Shih,Tanaka} 
we consider, for clarity, only the minimum terms to distinguish the material 
dependence, i.e., NN ($t$) and diagonal-neighbor ($t'$) hoppings: 
\begin{equation}
\varepsilon({\bf k})=-2t(\cos k_x+\cos k_y)-4t'\cos k_x\cos k_y. 
\label{eq:band}
\end{equation} 
We use $t$ and the lattice constant as the units of energy and length, 
respectively. 
As discussed in Appendix~\ref{sec:derivation}, the effective onsite 
repulsion $U/t$ may differ between electron-doped and hole-doped 
cases. 
For ease of calculation, we can restrict the electron density to
$n=N/N_{\rm s}\le 1$ ($N$: electron number; $N_{\rm s}$: site number), 
or the doping rate to $\delta=1-n\ge 0$ by a canonical transformation. 
Then, $t'/t$ becomes negative (positive) in hole- (electron-)doped 
cases (see Appendix~\ref{sec:derivation}). 
\par

Studies of this model for cuprates are roughly classified 
into weak- and strong-correlation theories. 
In the former,\cite{Weak} the stability and some properties of 
$d_{x^2-y^2}$-wave 
SC are discussed using RPA \cite{RPA}, FLEX approximations \cite{FLEX}, 
renormalization-group, \cite{NRG} and perturbative \cite{Nomura,Kondo} 
approaches. 
As we will show in \S\ref{sec:square} and \S\ref{sec:diag}, however, 
no substantial enhancement of the SC correlation function appropriate 
for high-$T_{\rm c}$ cuprates is found in a weak-correlation regime 
($U/t\lsim 5$). 
Such a result is not restricted to our studies. 
Early QMC studies\cite{QMC} drew negative conclusions for the appearance 
of SC in the weak-coupling regime ($U/t=2$-$4$). 
A recent QMC study\cite{Aimi} using an expansion with Gaussian bases 
also obtained a negative result for SC in the overdoped regime 
up to $U/t=7$ for $t'=0$. 
Another recent study\cite{YanagisawaQMC} concluded that 
the $d_{x^2-y^2}$-wave pair susceptibility is not enhanced for $t'=0$, 
but an SC Kosteritz-Thouless transition exists for moderately large 
values of $U/t$ and $t'/t=-0.2$. 
Thus, it is improbable that the small-$U/t$ Hubbard model with $t'=0$ gives 
such a high $T_{\rm c}$ that cuprates display; 
a doubt arises as to the reliability of the weak-correlation theories 
for cuprates, at least in a quantitative sense. 
In this connection, we will check whether or not the introduction of the 
$t'/t$ term can enhance SC in the weak-correlation regime. 
\par

In the strong-correlation limit, the Hubbard model is mapped to 
$t$-$J$-type models, in which robust $d_{x^2-y^2}$-wave SC is found 
by the VMC,\cite{YStJ2,Gros,YO} exact diagonalization\cite{Dagotto}, 
and projector Monte Carlo methods, \cite{Sorella} besides by various 
mean-field-type approximations.\cite{PALee,OF} 
In the corresponding large-$U/t$ regime of the Hubbard model, 
firm $d$-wave SC and its competition with AF orders were found near 
half filling by VMC studies \cite{Giamarchi,Yamaji,Yanagisawa} 
using trial functions similar to ours. \cite{YTOT,YOT,Oproc,Yproc}
Another promising quantitative method of treating this regime is the 
extension of the dynamical mean field theory (DMFT). 
First, using a dynamical cluster approximation, a finite-temperature 
instability in $d$-wave SC was found.\cite{Maier} 
Subsequently, the stability of $d$-wave SC and the competition with 
AF were studied using different variations of DMFT.
\cite{Senechal,Aichhorn,Capone} 
To our knowledge, there are no theories that reliably drew negative 
results for SC in this regime. 
It seems appropriate to assume that robust $d_{x^2-y^2}$-wave SC 
corresponding to high-$T_{\rm c}$ cuprates is realized in the 
Hubbard model with $U\gsim W$ near half filling. 
\par

Finally, we discuss the coexistence of $d$-wave SC and AF long-range 
orders. 
This problem has been repeatedly addressed for the 2D $t$-$J$ 
and Hubbard models, using, e.g., VMC methods\cite{co-Lee,Giamarchi,Himeda-co} 
and DMFT.\cite{Senechal,Kancharla} 
These studies found the stability of coexisting states against 
both pure $d$-wave and AF states near half filling; other 
studies\cite{Capone,Kobayashi-co,coexistcond} showed that such mixing 
states appear for a small $\delta$ only under certain conditions. 
Experimentally, no explicit evidence of microscopic coexistence has 
been obtained for most high-$T_c$ cuprates, \cite{Hosseini} 
except for multilayered Hg systems.\cite{Mukuda,Yamada,JPSJ-ML} 
It is probable that the coexistence can be realized in clean CuO$_2$ 
planes in multilayered systems, while the absence of the coexistence 
in usual cases will be due to disorder. 
Meanwhile, inhomogeneous phases such as stripes\cite{JPSJ-neutron} 
can compete or coexist 
with the $d_{x^2-y^2}$ wave SC.\cite{stripe,Kohsaka,JPSJ-STS}
Anyway, as a basis to tackle complicated states, it is vital to study 
the pure $d$-wave and AF states separately and their competition. 
\par

\subsection{Variational wave functions\label{sec:wf}}
We implement a series of VMC calculations. 
Removing irrelevant variational parameters, we elucidate basic 
behaviors of simple correlated $d$-wave SC and AF wave functions 
of the Jastrow type,\cite{Jastrow} 
\begin{equation}
\Psi={\cal P}\Phi, 
\label{eq:Jastrow}
\end{equation}
when it is applied to the Hubbard model, eq.~(\ref{eq:Hamil}).
Here, ${\cal P}$ denotes a many-body correlation (Jastrow) factor 
composed of projection operators, and $\Phi$ a one-body (mean-field) 
wave function given by a Slater determinant. 
\par 

In the many-body part, we consider two projection operators: 
${\cal P}={\cal P}_Q{\cal P}_{\rm G}$.\cite{YTOT,YOT} 
Although the onsite repulsive (Gutzwiller) projector\cite{Gutz} 
\begin{equation}
{\cal P}_{\rm G}=\prod_j\left[1-(1-g)d_j\right], 
\end{equation}
with $0\le g\le 1$, is of primary importance, intersite correlation 
factors are indispensable for a qualitatively correct description 
of Hubbard-type models.\cite{YS1}
Among them, an attractive factor between a doublon and a holon is crucial 
at half filling, which is defined as 
\cite{Kaplan,YS3} 
\begin{equation}
{\cal P}_Q=\prod_j\left(1-Q_j\right), 
\label{eq:PQ}
\end{equation}
where $Q_j$ (=$Q_j^{\rm S}$) is the D-H projector of site $j$: 
%
\begin{equation}
Q^{\rm S}_j=\mu\left[ 
d_j\prod_\tau(1-h_{j+\tau})+h_j\prod_\tau(1-d_{j+\tau})\right]. 
\label{eq:SymQ} 
\end{equation}
Here, $h_j=(1-n_{j\uparrow})(1-n_{j\downarrow})$ and $\tau$ runs 
over all the NN sites of site $j$. 
The projector ${\cal P}_Q$ with $Q_j^{\rm S}$ yields 
$(1-\mu)^{{\cal N}_{\rm D}+{\cal N}_{\rm H}}$ if isolated 
${\cal N}_{\rm D}$ doublons and ${\cal N}_{\rm H}$ holons exist 
in the electron configuration, 
where an isolated doublon (holon) indicates a doublon (holon) not 
accompanied by holons (doublons) in its four NN sites. 
The parameter $\mu$ ($\le 1$) controls the strength of the D-H 
correlation; for $\mu=1$, a doublon and a holon are completely 
bound in mutually NN sites, for $\mu=0$, they are 
completely free, and for $\mu<0$, they become repulsive to one another.
As we repeatedly showed, ${\cal P}_Q$ or its analog plays a leading 
role in inducing conductor (metal)-to-nonmagnetic insulator (Mott) 
transitions\cite{Wataorg,YOT,Miyagawa,YMO} and spin-gap transitions 
in attractive Hubbard models.\cite{YPTP,Tamura} 
Recent studies using more elaborate long-range D-H factors\cite{Miyagawa,YMO} 
showed that the simplest NN form $Q_j^{\rm S}$ works unexpectedly well. 
We use the form $Q_j^{\rm S}$ for $\delta=0$ in this study. 
We found that a D-H correlation factor is still important for 
$\delta\ne 0$\cite{YTOT} if $\delta$ is roughly within the range of 
SC for cuprates. 
An NN D-H pair yields, with a single hopping process, an antiparallel-spin 
pair, which contributes to the pairing in $d$-wave SC. 
Because the symmetry between a doublon and a holon is broken for 
$\delta\ne 0$, an asymmetric form of $Q_j$ seems reasonable. 
Thus, for $\delta>0$, we use the simple asymmetric form 
\begin{equation}
Q_j=Q_j^{\rm D}=\mu\ d_j\prod_\tau(1-h_{j+\tau}), 
\label{eq:DQ}
\end{equation} 
for ease of calculation. 
In Appendix~\ref{sec:dh}, we check different forms of $Q_j$ 
in detail. 
The main conclusion is that the details of $Q_j$ make no difference 
at least qualitatively; a simple form of $Q_j$ such as eq.~(\ref{eq:DQ}) 
preserves the essence of the D-H binding mechanism. 
In \S\ref{sec:D-H}, we will discuss the condition that the D-H binding 
factor becomes significant. 
\par

In the frustrated cases ($t'\ne 0$), the D-H correlation between diagonal 
sites may also play a certain role. 
Therefore, we similarly introduce a D-H projector $Q_j'$ with a variational 
parameter $\mu'$ for interdiagonal sites. 
Eventually, the D-H factor for $t'\ne0$ becomes 
\begin{equation}
{\cal P}_Q=\prod_j\left(1-Q_j\right)\left(1-Q_j'\right).  
\label{eq:PQD} 
\end{equation}
In most calculations for $t'=0$, we fixed $\mu'$ at zero, because 
the optimized $\mu'$ is small and the energy gain thereof is 
very small ($\sim 10^{-4}t$ even for a large $U/t$).\cite{note-errmu2}
\par

Now, we turn to the one-body function $\Phi$ in eq.~(\ref{eq:Jastrow}).  
As a continuation of preceding studies, we mainly study a fixed-$N$ 
$d_{x^2-y^2}$-wave singlet (BCS) state, 
$\Phi_d(\Delta_d,\zeta)$: \cite{Lhuillier} 
\begin{equation}
\Phi_d =\left(\sum_{\bf k}a_{\bf k}
c_{{\bf k}\uparrow}^\dagger c_{{\bf -k}\downarrow}^\dagger
\right)^\frac{N}{2}|0\rangle, 
\label{eq:Phi_d}
\end{equation}
with
\begin{equation}
a_{\bf k}=\frac{v_{\bf k}}{u_{\bf k}}=\frac{\Delta_{\bf k}}
{\varepsilon_{\bf k}-\zeta+
\sqrt{(\varepsilon_{\bf k}-\zeta)^2+\Delta_{\bf k}^2}}, 
\label{eq:BCSDelta}
\end{equation}
where $\zeta$ is a variational parameter that is reduced to the
chemical potential for $U/t\rightarrow 0$. 
Since the $d_{x^2-y^2}$-wave is the most stable among various gap 
shapes near half filling,\cite{YStJ2,Gros} here
we exclusively treat the basic $d_{x^2-y^2}$-wave gap: 
\cite{note-nonmono,Watat-J,Hirashima,nonmonotonic,Bi2212}
\begin{equation}
\Delta_{\bf k}=\Delta_d(\cos k_x-\cos k_y). 
\label{eq:gap}
\end{equation} 
We should emphasize that, even if the variational parameter $\Delta_d$ 
indicates the magnitude of the $d$-wave singlet gap, a many-body state 
($\Psi^d={\cal P}\Phi_d$) with a finite $\Delta_d$ does not necessarily 
mean an SC state. \cite{ZGRS,Yang} 
To confirm the SC order, one needs to calculate the order 
parameter\cite{YStJ2} or pair correlation functions such as $P_d({\bf r})$, 
which will be introduced later. 
For $\Delta_d=0$, $\Phi_d$ is reduced to the Fermi sea $\Phi_{\rm F}$; 
in this study, we use $\Psi^{\rm F}$ ($={\cal P}\Phi_{\rm F}$) 
as the reference normal state for the energy gain by $\Delta_d$ 
(or $\Delta_{\rm AF}$). 
Furthermore, we fix $t'/t$ in $\varepsilon_{\bf k}$ 
in the wave functions [eq.~(\ref{eq:BCSDelta})] at the same value as 
that in the Hamiltonian, eq.~(\ref{eq:band}), namely, we do not explicitly 
introduce a band renormalization effect due to correlation.  
Because the renormalization of $\varepsilon_{\bf k}$\cite{Himeda-t'} 
brings about a serious effect in the vicinity of half filling,
\cite{Wataorgco} we will treat this issue in upcoming publications. 
As $n$ moves away from half filling, the renormalization effect rapidly 
becomes modest. \cite{Shih,Watat-J}
\par 

In this study, we consider a correlated AF-ordered state 
$\Psi_Q^{\rm AF}(={\cal P}\Phi_{\rm AF})$ independently of 
the $d$-wave state; as a one-body state, we use a Hartree-Fock solution 
on the square lattice, $\Phi_{\rm AF}(\Delta_{\rm AF})$\cite{YS2}: 
\begin{equation}
\Phi_{\rm AF}=\prod_{\bf k}\alpha_{{\bf k}\uparrow}^\dag
                      \prod_{\bf k}\alpha_{{\bf k}\downarrow}^\dag,
\label{eq:AF1}
\end{equation}
with
\begin{equation}
\alpha_{{\bf k}\sigma}^\dag=u_{\bf k}c_{{\bf k}\sigma}^\dag
          +{\rm sgn}(\sigma)\ v_{\bf k}c_{{\bf k+Q}\sigma}^\dag,
\end{equation}
where ${\rm sgn}(\sigma)=1$ or $-1$ according to $\sigma=\uparrow$ or 
$\downarrow$, ${\bf Q}$ is the AF nesting vector $(\pi,\pi)$, and
\begin{equation}
u_{\bf k}(v_{\bf k})=
\sqrt{\frac{1}{2}\left[1-(+)\frac{\gamma_{\bf k}}
{\sqrt{\gamma_{\bf k}^2+\Delta_{\rm AF}^2}}\right]},
\end{equation} 
with $\gamma_{\bf k}=-2t(\cos k_x+\cos k_y)$. 
In the products of eq.~(\ref{eq:AF1}), we take {\bf k}-points in the 
Fermi sea determined by $\varepsilon_{\bf k}$ for convenience of 
optimization; namely, we do not allow for band renormalization, 
similarly to $\Phi_d$. 
\par

\subsection{Conditions of VMC calculations\label{sec:VMCtech}}
In optimizing parameters, we employ a simple method that repeatedly applies 
one-dimensional minimization (Brent method) to every parameter\cite{Numerical} 
with the others fixed, because the number of parameters is small ($\le 6$). 
We call the procedure in which every parameter is optimized once 
a `round' here. 
In most cases, the parameters and energy converge in a few 
rounds of loops except in special cases near 
Mott transitions. 
To reduce the statistical error, we continue the iteration 
for another $15$-$20$ rounds after the convergence, and average the data 
accumulated in this additional process. 
In each loop, we use $2.5\times10^5$ (typical)-$10^6$ fixed 
samples; consequently, the results in optimization are substantially 
averages of several million samples, with typical precision of the 
parameters and energy being on the order of $10^{-3}$ and $10^{-4}t$, 
respectively. 
In the calculations of physical quantities using the optimized parameter 
sets, we averaged at least $2.5\times10^5$ samples. 
\par

To check the system-size dependence, we use systems of 
$N_{\rm s}=L\times L$ sites typically with $L=10$-$16$, imposing 
the periodic($x$)-antiperiodic($y$) boundary conditions to reduce 
level degeneracy. 
In principle, we aim to satisfy the closed-shell condition, because 
open-shell systems bring about serious finite-size effects for small 
values of $U/t$, especially in the SC correlation function. 
However, when we treat the $t'/t$ dependence with other parameters fixed,  
we are obliged to use systems with open shells in addition to those 
with closed shells. 
In this connection, as we vary $t'/t$, the manner of ${\bf k}$-point 
occupation in $\Phi$, especially in $\Phi_{\rm AF}$ and $\Phi_{\rm F}$, 
undergoes discontinuous changes at the values of $t'/t$ specific 
to the system size. 
Consequently, a large and irregular system-size dependence on $t'/t$ 
occurs in most quantities even in large-$U/t$ cases, as we will encounter 
in \S\ref{sec:diag}. 
At any rate, we managed not to be affected by special cases by checking 
different systems. 
\par

\section{Properties for $t'=0$\label{sec:square}}
We discuss the results for $t'=0$ first to grasp the $U/t$ and $\delta$ 
dependences of various quantities. 
In \S\ref{sec:D-H}, we consider the relation of the D-H binding 
to the Mott physics. 
In \S\ref{sec:DelE4}, by analyzing energies, we study the crossover 
of the SC properties, the competition between SC and AF, and the instability 
of the AF state against phase separation. 
In \S\ref{sec:Udep4}, we consider the $U/t$ dependence of correlation 
functions. 
In \S\ref{sec:ndep4}, we argue that, in the underdoped regime, the $\delta$ 
dependence is different between the quantities derived from pairing and 
those from charge itinerancy, and that the strength of SC is represented 
by their product. 
In \S\ref{sec:holons}, a change in the mechanism of conduction is 
discussed. 
\par

\subsection{Relation of D-H binding to Mott physics\label{sec:D-H}}
In Fig.~2 in ref.~\citen{YTOT}, we showed that the D-H binding correlation 
is highly effective for large values of $U/t$ and small doping rates.
Before discussing the main results, here we summarize this feature 
for $t'=0$; the feature for finite $t'/t$ is basically the same as 
that for $t'=0$. 
\par

\begin{figure}[hob] 
\vspace{-0.2cm} 
\begin{center}
\includegraphics[width=6.5cm,clip]{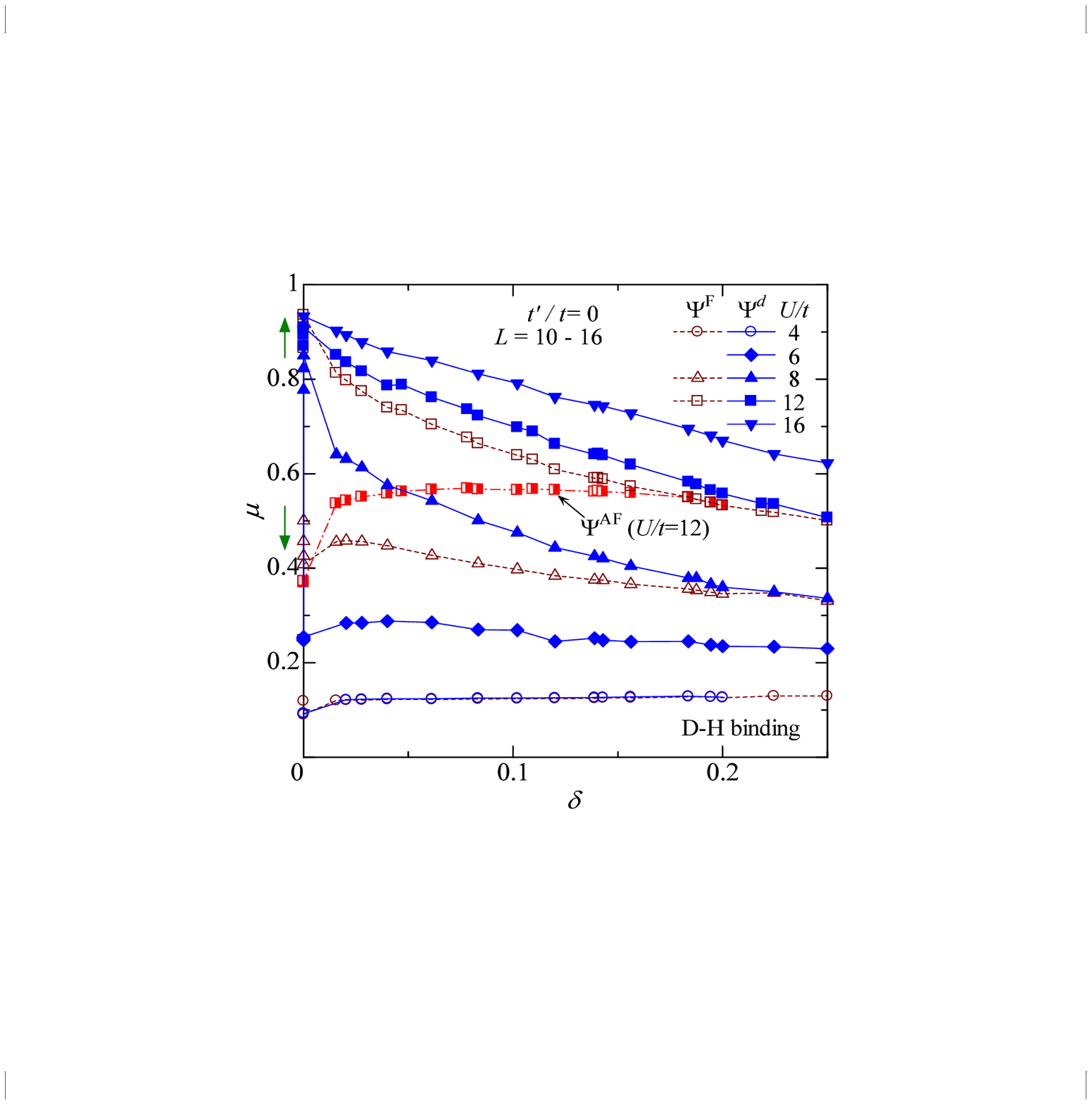} 
\end{center} 
\vskip -4mm 
\caption{(Color online) 
Optimized D-H binding parameters in the normal ($\Psi^{\rm F}_Q$), 
$d$-wave ($\Psi^d_Q$), and antiferromagnetic ($\Psi^{\rm AF}_Q$) states 
as functions of doping rate for some values of $U/t$. 
The arrows near the vertical axis indicate the directions of the system-size 
dependence (as $L$ increases) at $\delta=0$ for $U>U_{\rm c}$ and 
$U<U_{\rm c}$. 
The data for $L=10$-16 are plotted together. 
The Mott transition points are estimated at $U_{\rm c}/t\sim 7$ and $9$ 
for $\Psi_Q^d$ and $\Psi_Q^{\rm F}$, respectively.\cite{YOT} 
}
\label{fig:paramuvsn} 
\end{figure} 
%
Figure \ref{fig:paramuvsn} shows the optimized D-H correlation parameter 
$\mu$ [eqs.~(\ref{eq:SymQ}) and (\ref{eq:DQ})]. 
Generally, as $U/t$ increases, $\mu$ increases, namely, the D-H binding 
becomes tighter, for the normal and SC states. 
The behavior near half filling is sharply distinguished according 
as $U$ is smaller than $U_{\rm c}$ (Mott critical value) or not. 
For $U<U_{\rm c}$, $\mu$ remains small ($\mu\lsim 0.5$) and decreases 
as $L$ increases at $\delta=0$, as indicated by an arrow, whereas 
for $U>U_{\rm c}$, $\mu$ increases steadily as $\delta$ decreases, abruptly 
approaches 1 for $\delta\rightarrow 0$, and increases as $L$ increases. 
Thus, $\mu$ is still relevant to the Mott physics in doped systems. 
In the AF state, the D-H factor is redundant, as discussed 
in ref.~\citen{YTOT}; this redundancy is recognized from the behavior almost 
independent of $\delta$ in Fig.~\ref{fig:paramuvsn}. 
\par

\begin{figure*}[t!] 
\begin{center}
\includegraphics[width=12cm,clip]{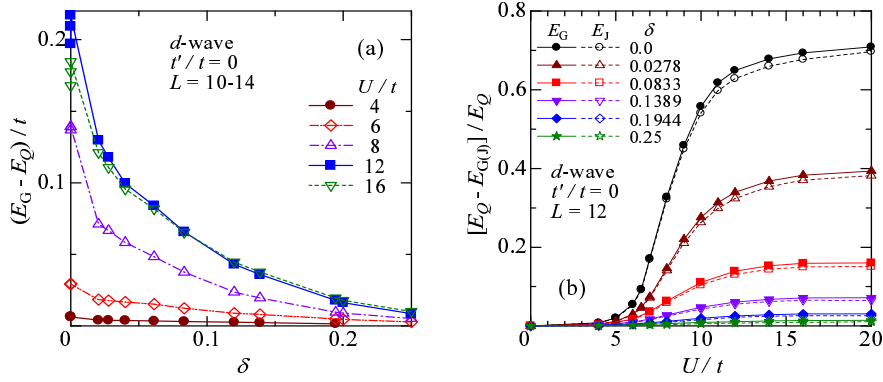} 
\end{center} 
\vskip -4mm 
\caption{(Color online) 
Improvement in energy by D-H binding factor on simple Gutzwiller 
projection in $d$-wave state. 
(a) Energy difference between $\Psi_Q^d$ and $\Psi_{\rm G}^d$ 
as function of doping rate for five values of $U/t$.  
(b) Energy difference between $\Psi_Q^d$ and $\Psi_{\rm G}^d$ normalized 
by $E_Q$ as function of $U/t$ for six doping rates. 
The dashed lines with open symbols indicate the values obtained 
using $\Psi_{\rm J}^d$ instead of $\Psi_{\rm G}^d$. 
}
\vspace{-0.2cm}
\label{fig:compE-GWF} 
\end{figure*}
%
\begin{table}
\caption{
Energy improvements by $\Psi_Q^d$ and $\Psi_{\rm J}^d$ over 
$E_{\rm G}$ of $\Psi_{\rm G}^d$ in percent, i.e., 
$
[E_{Q ({\rm J})}-E_{\rm G}]/E_{\rm G}\times 100,
$
are listed for three values of $U/t$ and $\delta$. 
} 
\vspace{1mm}
\begin{tabular}{c|c|r|r|r}
\hline
$\delta$ & $U/t\rightarrow$ & $4\ \ $ & $8\ \ $ & $12\ \ $ \\
\hline\hline
 0    & $\Psi_Q$       & 0.81 & 48.7 & 184.9  \\
      & $\Psi_{\rm J}$ & 0.08 &  0.7 &   5.8  \\
\hline
0.083 & $\Psi_Q$       & 0.33 & 17.1 &  16.1  \\
      & $\Psi_{\rm J}$ & 0.06 &  0.5 &   0.9  \\
\hline
0.194 & $\Psi_Q$       & 0.15 &  2.8 &   2.6  \\
      & $\Psi_{\rm J}$ & 0.08 &  0.2 &   0.4  \\
\hline
\end{tabular} 
\vspace{-0.5cm}
\label{table:G-Q-J}
\end{table}
%
The effect of the D-H factor is clearer in energy improvement. 
Figure \ref{fig:compE-GWF}(a) shows the difference in total energy 
per site ($E$) between 
$\Psi_Q^d={\cal P}_Q{\cal P}_{\rm G}\Phi_d$ ($E_Q$) and 
$\Psi_{\rm G}^d={\cal P}_{\rm G}\Phi_d$ ($E_{\rm G}$). 
The energy gain by ${\cal P}_Q $ increases as $\delta$ decreases.  
For $U<U_{\rm c}$, however, the gain remains small even at half filling 
(see also Table \ref{table:G-Q-J}). 
This feature is evident in Fig.~\ref{fig:compE-GWF}(b), where the $U/t$ 
dependence is shown; the improvement is negligible for $U/t\lsim 5$. 
In contrast, the energy gain becomes sizable for $U\gsim U_{\rm c}$, 
where $E$ is markedly improved by ${\cal P}_Q$ at $\delta=0$
[Fig.~\ref{fig:compE-GWF}(a)]. 
Although the energy gain by ${\cal P}_Q$ decreases as $\delta$ increases, 
${\cal P}_Q$ still brings appreciable improvement for $0\le\delta\lsim 0.15$. 
This is roughly consistent with experiments on cuprates.\cite{Zheng,Fournier}
For large values of $U/t$, the contribution of ${\cal P}_Q$ amounts to 
$2/3$ of the total $E_Q$ at half filling, and remains at about 7\% even for 
$\delta\sim 0.15$. 
\par

The D-H binding correlation has a close relation to the spin exchange 
coupling, because an NN D-H pair yields an NN antiparallel spin pair 
with a single hopping. 
Correspondingly, the behavior of energy improvement 
in Figs.~\ref{fig:compE-GWF}(a) and \ref{fig:compE-GWF}(b) becomes 
highly similar to that of the spin structure factor $S({\bf q})$ 
at ${\bf q}=(\pi,\pi)$ as will be shown later in Figs.~\ref{fig:sqvsn} and 
\ref{fig:sqnivsu}(a), respectively.\cite{ISS2011} 
Then, the effective range of ${\cal P}_Q$ in the $U/t$-$\delta$ 
space coincides with the range of robust $d$-wave pairing 
[see $\Delta_d$ as will be shown later in Figs.~\ref{fig:delDEvsn}(b) and 
\ref{fig:paradeld-jpsj}], because the $d$-wave pairing is mainly 
mediated by the AF spin correlation. 
Thus, the D-H binding correlation is crucial to the $d$-wave SC, and 
$\Psi_Q^d$ literally represents a doped Mott insulator for $U>U_{\rm c}$. 
\par

Having focused on attractive correlation as intersite factors, 
here we check that intersite repulsive correlation factors are irrelevant 
near half filling. 
For this purpose, we consider the typical NN repulsive Jastrow factor 
\begin{equation}
{\cal P}_{\rm J}=
\rho_{\rm D}^{\hat{\cal N}_{\rm DD}} 
\rho_{\rm H}^{\hat{\cal N}_{\rm HH}}, 
\end{equation}
with $\rho_{\rm D}$ and $\rho_{\rm H}$ being variational parameters, 
and
\begin{equation}
\hat{\cal N}_{\rm DD}=\sum_{j,\tau}d_jd_{j+\tau}, \quad
\hat{\cal N}_{\rm HH}=\sum_{j,\tau}h_jh_{j+\tau},
\end{equation}
where $j$ runs over all the lattice sites, and $\tau$ NN sites 
of site $j$. 
We compare the improvement in energy by 
$\Psi_{\rm J}^d={\cal P}_{\rm J}{\cal P}_{\rm G}\Phi_d$ ($E_{\rm J}$)
with that of $\Psi_Q^d$ over $E_{\rm G}$ in Table \ref{table:G-Q-J}. 
In every case, the energy improvement by $\Psi_{\rm J}^d$ is 
much smaller than that by $\Psi_Q^d$. 
In Fig.~\ref{fig:compE-GWF}(b), we plot $(E_Q-E_{\rm J})/E_Q$ for comparison 
with the case of $E_{\rm G}$; the insignificance of ${\cal P}_{\rm J}$ is 
evident in the present parameter regime. 
Long-range Jastrow factors are also considered irrelevant for 
$\delta\sim 0$, by inferring from the study 
at half filling.\cite{Miyagawa,YMO} 
Henceforth, we disregard intersite repulsive factors. 
\par

\subsection{Energy gain by $d$-wave gap\label{sec:DelE4}}
%
Now, we consider SC properties. 
We begin with the energy gain ($\Delta E$) by the $d$-wave gap 
parameter $\Delta_d$: 
\begin{equation}
\Delta E=E(0)-E(\Delta_d^{\rm opt}), 
\label{eq:DE}
\end{equation}
where $E(\Delta_d)$ denotes the total energy for which the variational 
parameters other than $\Delta_d$ are optimized, and $\Delta_d^{\rm opt}$ 
indicates the optimized $\Delta_d$. 
Although $\Delta E$ has often been called ``condensation energy", 
we call it ``energy gain" or ``energy difference" in this paper, 
because $\Delta E$ does not necessarily correspond to the SC condensation 
energy measured experimentally, as will be discussed in \S\ref{sec:ndep4}. 
\par

\begin{figure}[hob] 
\vspace{-0.2cm}
\begin{center}
\includegraphics[width=7.0cm,clip]{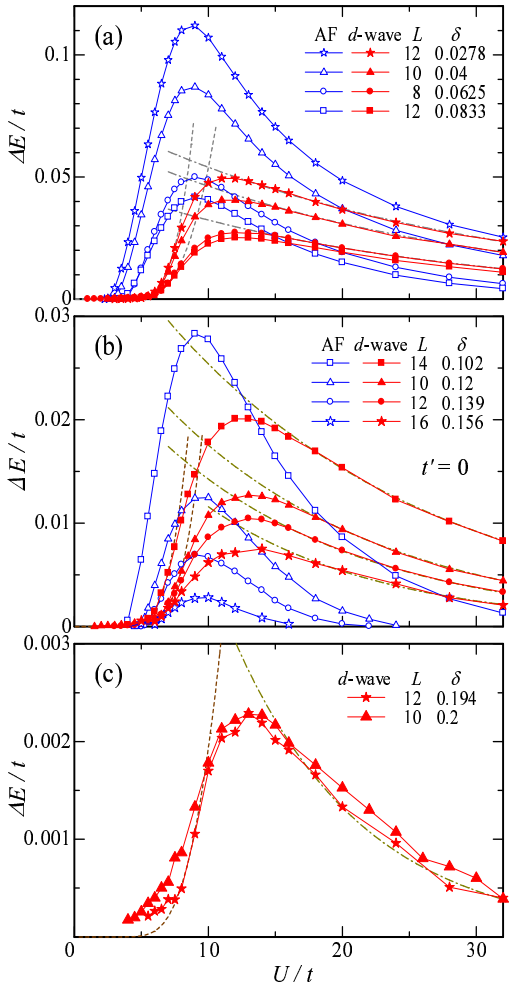} 
\end{center} 
\vskip -4mm 
\caption{(Color online) 
$U/t$ dependences of energy gain $\Delta E/t$ for $d$-wave (solid 
symbols) and AF (open symbols) states for $t'=0$. 
The three panels show different ranges of doping rates: 
(a) underdoped ($0<\delta<0.1$), 
(b) lightly underdoped to optimum-doped ($0.1<\delta<0.16$) and 
(c) overdoped ($\delta\sim 0.2$) regimes. 
The AF state is not stabilized for the values of $\delta$ in (c). 
Note that the energy scales are different among the panels. 
In each panel, the data of the $d$-wave state are fitted with the functions 
$\alpha'\exp(-\beta' t/U)$ [dashed lines] on the weak-correlation side 
($U<U_{\rm co}$) and $\alpha\exp(-U/\beta t)$ [dash-dotted lines] 
on the strong-correlation side ($U>U_{\rm co}$). 
}
\label{fig:DEvsU} 
\end{figure}

\begin{figure}[hob] 
\vspace{-0.2cm}
\begin{center} 
\includegraphics[width=7.5cm,clip]{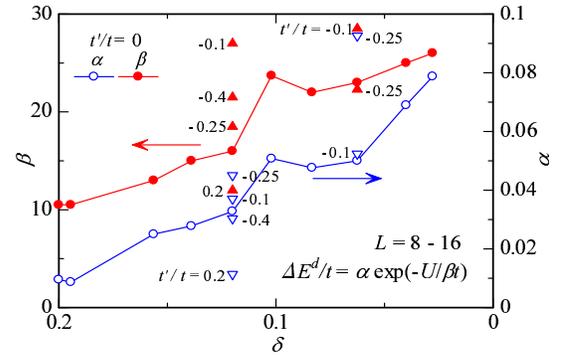} 
\end{center} 
\vskip -5mm 
\caption{(Color online) 
Coefficients $\alpha$ (open symbols) and $\beta$ (closed symbols) 
in fitting function of $\Delta E^d/t$ for $U>U_{\rm co}$ 
[eq.~(\ref{eq:DEfit})] as functions of hole density. 
The circles denote the data for $t'=0$. 
The data for $t'\ne 0$ (triangles) are included with the values of 
$t'/t$. 
}
\label{fig:coeff} 
\end{figure}
%
Figure \ref{fig:DEvsU} shows the $U/t$ dependences of $\Delta E$ 
in the $d$-wave state ($\Delta E^d$) and AF state 
($\Delta E^{\rm AF}$) defined for $\Delta_{\rm AF}$ similarly 
to eq.~(\ref{eq:DE}); panels (a), (b), and (c) show the data 
for the underdoped, optimum-doped, and overdoped values of $\delta$, 
respectively. 
As discussed in ref.~\citen{YTOT}, $\Delta E^d$ is negligible for 
$U\lsim 6t$, but abruptly increases as $U/t$ increases, and is roughly 
fitted as $\Delta E^d/t\sim \alpha'\exp(-\beta' t/U)$\cite{note-Kondo}
at approximately $U=W(=8t)$ ($\alpha'$, $\beta'$: constants). 
$\Delta E^d$ has a maximum at $U\sim U_{\rm co}=10t$-$12t$, then 
decreases slowly as it is fitted well with the curve 
\begin{equation}
\Delta E^d/t=\alpha\exp\left(-\frac{U}{\beta t}\right), 
\label{eq:DEfit}
\end{equation}
for $2W\lsim U<4W$. 
The estimated values of the constants $\alpha$ and $\beta$ are shown in 
Fig.~\ref{fig:coeff} as functions of $\delta$. 
From the form of eq.~(\ref{eq:DEfit}), we consider that the attractive 
pairing potential in this regime is proportional to $t^2/U (=J/4)$; 
this indicates that the low-energy physics in this regime is parallel 
to that of the corresponding $t$-$J$-type models.\cite{YTOT,note-strong}
Thus, we conclude that the properties of SC undergo a crossover 
from a BCS type to an unconventional $t$-$J$ type at $U\sim U_{\rm co}$ 
(this is a loose definition of $U_{\rm co}$). 
As long as $\Psi_Q^d$ is SC, this behavior is qualitatively independent 
of $\delta$, although the magnitude of $\Delta E^d$ decreases with $\delta$. 
\par 

\begin{figure}[htb]
\begin{center}
\includegraphics[width=7.5cm,clip]{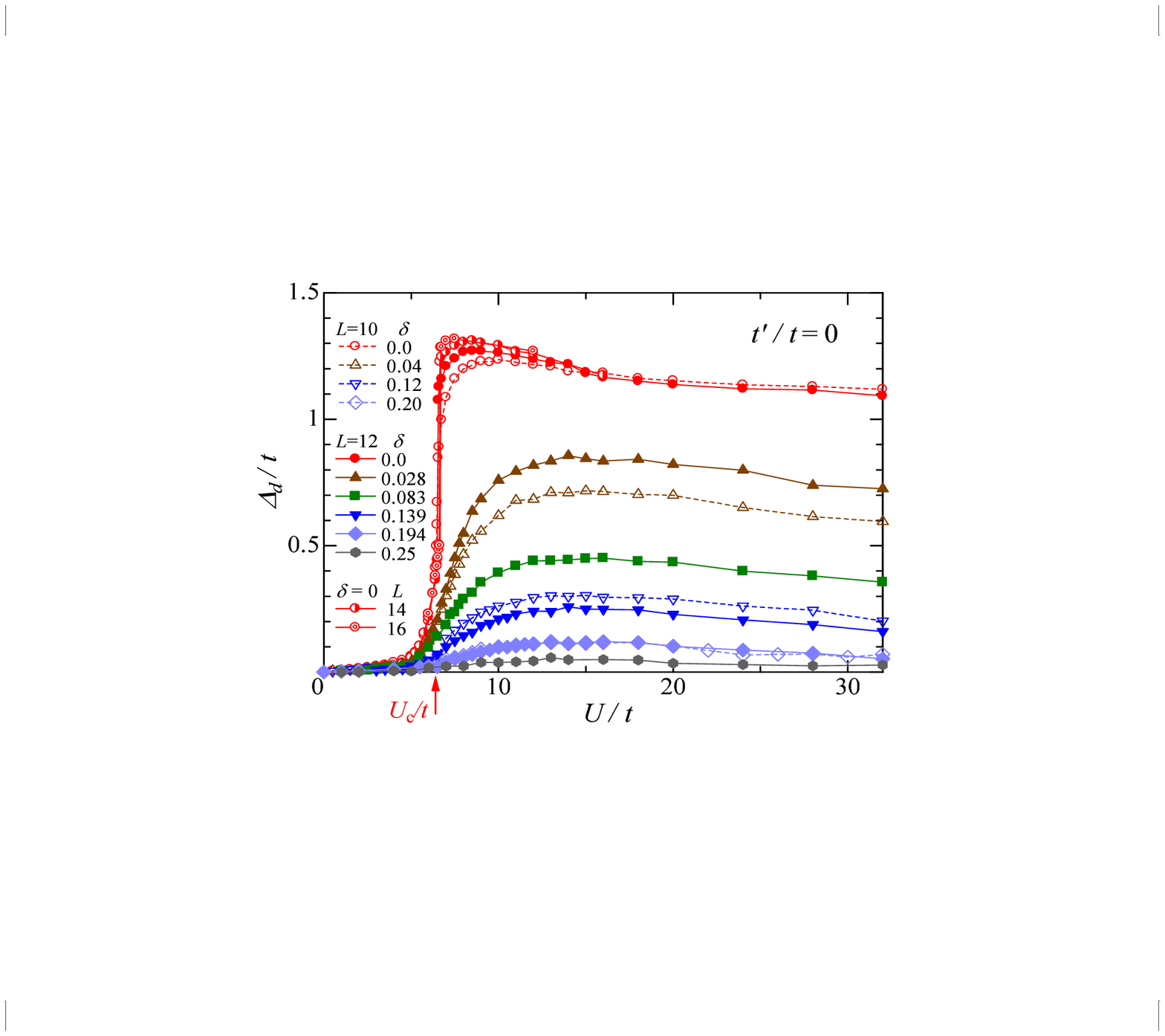}
\end{center} 
\vskip -3mm 
\caption{(Color online)
Optimized $d$-wave gap parameter in $\Psi_Q^d$ as function of $U/t$ 
for different values of $\delta$ (and $L$). 
}
\vskip -3mm 
\label{fig:paradeld-jpsj} 
\end{figure} 
%
As is evident from eq.~(\ref{eq:DE}), $\Delta E^d$ originates 
from a finite $d$-wave gap. 
Actually, the behavior of $\Delta E^d$ closely corresponds to that 
of the optimized $\Delta_d$ shown in Fig.~\ref{fig:paradeld-jpsj}; 
$\Delta_d$ abruptly starts to increase at $U\sim U_{\rm c}$, 
where D-H binding also becomes effective, as mentioned. 
\par 

Now we compare $\Delta E^d$ with $\Delta E^{\rm AF}$. 
As discussed in refs.~\citen{YTOT} and \citen{YOT}, $\Delta E^{\rm AF}$ 
is also maximum 
at $U\sim W$ ($U<U_{\rm co}$) and has a tail for large $U/t$. 
At half filling, $\Delta E^{\rm AF}$ is always larger than 
$\Delta E^d$ [Fig.~3 in ref.~\citen{YTOT}], although $\Delta E^{\rm AF}$ 
decreases rapidly for large $U/t$. 
Almost as soon as carriers are doped, $\Delta E^{\rm AF}$ is surpassed 
for large $U/t$ [Fig.~\ref{fig:DEvsU}(a)], and tends 
to vanish at a certain $U/t$ in the optimally doped region 
[Fig.~\ref{fig:DEvsU}(b)]. 
As $\delta$ increases, $\Delta E^{\rm AF}$ becomes smaller than 
$\Delta E^d$ ($\delta=0.156$) for any $U/t$, and finally vanishes 
for $\delta\sim 0.2$ [Fig.~\ref{fig:DEvsU}(c)]. 
The phase diagram constructed from these data [Fig.~4(a) in ref.~\citen{YTOT}] 
shows that the domain of $d$-wave SC tends to expand to $\delta=0$ 
and that of the AF state shrinks as $U/W$ increases, supporting 
the result of the $t$-$J$ model, in which the stable state switches 
from an AF state to a $d$-wave state as soon as holes are doped.\cite{YO} 
On the other hand, the AF state seems predominant at $U\sim W$. 
In fact, however, the AF domain in the phase diagram probably vanishes 
for $\delta\ne 0$ (unless a $d$-wave SC order 
coexists\cite{Himeda-co,Kobayashi-co}) 
in accordance with the actual behavior of cuprates, because the AF state 
for a finite $\delta$ is unstable against phase separation, as we will 
discuss shortly. 
\par

\begin{figure*}[!t]
\begin{center}
\includegraphics[width=13cm,clip]{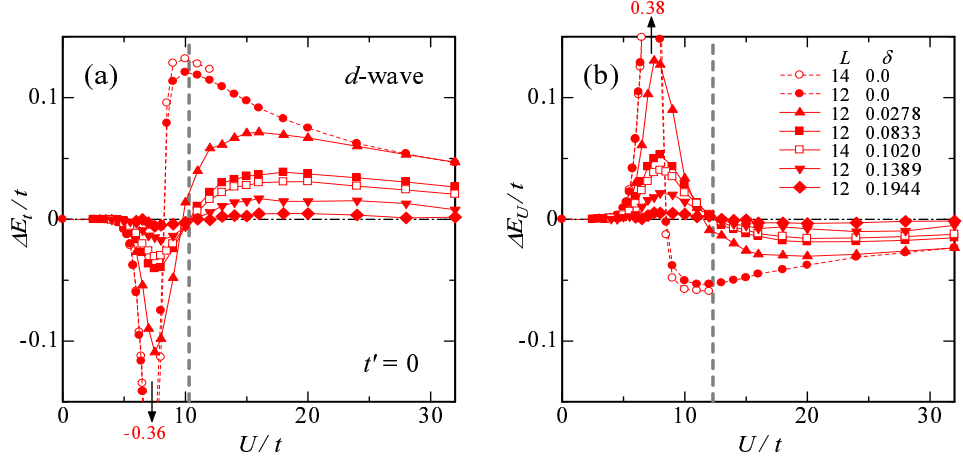} 
\vskip -4mm
\caption{(Color online) 
Components [(a) kinetic and (b) interaction parts] of energy gain 
$\Delta E^d$ as functions of $U/t$. 
The broad gray dashed lines roughly indicate the points where the signs 
of $\Delta E_t$ and $\Delta E_U$ are reversed.
\label{fig:deteudwave}} 
\end{center} 
\end{figure*} 
\begin{figure*}[!t] 
\vspace{-0.2cm}
\begin{center} 
\includegraphics[width=13cm,clip]{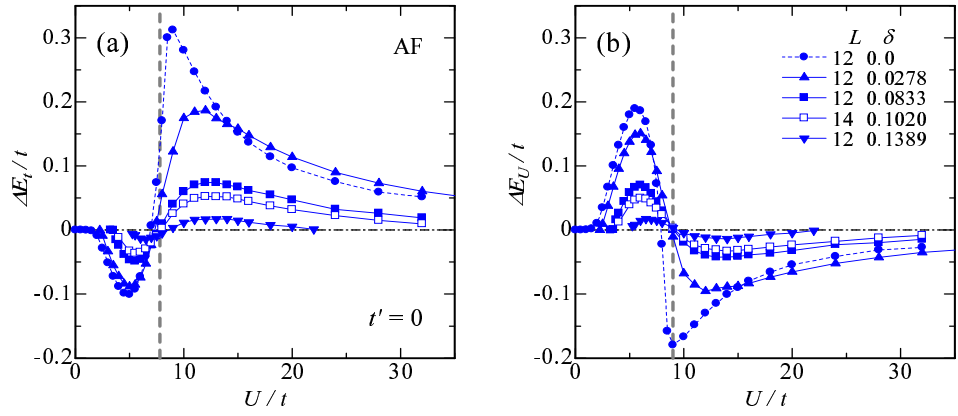}
\end{center} 
\vskip -4mm
\caption{(Color online) 
Components [(a) kinetic and (b) interaction parts] of energy gain 
$\Delta E^{\rm AF}$ as functions of $U/t$. 
The broad gray dashed lines show the same as those in 
Fig.~\ref{fig:deteudwave}.
}
\vskip -4mm
\label{fig:deteuAF} 
\end{figure*} 

Let us turn to the behavior of the components of $\Delta E$ 
($=\Delta E_t+\Delta E_U$). 
In Figs.~\ref{fig:deteudwave}(a) and \ref{fig:deteudwave}(b), 
the kinetic and interaction parts of $\Delta E^d$ are shown 
as functions of $U/t$. 
For small $U/t$ ($\lsim 10$), the energy gain is derived from the 
interaction part $\Delta E_U$ with a loss in the kinetic part 
$\Delta E_t$. 
This mechanism is identical to that of the BCS theory. 
On the other hand, for large $U/t$ ($\gsim 12$), the SC state is 
stabilized by the reduction in $\Delta E_t$. 
As discussed in ref.~\citen{YTOT}, in this regime of $U/t$, 
the quasi-particle renormalization factor $Z$ in the nodal direction 
is enhanced by introducing $\Delta_d$, suggesting that 
the SC coherence promotes the motion of electrons. 
In particular, local hoppings that create or annihilate doublons 
(cf. \S\ref{sec:holons}) are more enhanced in the SC state. 
Thus, we conclude that the so-called ``kinetic-energy-driven SC" is 
realized only for $U>U_{\rm co}$. 
A FLEX calculation for the Hubbard model also supports this behavior, 
although $U_{\rm co}/t$ is small and a sharp crossover is not 
obtained.\cite{Yanase} 
Because $E_t$ is proportional to the sum of optical conductivity 
$\sigma_1(\omega)$ (exactly for $t'=0$),\cite{Maldague} it is 
possible that precise optical measurements can determine whether or not 
cuprates belong to the regime of the $t$-$J$ model. 
\par

Incidentally, similar behaviors of $\Delta E_U$ and $\Delta E_t$ 
are found for the AF state as shown in Fig.~\ref{fig:deteuAF}, and 
for an $s$-wave SC state in the 2D attractive Hubbard 
model.\cite{Tamura} 
Accordingly, it is possibly a common tendency that nonorder-to-order 
transitions are induced by the reduction in kinetic energy in 
strong-correlation regimes.\cite{Nagaoka} 
\par

\begin{figure*}[!t]
\vspace{0.2cm} 
\begin{center}
\includegraphics[width=13.0cm,clip]{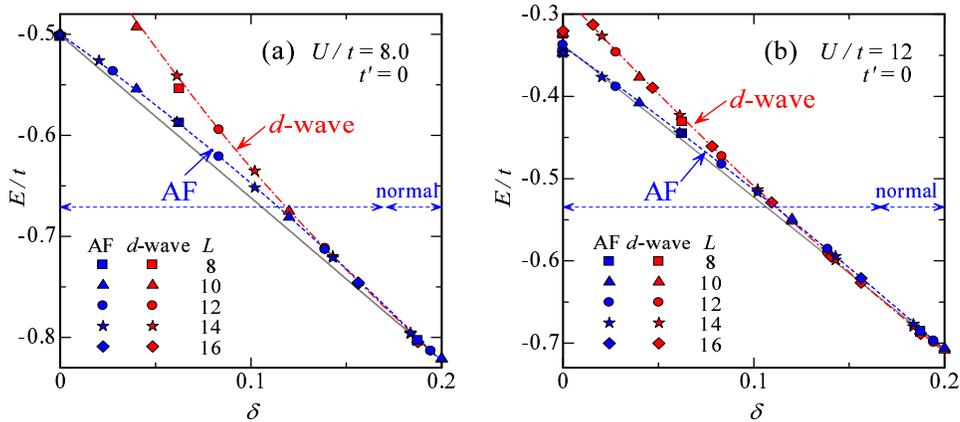}
\vskip -2mm
\caption{(Color online)
Total energy as function of doping $\delta$ of AF and $d$-wave states
for (a) $U/t=8$ and (b) $U/t=12$. 
The AF state is reduced to a normal state for $\delta\gsim 0.18$, 
as shown by horizontal dashed arrows. 
The solid lines are straight guide lines. 
}
\vskip -2mm
\label{fig:EvsnAFt2-0} 
\end{center} 
\end{figure*}

Finally, we discuss possible phase separation. 
This subject is important in relation to the density modulations or 
phase separation reported in underdoped cuprates,\cite{inhomogeneous} 
and has been theoretically pursued mainly using the $t$-$J$ model 
with small $J/t$.\cite{PSt-J}
A homogeneous state is unstable against phase separation if the charge 
compressibility 
\begin{equation}
\kappa=\frac{1}{n^2}\chi_{\rm c}
=\left(n^2\frac{\partial^2E}{\partial n^2}\right)^{-1} 
\label{eq:kappa}
\end{equation} 
is negative, namely, the total energy $E/t$ is convex as a function of 
electron density $n$. 
In order to check this possibility, we show, in Fig.~\ref{fig:EvsnAFt2-0}, 
the total energies of $\Psi_Q^d$ and $\Psi_Q^{\rm AF}$ as functions 
of $\delta$. 
For both $U/t=8$ and 12, the $E/t$ of the $d$-wave state exhibits a concave 
curve, namely, $\kappa>0$, meaning the $d$-wave state has intrinsic stability. 
In contrast, the AF state exhibits a convex curve, $\kappa<0$, 
indicating that the AF state of partial filling gives rise to a phase 
separation of the insulating AF state with local filling $\delta=0$ 
and a SC (or normal) state with local filling $\delta\gsim 0.2$, 
according to the Maxwell construction. 
A similar result was reached by a recent QMC study 
for $t'=0$.\cite{Chang}
This instability does not occur in electron-doped cases as will 
be discussed for $t'\ne 0$ in \S~\ref{sec:dandAF} or in the 
coexistence of the SC and AF orders.\cite{Himeda-co,Kobayashi-co} 
\par

\subsection{$U/t$ dependence of correlation functions\label{sec:Udep4}} 
To measure the strength of SC, the $d_{x^2-y^2}$-wave pairing correlation 
function 
\begin{eqnarray}
P_d({\bf r})=&&\frac{1}{N_{\rm s}}
\sum_{i}\sum_{\tau,\tau'=\hat {\bf x},\hat {\bf y}}
(-1)^{1-\delta(\tau,\tau')}\times\qquad \nonumber\\ 
&& \left\langle{\Delta _\tau^\dag({\bf R}_i)\Delta_{\tau'}
({\bf R}_i+{\bf r})}\right\rangle 
\label{eq:pd}
\end{eqnarray}
is appropriate for the present method. 
Here, $\hat{\bf x}$ and $\hat{\bf y}$ denote the lattice vectors 
in the $x$- and $y$-directions, respectively, $\delta(\tau,\tau')$ 
indicates the Kronecker delta, and 
$\Delta_\tau^\dag({\bf R}_i)$ is the creation operator of an 
NN singlet pair at site ${\bf R}_i$, 
\begin{equation}
\Delta_\tau^\dag({\bf R}_i)=
(c_{{i}\uparrow}^\dag c_{{i}+\tau\downarrow}^\dag+ 
 c_{{i}+\tau\uparrow}^\dag c_{{i}\downarrow}^\dag)
 /{\sqrt 2}. 
\label{eq:singlet}
\end{equation}
If $P_d({\bf r})$ remains finite for 
$|{\bf r}|\rightarrow\infty$, a $d$-wave off-diagonal long-range order 
exists. 
As explained in Appendix~\ref{sec:pairfunc}, we found a suitable scheme 
for reliably estimating $P_d({\bf r)}$ for $|{\bf r}|\rightarrow\infty$
($P_d^\infty$) from the data of finite $L$. 
We confirmed that $P_d^\infty$ converges to zero for $\Psi_Q^{\rm F}$ 
and $\Psi_Q^{\rm AF}$ as $L$ increases. 
Henceforth, we discuss $P_d^\infty$. 
\par

\begin{figure}[htb]
\begin{center}
\includegraphics[width=8.5cm,clip]{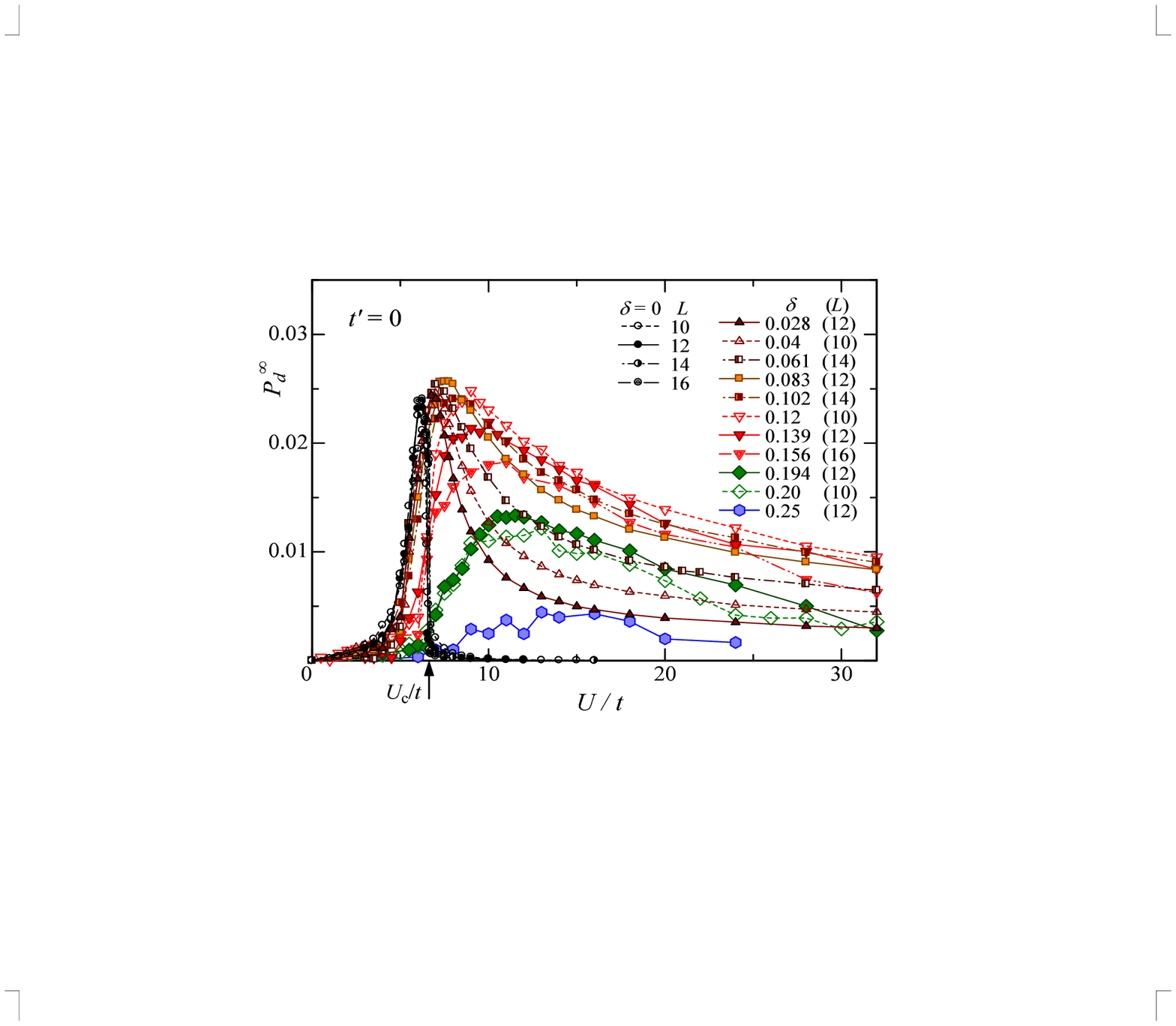}
\end{center} 
\vskip -3mm 
\caption{(Color online) 
$d$-wave SC correlation function as function of $U/t$ computed 
with $d$-wave states for various doping rates. 
Circles indicate half filling ($\delta=0$); similarly, 
upward triangles ($\bigtriangleup$) $\delta\sim 0.03$,
squares ($\Box$) $\delta\sim 0.09$, 
downward triangles ($\bigtriangledown$) $\delta\sim 0.14$, 
diamonds $\delta\sim 0.20$, and hexagons $\delta=0.25$. 
The arrow indicates the Mott critical value at half filling. 
}
\vskip -3mm 
\label{fig:pdvsut2-0} 
\end{figure}

Figure \ref{fig:pdvsut2-0} shows the $U/t$ dependence of $P_d^\infty$ 
calculated with the optimized $\Psi_Q^d$ for some values of $\delta$. 
The overall features in the doped cases ($\delta>0$) resemble those of 
$\Delta E/t$ (Fig.~\ref{fig:DEvsU}), especially for the optimal and 
overdoped regimes ($\delta\gsim 0.15$). 
A large difference between $P_d^\infty$ and $\Delta E/t$ emerges 
as $\delta$ approaches 0 (e.g., $\delta=0.028$); $P_d^\infty$, compared 
with $\Delta E/t$, decreases rapidly as $U/t$ increases. 
The extreme case is half filling, at which $P_d^\infty$ vanishes for 
$U>U_{\rm c}$, corresponding to the Mott transition, as discussed 
in ref.~\citen{YOT}. 
Thus, the value of $U/t$ at the maximum $P_d^\infty$, which is a better 
definition 
of $U_{\rm co}/t$, decreases to the Mott critical value $U_{\rm c}/t$ 
as $\delta$ decreases. 
We leave the analysis of the $\delta$ dependence for the next subsection. 
\par

\begin{figure}[hob]
\begin{center}
\includegraphics[width=8.5cm,clip]{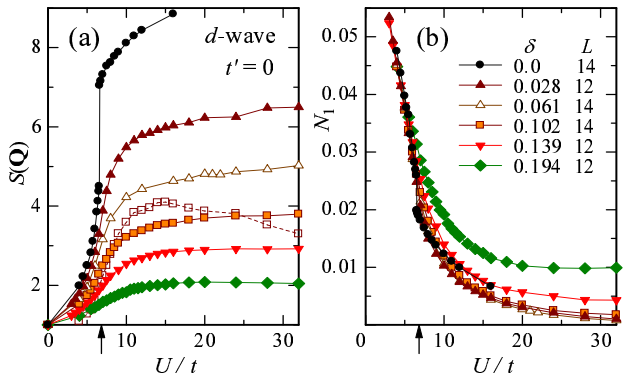}
\vskip -2mm
\caption{(Color online)
(a) Spin structure factor $S({\bf q})$ at ${\bf q}=(\pi,\pi)$ 
for six electron densities. 
For comparison, the behavior of the optimized $d$-wave gap parameter 
$\Delta_d/t$ (exactly $8\Delta_d/t+1$) is also shown for $\delta=0.102$ 
(open squares and dashed line). 
(b) Nearest-neighbor charge correlation function as function of $U/t$.
The symbols for $\delta$ and $L$ are common between the two panels. 
The arrows indicate the Mott transition point at half filling. 
} 
\label{fig:sqnivsu} 
\end{center} 
\end{figure} 

For SC in the present case, the spin structure factor 
\begin{equation} 
S({\bf q})=\frac{1}{N_{\rm s}}\sum_{ij}{e^{i{\bf q}
\cdot({\bf R}_i-{\bf R}_j)} 
\left\langle{S_{i}^zS_{j}^z}\right\rangle} 
\label{eq:sq}
\end{equation} 
is an important quantity, because the electron pair scattering with 
${\bf Q}=(\pi,\pi)$ will contribute to the $d_{x^2-y^2}$-wave pairing. 
In Fig.~\ref{fig:sqnivsu}(a), the $U/t$ dependence of $S({\bf Q})$ 
calculated with $\Psi_Q^d$ is plotted for six values of $\delta$. 
As discussed in ref.~\citen{YOT}, $S({\bf Q})$ increases with $U/t$ 
and, in particular, increases discontinuously at $U_{\rm c}/t$ at half filling 
in accordance with the first-order Mott transition. 
For $\delta>0$, $S({\bf Q})$ still rapidly increases for 
$U_{\rm c}\lsim U\lsim U_{\rm co}$, although the discontinuity vanishes. 
The increase in $S({\bf Q})$ for $U<U_{\rm co}$ coincides with the 
behavior of $P_d^\infty$, but a further increase in $S({\bf Q})$ 
for $U>U_{\rm co}$ is opposed to the decrease in $P_d^\infty$. 
This is because SC also depends on the mobility of electrons, 
represented, for example, by the quasi-particle renormalization factor 
$Z$ [Fig.~6 in ref.~\citen{YTOT}]. 
\par

Similar decreasing behavior is found in the charge correlation function 
between NN sites, 
\begin{equation}
N_1=\left|\frac{1}{4N_{\rm s}}
\sum_{j,\tau}\left\langle{n_{j} n_{j+\tau}}\right\rangle 
- n^2\right|, 
\label{eq:N1}
\end{equation} 
as shown in Fig.~\ref{fig:sqnivsu}(b). 
Here, $\tau$ runs over the four NN sites of site $j$. 
$N_1$ represents NN charge fluctuation and is related to the mobility 
of electrons. 
The decrease in $N_1$ and $Z$, showing the suppression of charge fluctuation 
by $U/t$, resembles the behavior of $P_d^\infty$ for $U>U_{\rm c}$. 
\par

The above argument implies that the strength of SC requires two factors, 
namely, pair formation owing to AF spin correlation and the fluidity 
of pairs owing to charge fluctuation. 
We will argue this topic again in the light of the $\delta$ dependence, 
which is 
controllable in experiments on cuprates, in the next subsection. 
\par

\subsection{Doping-rate dependence of various quantities\label{sec:ndep4}} 
\begin{figure}[htb]
\begin{center} 
\includegraphics[width=7.0cm,clip]{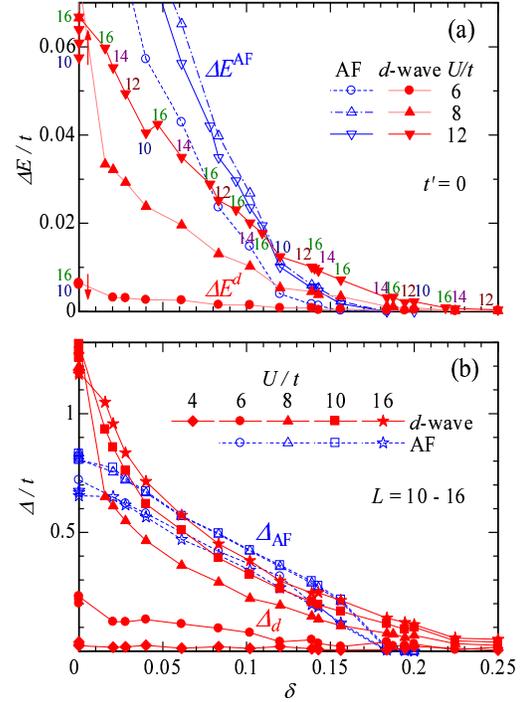}
\end{center}
\vskip -2mm
\caption{(Color online)
(a) Energy gain $\Delta E/t$ in $d$-wave and AF states versus doping 
rate. 
The numbers near the data points indicate the system sizes ($L$) used. 
The arrows alongside the vertical axis denote the direction of the 
system-size dependence at $\delta=0$ 
(downward: $U/t=6$; upward: $U/t=8$ and 12). 
(b) Optimized gap parameters in $d$-wave ($\Delta_d$) and AF 
($\Delta_{\rm AF}$) states as functions of doping rate. 
} 
\vskip -2mm
\label{fig:delDEvsn} 
\end{figure}

First, we examine the energy gain of the $d$-wave state $\Delta E^d$ 
[eq.~(\ref{eq:DE})]. 
In Fig.~\ref{fig:delDEvsn}(a), we show the $\delta$ dependence of 
$\Delta E^d$; $\Delta E^d$ is largest at half filling and monotonically 
decreases as $\delta$ increases for any $U/t$.
Figure \ref{fig:delDEvsn}(b) shows the $\delta$ dependence of the optimized 
$d$-wave gap $\Delta_d$, which is the sole parameter directly controlling 
the energy scale of a singlet gap in $\Psi_Q^d$. 
It is natural to consider that $\Delta_d$ corresponds to the 
pseudogap.\cite{ZGRS,Randeria} 
$\Delta_d$ is highly similar to $\Delta E^d$ as in the $U/t$ 
dependence. 
Thus, we again find that $\Psi_Q^d$ is stabilized by the $d$-wave 
singlet formation. 
We will see the relation of these quantities to the spin correlation 
shortly.   
\par

\begin{figure}[htb]
\begin{center}
\includegraphics[width=8.0cm,height=5.5cm]{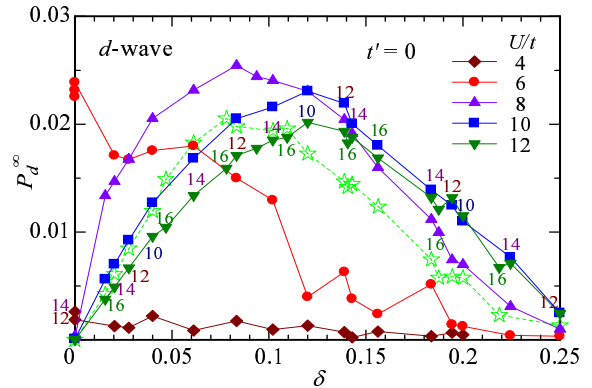} 
\end{center} 
\vskip -3mm
\caption{(Color online)
$d$-wave SC correlation function vs doping rate for several 
values of $U/t$ in $\Psi_Q^d$. 
The numbers (10-16) denote the system sizes $L$ used. 
The stars denote $\Delta_{\rm SC}^2$ for $U/t=12$ calculated 
with eq.~(\ref{eq:DeltaSC}) using $\Delta_d$ obtained by VMC. 
The magnitude of $\Delta_{\rm SC}^2$ is adjusted to be equal to 
the corresponding $P_d^\infty$. 
} 
\vskip -2mm
\label{fig:Pdvsn} 
\end{figure} 
%
Next, we consider the $d$-wave SC correlation $P_d^\infty$; its $\delta$ 
dependence is shown in Fig.~\ref{fig:Pdvsn}. 
For $U/t=4$, the magnitude of $P_d^\infty$ is very small and has relatively 
strong $L$ and $\delta$ dependences, indicating that firm SC is unlikely 
to appear for $U/t=4$. 
This result is consistent with those of QMC 
calculations,\cite{QMC,Aimi,Yanagisawa} in which the increase 
in SC correlation as $U/t$ increases is not found for small values of $U/t$. 
When $U/t$ is slightly below the Mott transition point $U_{\rm c}/t$ 
$(\sim 7)$, 
as in $U/t=6$, $P_d^\infty$ has the maximum at half filling, and is 
basically a decreasing function of $\delta$, except for the fluctuation 
in $L$ (and $\delta$). 
This strong system-size dependence suggests that the SC is still fragile. 
These results for $U<U_{\rm c}$ indicate that $T_{\rm c}$ does not 
exhibit a dome shape in the weakly correlated Hubbard model, although 
the approximate methods claim such behavior.\cite{Weak,Moriya}
\par

In contrast, $P_d^\infty$ for $U>U_{\rm c}$ displays a dome shape 
as a function of $\delta$, which is caused by the vanishing of 
$P_d^\infty$ at half filling as a Mott insulator. 
The shape of $P_d^\infty$, especially for $U/t=12$, closely resembles 
that of the SC order parameter for the NN pairing 
$|\langle c_{0\uparrow}^\dag c_{\tau\downarrow}^\dag\rangle|$ 
obtained in the $t$-$J$ model,\cite{YStJ2} and is consistent with 
the experimental results of $T_{\rm c}$ and condensation energy 
in cuprates. 
For these values of $U/t$, SC is considered robust owing to the weak 
system-size dependence. 
This result indicates that the effective correlation strength in cuprates 
is high, i.e., $U>U_{\rm c}$. 
Thus, high-$T_{\rm c}$ cuprates are literal ``doped Mott 
insulators".\cite{Anderson,PALee,OF}
\par

\begin{figure}[htb]
\begin{center}
\vskip 2mm
\includegraphics[width=8.7cm,clip]{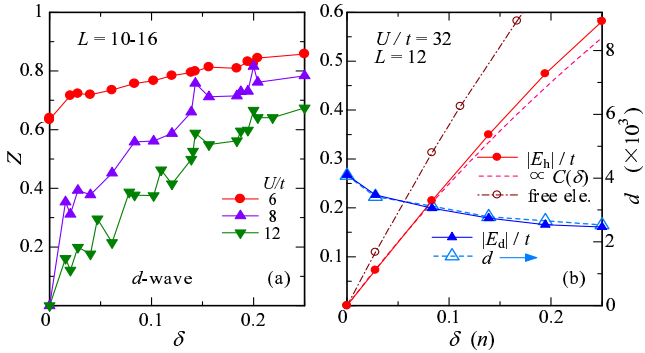} 
\end{center} 
\vskip -4mm
\caption{(Color online)
(a) Quasi-particle renormalization factor in nodal direction versus doping 
rate for $d$-wave state. 
(b) Absolute values of $E_{\rm d}$ and $E_{\rm h}$ are shown as functions 
of $\delta$ by solid symbols in a strongly correlated case ($U/t=32$). 
For comparison, we add the doublon density (open triangles, right axis), 
a guide line $\propto C(\delta)=2\delta/(1+\delta)$ (dashed line), 
and the energy of free electrons as a function of electron density $n$ 
(open circles).
}
\vskip -4mm
\label{fig:Zvsneci} 
\end{figure}
%
For $U>U_{\rm c}$, $P_d^\infty$ increases almost linearly with $\delta$ 
in the underdoped regime, in contrast to quantities such as $\Delta_d/t$ 
and $\Delta E^d/t$, which are monotonically decreasing functions 
of $\delta$. 
As mentioned, this is because the strength of SC depends on two factors, 
singlet-pair formation ($\Delta_d$) and quantities related to charge 
transport such as carrier density and the mobility of carriers. 
The latter quantities are bound to increase as $\delta$ increases for 
$U>U_{\rm c}$, being released from the suppression of charge fluctuation 
in Mott insulators. 
As an example, in Fig.~\ref{fig:Zvsneci}(a), we show the quasi-particle 
renormalization factor $Z$ estimated from the jumps in $n({\bf k})$ 
in the nodal direction. 
Since $Z$ roughly represents the inverse effective mass, $Z$ is zero 
at half filling and monotonically increases as $\delta$ increases 
for $U>U_{\rm c}$, in accordance with the result for the $t$-$J$ 
model.\cite{Randeria} 
$N_1$ [eq.~(\ref{eq:N1})] and the conductive part of $E_t$ discussed 
in \S\ref{sec:holons} [$|E_h|$ in Fig.~\ref{fig:Zvsneci}(b)] are 
also increasing functions of $\delta$. 
\par

To consider the $\delta$ dependence of SC strength, it is useful to mention 
a simple analytic calculation. 
In a pioneering study using a Gutzwiller-type approximation for the $t$-$J$ 
model,\cite{ZGRS} the relation 
\begin{equation}
\Delta_{\rm SC}=\frac{2\delta}{1+\delta}\Delta_d 
\label{eq:DeltaSC}
\end{equation}
was derived; here, the energy scale of SC ($\Delta_{\rm SC}$) originates 
solely from $\Delta_d$ ($d$-wave singlet gap) but is modified 
by the factor of the carrier density. 
As an example, we estimate $\Delta_{\rm SC}^2$ for $U/t=12$ using 
$\Delta_d$ calculated by VMC, and the result is plotted 
in Fig.~\ref{fig:Pdvsn} by open stars. 
$\Delta_{\rm SC}^2$ qualitatively agrees with $P_d^\infty$ for $U/t=12$. 
A similar dome shape is also obtained from $Z\Delta_d$, instead of 
eq.~(\ref{eq:DeltaSC}). 
Note that in slave-boson-mean-field theory,\cite{SBMFT,PALee} 
$T_{\rm c}$ in the underdoped regime is determined by the Bose condensation 
temperature $T_{\rm B}$ ($\propto\delta$) of holons, which represent 
the charge part of carriers. 
\par

\begin{figure}[hob]
\begin{center} 
\vspace{-4mm}
\includegraphics[width=8.0cm,clip]{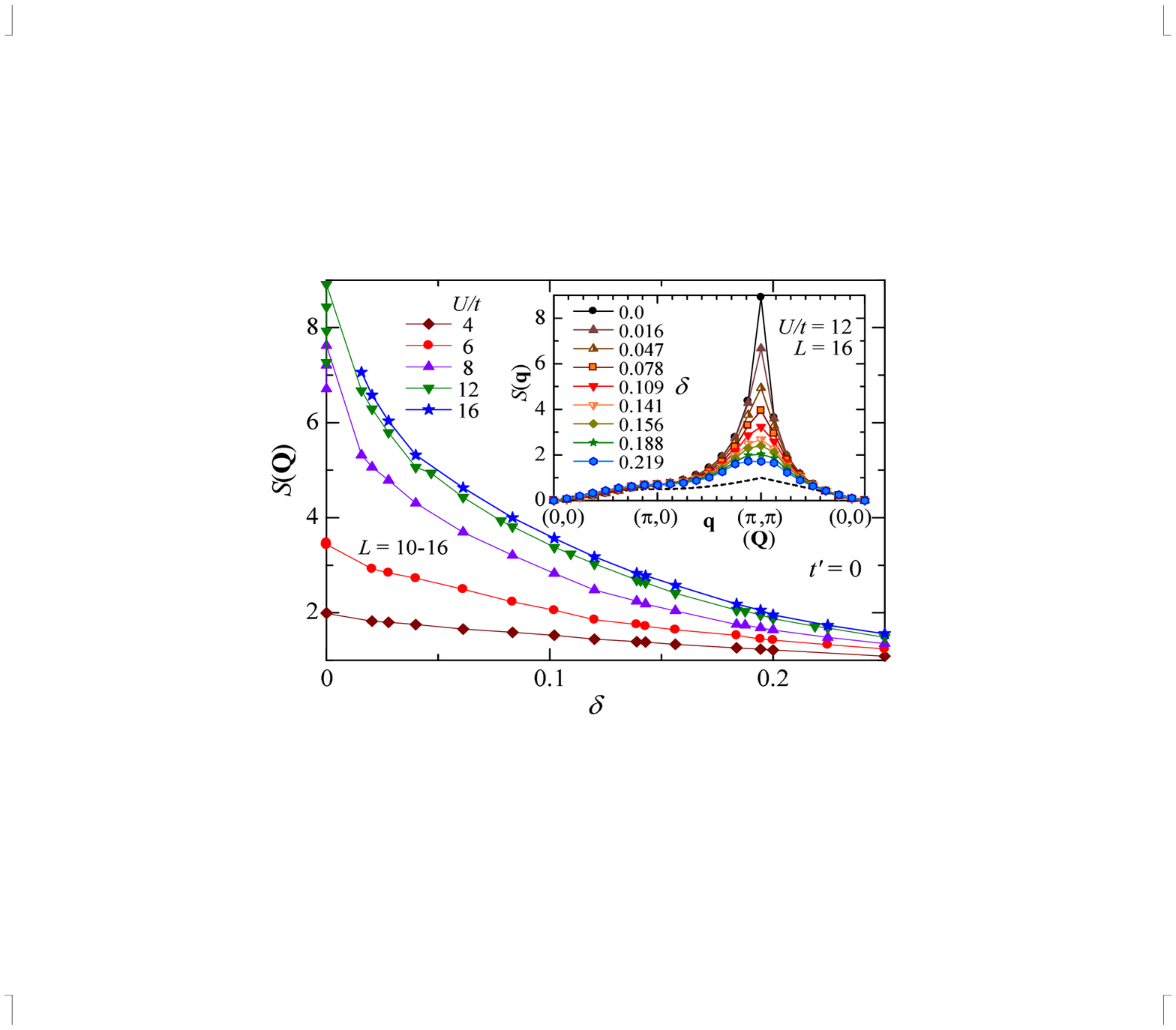} 
\vskip -2mm
\caption{(Color online)
Spin structure factor at ${\bf q}=(\pi,\pi)$ in $d$-wave state as function 
of doping rate for five values of $U/t$. 
The inset represents $S({\bf q})$ at $U/t=12$ for various doping rates 
along the path $(0,0)\rightarrow(\pi,0)\rightarrow(\pi,\pi)\rightarrow(0,0)$. 
The dashed line shows the case of $U=0$ and $\delta=0$. 
} 
\vskip -2mm
\label{fig:sqvsn} 
\end{center}
\end{figure}
%
We turn to the spin structure factor $S({\bf q})$. 
In the inset of Fig.~\ref{fig:sqvsn}, $S({\bf q})$ of $\Psi_Q^d$ 
at $U/t=12$ is depicted for various $\delta$. 
The sharp peak at ${\bf q}={\bf Q}$ near half filling confirms a predominant 
AF spin correlation, and $S({\bf q})$ preserves the maximum at ${\bf Q}$ 
for $\delta\lsim 0.2$. 
The main panel of Fig.~\ref{fig:sqvsn} shows the $\delta$ dependence 
of $S({\bf Q})$ for five values of $U/t$; $S({\bf Q})$ is a decreasing 
function of $\delta$ for any correlation strength.
The small system-size dependence indicates the short-range nature 
of $S({\bf Q})$. 
We here reiterate the close connection between the spin correlation and 
the D-H binding correlation (Fig.~\ref{fig:compE-GWF}); the $\delta$ 
dependences 
of $\Delta E^d$, $\Delta_d$, and $S({\bf Q})$ closely resemble one another. 
This suggests a strong correlation between $\Delta_d$ and the AF spin 
correlation, namely, the energy of $\Psi_Q^d$ is probably reduced 
by the formation of $d$-wave singlet pairs through the AF spin correlation. 
\par

\begin{figure}[hob]
\begin{center}
\includegraphics[width=8.7cm,clip]{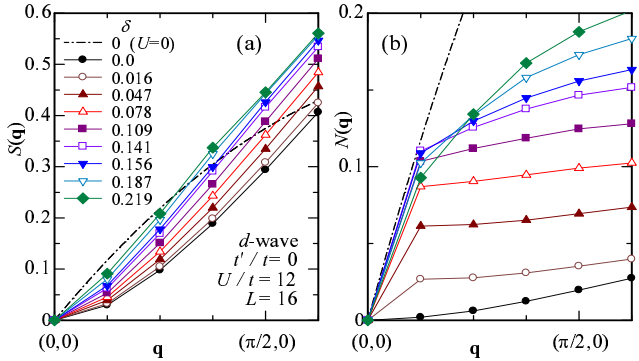}
\caption{(Color online) 
Small-$|{\bf q}|$ behaviors of (a) spin and (b) charge density structure 
factors in (0,0)-($\pi$,0) direction for $U/t=12$ and some doping rates. 
In (b), $N(|{\bf q}|)\propto|{\bf q}|^\gamma$ with $\gamma\le 1$ 
for ${\bf q}\rightarrow 0$ seems to hold for $\delta>0$. 
The system used and symbols of $\delta$ are common to the two panels. 
} 
\label{fig:nqsqu12} 
\end{center} 
\end{figure} 
%
Now, we discuss how spin and charge density gaps evolve when 
$\delta$ is introduced. 
It was revealed in ref.~\citen{YOT} for half filling that $\Psi_Q^d$ 
has a finite spin (SC) gap for any positive $U/t$, but is gapless for 
$U<U_{\rm c}$ and gapped for $U>U_{\rm c}$ in the charge sector. 
It is known within the single-mode approximation\cite{SMA} that 
a charge density gap opens [closes] if the charge density structure 
factor 
\begin{equation}
N({\bf q})=\frac{1}{N_{\rm s}} 
\sum_{i,j}e^{i{\bf q}\cdot({\bf R}_i-{\bf R}_j)} 
\left\langle{n_{i} n_{j}}\right\rangle - n^2 
\label{eq:nq}
\end{equation} 
behaves as $N({\bf q})\propto |{\bf q}|^2$ [$\propto|{\bf q}|$] for 
$|{\bf q}|\rightarrow 0$. 
For the spin sector, a similar treatment with $S({\bf q})$ is available. 
In Fig.~\ref{fig:nqsqu12}, we show small-$|{\bf q}|$ behaviors of 
$S({\bf q})$ and $N({\bf q})$ at $U=12t$ ($>U_{\rm c}$) for various 
doping rates. 
An SC (singlet) gap survives up to the overdoped regime ($\delta>0.15$), 
although the quadratic feature of $S({\bf q})$ becomes less distinctive 
as $\delta$ increases. 
On the other hand, in $N({\bf q})$, the quadratic behavior immediately 
disappears upon doping, namely, the charge density gap vanishes 
for $\delta>0$, because $|{\bf q}|^2/N({\bf q})$ seems to vanish 
for $|{\bf q}|\rightarrow 0$. 
\par 

\begin{figure}[htb]
\begin{center}
\vskip 4mm
\includegraphics[width=5.5cm,clip]{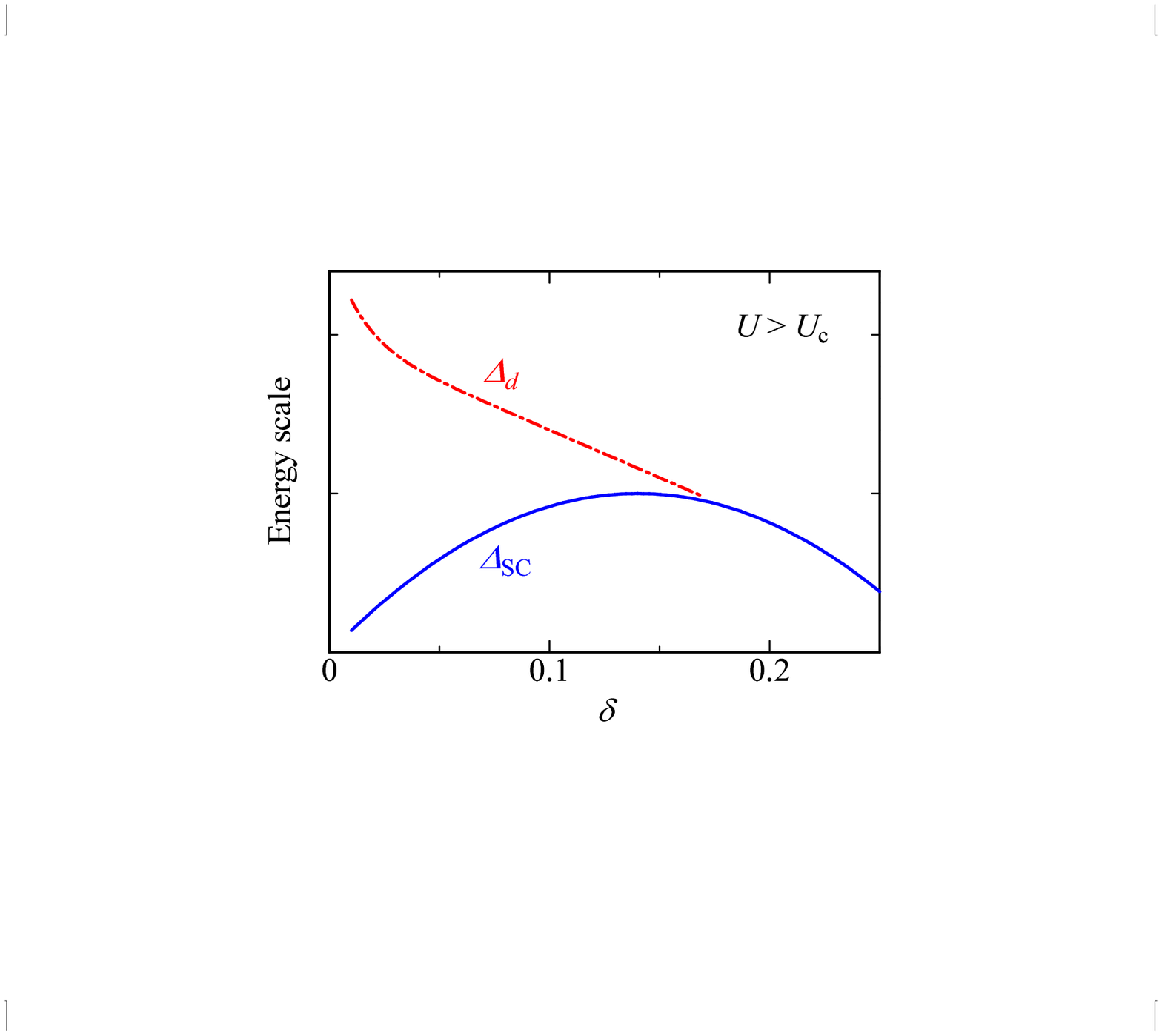} 
\end{center} 
\vskip -3mm
\caption{(Color online)
Schematic $\delta$ dependence of two kinds of gap scales deduced from 
present calculations for $U>U_{\rm c}$. 
}
\label{fig:S-Phased} 
\end{figure}
%
Let us summarize the energy scales deduced from the present calculations 
(Fig.~\ref{fig:S-Phased}). 
The gap scales representing the features of SC for $U>U_{\rm c}$ are 
classified into two kinds: 
(i) Quantities exclusively related to the singlet formation, which is 
derived from the superexchange interaction, monotonically decrease 
by doping, which weakens the AF correlation. 
They are symbolically indicated by $\Delta_d$ in Fig.~\ref{fig:S-Phased}.
(ii) Quantities directly related to SC, which is affected by both the 
singlet gap $\Delta_d$ and charge transportability, increase as $\delta$ 
increases in the underdoped regime, as shown by $\Delta_{\rm SC}$. 
It is natural to substitute $T_{\rm c}$ ($T_{\rm pair}$)
for $\Delta_{\rm SC}$ ($\Delta_d$) in Fig.~\ref{fig:S-Phased}.
\par

Finally, we touch on the relation between $\Delta E^d$ in eq.~(\ref{eq:DE}) 
and the SC condensation energy experimentally observed 
($\Delta E_{\rm cond}$). 
In the BCS theory, $\Delta E^d$ coincides with $\Delta E_{\rm cond}$ 
at $T=0$. 
However, this relation is not applicable to the present case, because 
only a part of $\Delta_d$ contributes to $\Delta_{\rm SC}$ and 
the residual part remains as an incoherent singlet gap. 
Actually, values $\Delta E_{\rm cond}$ estimated from the specific heat 
measurement\cite{Tallon,Matsuzaki} exhibit a dome shape as a function of 
$\delta$, similarly to $T_{\rm c}$, which is completely different from 
the monotonic 
behavior of $\Delta E^d$ shown in Fig.~\ref{fig:delDEvsn}(a). 
Ideally, the above incoherent part should cancel with 
the corresponding part of a proper normal state; thus, an improvement of 
the normal state is necessary. 
Anyway, we should be prudent in comparing $\Delta E^d$ with 
$\Delta E_{\rm cond}$. 
\par

\subsection{Mechanism of conduction\label{sec:holons}}
\begin{figure}[htb]
\begin{center}
\hskip 7mm
\includegraphics[width=6.5cm,height=2.8cm]{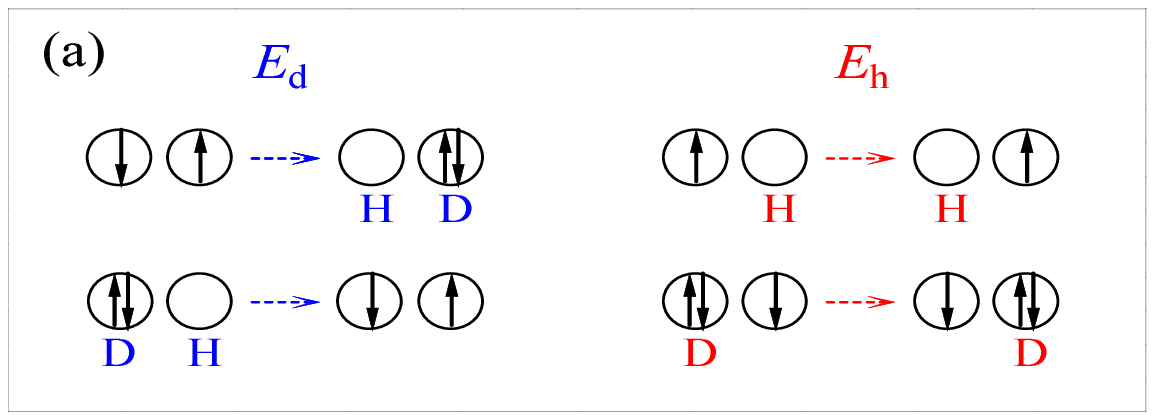} 
\includegraphics[width=7.8cm,clip]{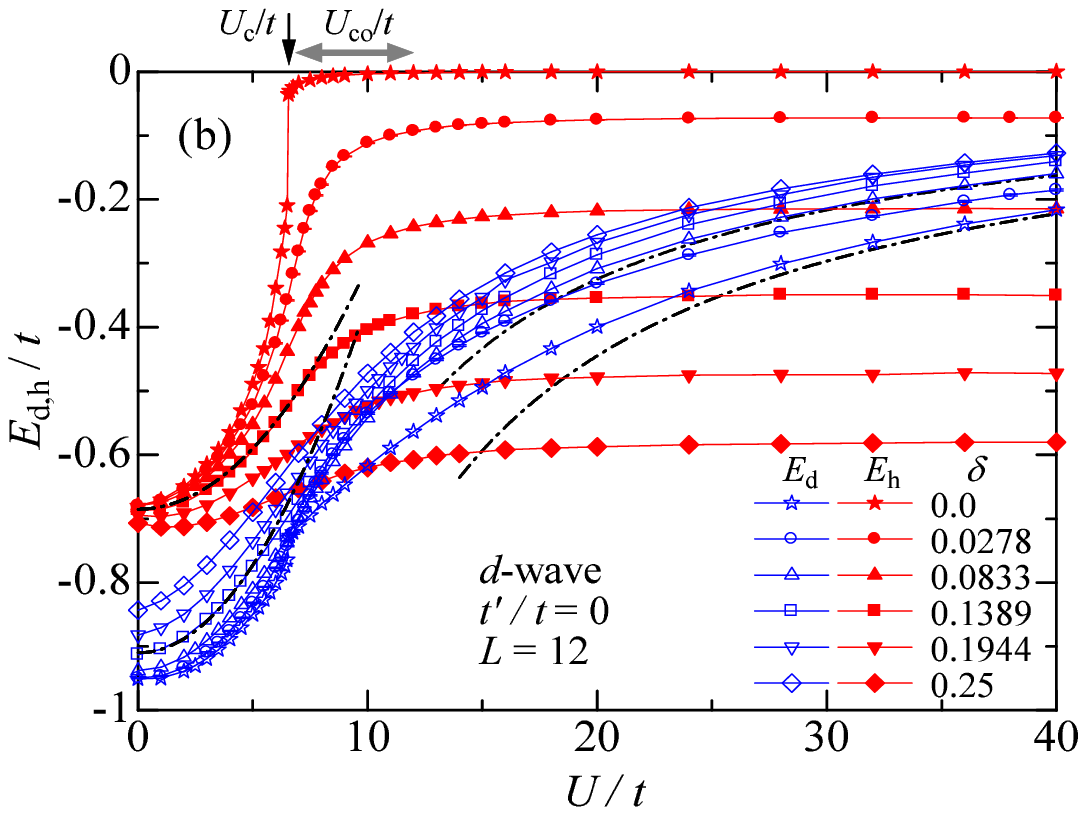} 
\end{center} 
\vskip -4mm
\caption{(Color online) 
(a) Hopping processes that contribute to the two components of kinetic 
energy, $E_{\rm d}$ and $E_{\rm h}$, are schematically shown. 
(b) The $U/t$ dependences of $E_{\rm d}$ and $E_{\rm h}$ are shown for six 
doping rates. 
The two dash-dotted lines in the small-$U/t$ [large-$U/t$] regime are 
guides as $\sim(U/t)^2+\mbox{const.}$ for $\delta=0.1389$ 
[$\sim-t/U=-J/(4t)$ in $E_{\rm d}$ for $\delta=0$ and $0.0833$].
}
\vskip -4mm 
\label{fig:ecicomp} 
\end{figure}

In this subsection, we argue that a feature of DC conduction undergoes 
a marked change through crossover. 
Here, we aim to obtain an intuitive picture of conduction within the scope 
of kinetic energy $E_t$ rather than to have a quantitative discussion 
using a direct measure such as the Drude weight. 
To this end, it is useful to analyze the kinetic energy, by dividing it 
into two components, 
$E_t=E_{\rm d}+E_{\rm h}$, as the hopping 
process varies ($E_{\rm d}$) or does not vary ($E_{\rm h}$) the number 
of doublons,\cite{Tocchio} as shown in Fig.~\ref{fig:ecicomp}(a). 
$E_{\rm d}$ is derived from the hopping processes that create and 
destroy D-H pairs, and corresponds to the $J$ term in the $t$-$J$ 
model. 
$E_{\rm h}$ comes from the direct hopping of holons and doublons, 
which corresponds to the $t$ term in the $t$-$J$ model.\cite{note-Eh} 
In Fig.~\ref{fig:ecicomp}(b), we plot $E_{\rm d}$ and $E_{\rm h}$ 
as functions of $U/t$ for several $\delta$. 
The behavior of the two components clearly changes at approximately 
$U_{\rm co}/t$. 
\par 

\begin{figure}[htb]
\begin{center}
\includegraphics[width=8.5cm,clip]{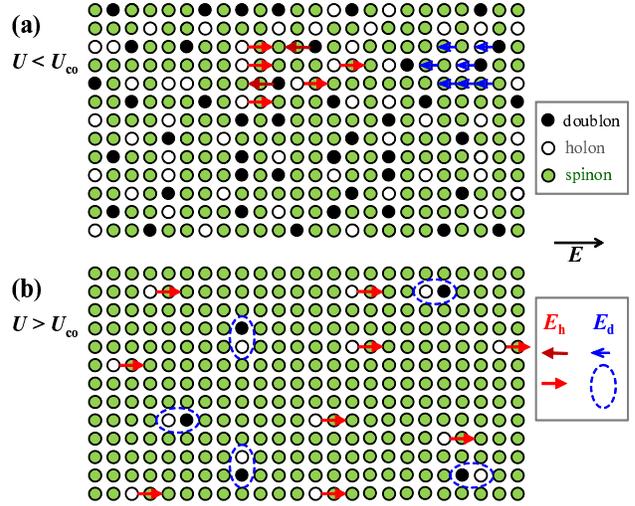} 
\end{center} 
\caption{(Color online) 
Mechanism of conduction is schematically compared between (a) the weakly 
correlated [Fermi liquid or BCS-type SC] regime and (b) the strongly 
correlated [doped Mott insulator] regime. 
Small arrows represent the motion of particles if a field ($\vec E$) 
is applied; in (a) arrows are drawn only for a small number of particles 
for clarity. 
The dashed ellipses in (b) indicate the local creation and annihilation 
processes of bound D-H pairs, which do not contribute to current.
}
\label{fig:conduct} 
\end{figure}

%
We begin with the strongly correlated regime ($U>U_{\rm co}$). 
At half filling, $E_{\rm h}$ substantially vanishes in the Mott insulating 
regime ($U>U_{\rm c}$); $E_{\rm d}$ remains finite and behaves proportionally 
to $-4t^2/U$ ($=-J$). 
Since the state is insulating, $E_{\rm d}$ here does not contribute 
to current, namely, local processes that create and annihilate 
D-H pairs are only repeated. 
They correspond to the large-$\omega$ part in $\sigma_1(\omega)$: transitions 
between the lower and upper Hubbard bands.
When carriers are doped, this behavior of $E_{\rm d}$ ($\propto -J$) 
is basically unchanged; its magnitude decreases slowly as $\delta$ 
increases, accurately corresponding to a decrease in doublon number, 
as shown in Fig.~\ref{fig:Zvsneci}(b). 
Thus, the local D-H processes at $\delta=0$ remain intact for $\delta>0$, 
meaning that $E_{\rm d}$ is not involved in conduction 
or itinerancy. 
On the other hand, $E_{\rm h}$ becomes finite and is still almost constant 
for $U>U_{\rm co}$ with the magnitude linear in $\delta$ or 
$\propto 2\delta/(1+\delta)$ in eq.~(\ref{eq:DeltaSC}), as depicted 
in Fig.~\ref{fig:Zvsneci}(b). 
This indicates that the independent motion of doped holons is realized 
for $U>U_{\rm co}$, although their mass is somewhat heavier than that 
of the free electrons.\cite{Drude} 
It follows that two kinds of holons play entirely different roles in SC, 
namely, the holons created as D-H pairs devote themselves to forming 
local singlet pairs, whereas the holons introduced by doping act 
as current carriers, as schematically sketched in Fig.~\ref{fig:conduct}(b). 
Thus, the number of doped holes is equal to the number of carriers, 
which will make a small quasi-Fermi surface such as a Fermi arc or pocket. 
This feature is simply that of the $t$-$J$ model,\cite{OF} 
and consistent with various experiments on cuprates,\cite{Condprop} 
and supports the very low superfluid densities.\cite{Uemura} 
The residual (background) spinons (singly occupied sites) remain localized 
unless the doped holons collide with them; this fact is possibly related 
to the recent observed neutron scattering for Bi2212,\cite{Xu} which 
indicated that 
the source of the magnetic response in doped cuprates is localized spins. 
\par

%
In contrast, for $U<U_{\rm co}$, both $E_{\rm d}$ and $E_{\rm h}$ 
behave as 
\begin{eqnarray}
E_{\rm d}&\sim&c_{\rm d} U^2/t+E_{\rm d}(0), \nonumber \\
E_{\rm h}&\sim&c_{\rm h} U^2/t+E_{\rm h}(0), \nonumber
\end{eqnarray}
with $c_{\rm d}$ and $c_{\rm h}$ being constants, as shown 
in Fig.~\ref{fig:ecicomp}(b). 
This common behavior indicates that every hopping process contributes 
to $E_t$ in the same way. 
Since the D-H binding is ineffective in this regime, holons cannot be 
classified into the two kinds, and all electrons can contribute to 
conduction, as in Fig.~\ref{fig:conduct}(a), resulting in a large 
(ordinary) Fermi surface with a carrier number of $N$. 
Correspondingly, the number of doped holes is not in agreement with 
the carrier number, in 
contrast to the experimental results on cuprates. 
\par

We have also obtained results similar to those in Fig.~\ref{fig:ecicomp}(b) 
for the normal and AF states, although they are not shown here. 
The mechanism of conduction in doped Mott insulators is entirely different 
from that in weakly correlated systems. 
\par

\section{Effect of Diagonal Transfer $t'$\label{sec:diag}}
In this section, we study the effect of next-nearest-neighbor hopping 
($t'$), which is the principal term for characterizing the individuality 
of each high-$T_{\rm c}$ cuprate. 
In \S\ref{sec:dandAF}, we consider the stability of $\Psi_Q^d$ 
against $\Psi_Q^{\rm AF}$ in introducing $t'$and possible phase separation 
near half filling. 
In \S\ref{sec:origin}, we discuss the origin of the stability of the 
$d$-wave and AF states on the basis of the effect of $t'$. 
In \S\ref{sec:Pair}, we study the effect of $t'/t$ on $P_d^\infty$, 
whereby we discuss the mechanism of enhancing $P_d^\infty$.
\par
\subsection{Energy gain by $d$-wave gap\label{sec:dandAF}}
\begin{figure}[hob]
\vspace{-0.4cm} 
\begin{center}
\includegraphics[width=8.0cm,clip]{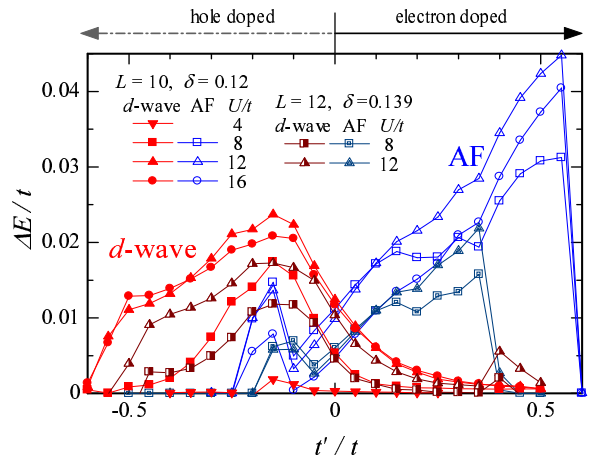}
\vskip -2mm
\caption{(Color online)
Energy gains for $d$-wave (solid and half-solid symbols) and AF (open 
and double-line symbols) states as functions of $t'/t$ for optimally doped 
densities. 
For comparison, data for some values of $U/t$ and two values of $L$ 
are plotted. 
}
\vskip -4mm
\label{fig:condedAF} 
\end{center} 
\end{figure}
%
At half filling, we may focus on the range $t'/t\le 0$, because 
there is symmetry between $t'$ and $-t'$, as discussed 
in ref.~\citen{YOT} and Appendix \ref{sec:derivation}. 
Upon doping carriers, however, this symmetry is broken. 
Figure \ref{fig:condedAF} shows $\Delta E^d$ and $\Delta E^{\rm AF}$ 
for some values of $U/t$ as functions of $t'/t$ at roughly optimally doped 
densities $\delta=0.12$ and $0.139$. 
In the hole-doped cases ($t'/t<0$), $\Delta E^d$ tends to be
enhanced, and $\Delta E^{\rm AF}$ is suppressed. 
For $U/t\ge 8$, $\Delta E^d$ has a broad maximum in the range 
$-0.3\lsim t'/t\lsim -0.1$, and tends to vanish relatively rapidly 
for $t'/t\lsim-0.45$. 
In contrast, $\Delta E^{\rm AF}$ vanishes except for a small 
peak at $t'/t\sim -0.15$. 
This peak is caused by the overlap of the Fermi surface with flat-band 
parts near ($\pi,0$), as we will discuss later for $\Psi_Q^d$. 
In the electron-doped cases ($t'/t>0$), the situation is opposite. 
$\Delta E^d$ gradually and monotonically decreases as $t'/t$ increases, 
whereas $\Delta E^{\rm AF}$ monotonically increases until the sudden 
breakdown at a large $t'/t$. 
\par

The $U/t$ dependences (not shown) of $\Delta E^d$ and $\Delta E^{\rm AF}$ 
are qualitatively similar to those for $t'=0$ in Fig.~\ref{fig:DEvsU}, 
although the overall amplitude depends on $t'/t$ as 
in Fig.~\ref{fig:condedAF}. 
\par 

\begin{figure}[hob]
\begin{center}
\includegraphics[width=8.7cm,clip]{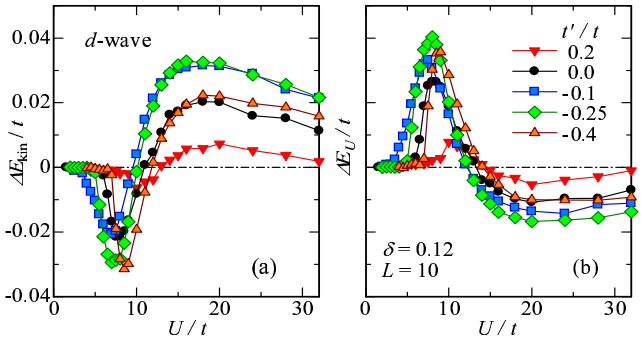} 
\vskip -2mm
\caption{(Color online)
(a) Kinetic and (b) interaction parts of energy gain $\Delta E^d$ for 
various values of $t'/t$ as functions of $U/t$. 
} 
\vskip -4mm 
\label{fig:deteut2} 
\end{center} 
\end{figure} 
%
We touch on the components of the energy gain $\Delta E$ for finite $t'/t$. 
In Figs.~\ref{fig:deteut2}(a) and \ref{fig:deteut2}(b), we plot 
the kinetic ($t$ and $t'$ terms) and interaction parts of $\Delta E$, 
respectively, as functions of $U/t$. 
The behaviors of the two components are qualitatively identical to 
those for $t'=0$ (Fig.~\ref{fig:deteudwave}): SC is induced by the 
reduction in interaction (kinetic) energy for small (large) 
values of $U/t$. 
Here, the doping rate is fixed at $\delta=0.12$, but the tendency is 
unchanged for other $\delta$. 
For small $|t'/t|$, the sum of $\sigma_1(\omega)$ becomes approximately 
proportional to the kinetic energy ($-E_{\rm kin}$).
\par

\begin{figure*}[!t]
\begin{center}
\includegraphics[width=14cm,clip]{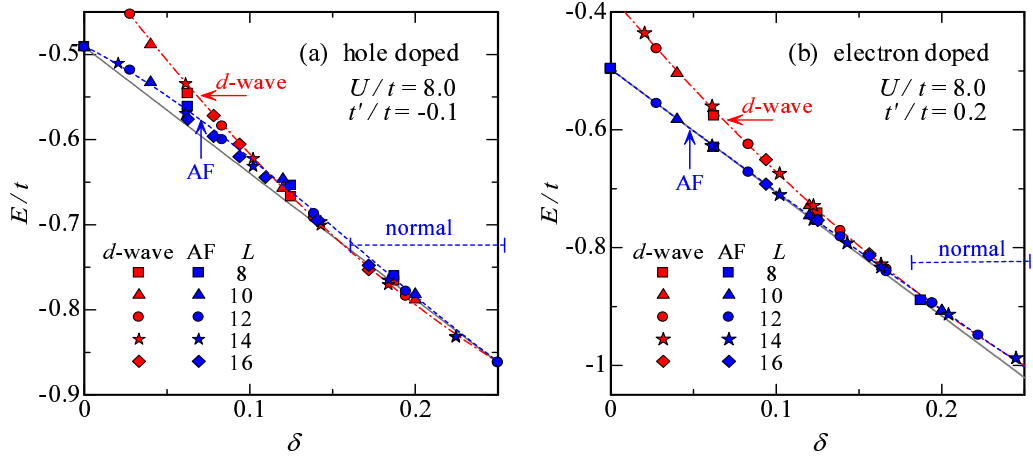}
\vskip -3mm
\caption{(Color online) 
Total energy of $d$-wave ($\Psi_Q^d$) and AF ($\Psi_Q^{\rm AF}$) 
states as functions of doping rate. 
In (a) and (b), hole-doped ($t'/t<0$) and electron-doped ($t'/t>0$)
cases are plotted, respectively. 
The data for several system sizes are fitted by the method of least 
squares, and shown with dashed (AF) and dash-dotted ($d$-wave) lines. 
The gray solid lines are straight guide lines in the AF case. 
In the region indicated by `normal', the optimized AF state is reduced 
to the normal state, namely, $\Delta_{\rm AF}=0$. 
In (b), we omit the data of $\Psi_Q^{\rm AF}$ in the vicinity 
of half filling owing to the breakdown of the AF phase
(see text). 
Similar data for $t'=0$ are given in Fig.~\ref{fig:EvsnAFt2-0}. 
}
\vskip -4mm
\label{fig:evsnAF} 
\end{center}
\end{figure*}
%
In \S\ref{sec:DelE4}, we showed that $\Psi_Q^{\rm AF}$ is unstable 
against phase separation for $t'=0$ (Fig.~\ref{fig:EvsnAFt2-0}). 
Here, we discuss the tendency toward the phase separation for $t'\ne 0$. 
In Fig.~\ref{fig:evsnAF}, we plot the total energy of the AF and 
$d$-wave states as functions of $\delta$ for hole- and electron-doped 
cases. 
For $\Psi_Q^d$, $E^d/t$ is concave ($\kappa>0$) for both values of 
$t'/t$, indicating that the $d$-wave state is still stable if $t'$ 
is added, regardless of the kind of doped carriers. 
For the AF state, it becomes slightly concave for $t'/t=0.2$ 
(electron-doped case) [Fig.~\ref{fig:evsnAF}(b)], indicating that  
the commensurate AF state is intrinsically stable for electron 
doping. 
On the other hand, $E^{\rm AF}/t$ is still convex ($\kappa<0$) 
for $t'/t=-0.10$ (hole-doped cases) in the whole range of finite 
optimized $\Delta_{\rm AF}$ ($0<\delta\lsim 0.15$) 
[Fig.~\ref{fig:evsnAF}(a)]. 
Thus, for hole doping, a phase separation occurs in the underdoped 
regime as in $t'=0$. 
Such behavior is almost independent of $U/t$ when 
$U/t$ is sufficiently large; the same behavior has been found 
for the $t$-$J$ model.\cite{WatatJ} 
\par

The reason for the instability of the AF state seems trivial; 
doping rapidly deteriorates the nesting of the Fermi surface 
for $t'/t<0$, in contrast to the case of $t'/t>0$. 
Although we have to be careful in directly comparing the present result 
with experiments, because the band renormalization effect is not 
explicitly introduced in $\Psi_Q^d$ or $\Psi_Q^{\rm AF}$, 
there are some suggestive experiments: 
A neutron scattering experiment in very lightly doped 
La$_{2-x}$Sr$_x$CuO$_4$ with $x<0.02$\cite{Matsuda} found a phase 
separation to a commensurate AF ordered phase and a spin-glass phase. 
On the other hand, such a phase separation does not appear 
in electron-doped Pr$_{1-x}$LaCe$_x$CuO$_4$.\cite{Matsuda,Fujita} 
\par

\subsection{Origin of $t'$-dependence regarding stability of $d$-wave 
and AF states\label{sec:origin}}
\begin{figure}[hob]
\vspace{-0.4cm} 
\begin{center}
\includegraphics[width=7.0cm,clip]{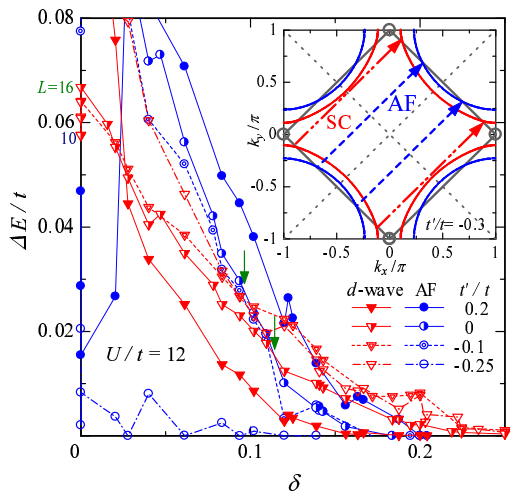} 
\vskip -2mm
\caption{(Color online)
Comparison of energy gain between $d$-wave (triangles) and AF (circles) 
states as function of doping rate for four $t'/t$ and $U/t=12$. 
The arrows indicate the crossing points of $\Delta E^{\rm AF}$ and 
$\Delta E^d$ for $t'/t=0$ and $-0.1$. 
Systems of $L=10$-16 are used. 
The inset shows the Fermi surface in the electron picture for $n=0.85$ 
(favored by $d$-wave) and $1.10$ (favored by AF) for $t'/t=-0.3$. 
The arrows are the {\bf Q} vectors connecting the hot spots [intersections 
of the Fermi surface and magnetic Brillouin zone boundary indicated 
by the gray bold line]. 
The gray dotted lines indicate the nodes of the $d$-wave state, and 
the gray circles the van Hove singularity points near the Fermi surface. 
}
\label{fig:condvsnAF}
\end{center}
\end{figure}

%
The $t'/t$ dependences of $\Delta E^d$ and $\Delta E^{\rm AF}$ in the 
optimally doped regime (Fig.~\ref{fig:condedAF}) are basically unchanged 
even if $\delta$ varies, as shown in Fig.~\ref{fig:condvsnAF} for $U/t=12$. 
For $t'/t=0.2$ (solid symbols), we have $\Delta E^{\rm AF}>\Delta E^d$ 
for any $\delta$; for $t'/t=0$ (half solid symbols) and $-0.1$ (double-line 
symbols), $\Delta E^{\rm AF}$ and $\Delta E^d$ interchange 
at $\delta\sim 0.89$ and $0.91$, respectively, whereas for $t'/t=-0.25$ 
(open symbols), we always have $\Delta E^{\rm AF}<\Delta E^d$. 
Thus, the range of $\delta$ where $\Psi_{\rm AF}$ ($\Psi_d$) is predominant 
rapidly shrinks as $|t'/t|$ increases on the hole- (electron-)doped side. 
Incidentally, $\Delta E^{\rm AF}$ remains small near half filling 
for $t'/t=0.2$ and $-0.25$.
This is probably because the renormalization of the Fermi surface is not 
explicitly introduced in $\Psi_Q^{\rm AF}$; this effect becomes decisive 
for $\delta\rightarrow 0$ to retrieve the nesting 
condition.\cite{WataOrg2,Kobayashi} 
We expect $\Delta E^{\rm AF}>\Delta E^d$ for $\delta\sim 0$ even 
for large $|t'/t|$ in improved wave functions. 
\par 

\begin{figure}[hob]
\begin{center}
\includegraphics[width=7.0cm,clip]{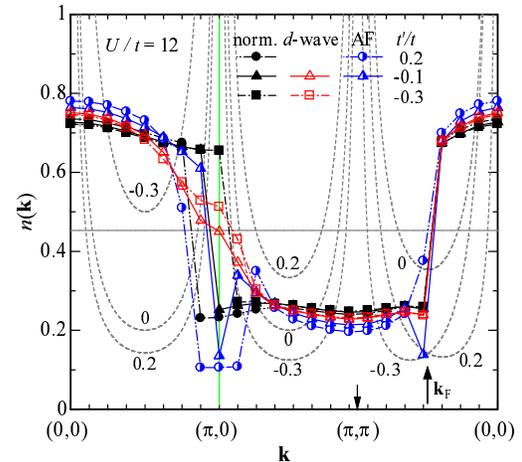}
\vskip -2mm
\caption{(Color online)
The momentum distribution functions are compared among the $d$-wave 
(open symbols), AF (half closed), and normal (closed) states for three 
values of $t'/t$, $\delta=0.094$ and $L=16$. 
Firm orders exist for the $d$-wave and AF states. 
The gray horizontal line shows $n/2$. 
The gray dashed curves represent the relative magnitude of the noninteracting 
{\bf k}-dependent density of states $1/|\nabla \varepsilon_{\bf k}|$ 
for three values of $t'/t$. 
}
\vskip -4mm
\label{fig:nkcomp} 
\end{center} 
\end{figure} 

On the basis of Figs.~\ref{fig:condedAF} and \ref{fig:condvsnAF}, 
let us consider the stability of the $d$-wave and AF states 
with respect to $t'/t$. 
As $|t'/t|$ increases, the curvature of the Fermi surface in $\Phi_{\rm F}$ 
becomes more concave in the nodal direction $(0,0)$-$(\pi,\pi)$, 
as shown in the inset of Fig.~\ref{fig:condvsnAF}. 
Furthermore, it is expected that the scattering with ${\bf q}={\bf Q}$ 
is activated in the area of doped Mott insulators, as discussed 
in \S\ref{sec:Udep4} and \S\ref{sec:ndep4}. 
\par 

First, we discuss the origin of stability of $\Psi_Q^{\rm AF}$. 
In the electron-doped cases, the nesting does not deteriorate rapidly 
when $\delta$ increases away from half filling, namely, the Fermi surface 
continues to largely overlap with the magnetic Brillouin zone boundary 
as a whole, as shown in the inset of Fig.~\ref{fig:condvsnAF}. 
The effect of a large density of states near ($\pi,0$) 
[see Fig.~\ref{fig:nkcomp}] is subsidiary 
for $\Psi_Q^{\rm AF}$. 
On the other hand, the N\'eel order disappears rather suddenly when
$t'/t$ becomes excessively large or when holes are doped, because the 
above overlap suddenly vanishes. 
Incidentally, the band renormalization effect becomes important for the AF 
order for $U>W$;\cite{Watat-J,Kobayashi} we will reconsider 
this subject in future publications. 
\par

In contrast, conceivable reasons why the $d$-wave favors finite negative 
values of $t'/t$ are as follows: 
(i) 
Because the $d$-wave gap is minimum in the nodal directions and 
maximum in the antinodal ($k_x$ and $k_y$) directions, the energy 
gain is large when the electrons near $(0,\pi)$ and $(\pi,0)$ are 
occupied. 
(ii) 
In the noninteracting case, the ${\bf k}$-dependent density 
of states, $1/|\nabla\varepsilon({\bf k})|$, is minimum in the nodal 
direction including ($\pi/2,\pi/2$), but becomes large near the antinodal 
points $(\pi,0)$ and $(0,\pi)$ owing to the band flatness. 
Actually, as shown by the dashed line in 
Fig.~\ref{fig:nkcomp}, 
$1/|\nabla\varepsilon({\bf k})|$ becomes large especially 
for $t'/t=-0.3$ as ${\bf q}$ approaches ($\pi,0$) from (0,0). 
(iii) 
The vector ${\bf Q}$ connects two {$\bf k$} points near the 
magnetic Brillouin zone boundary including the antinodal 
points with mutually inverse signs of $\Delta_{\bf k}$. 
Thus, $\Psi_Q^d$ can take advantage of the $d$-wave pair scattering of {\bf Q} 
near the antinodal points.
We will pursue this topic in \S\ref{sec:Pair} and \S\ref{sec:discussions}. 
\par

This feature of $\Psi_Q^{\rm AF}$ and $\Psi_Q^d$ is 
reflected in the behavior of the momentum distribution function 
\begin{equation}
n({\bf k})=
\frac{1}{2}\sum_\sigma\langle c^\dag_{\bf k\sigma}c_{\bf k\sigma}\rangle. 
\label{eq:nk}
\end{equation}
Figure \ref{fig:nkcomp} shows a comparison of $n({\bf k})$ among the 
optimized 
$d$-wave, AF, and normal states for an underdoped density. 
The $n({\bf k})$ of the normal state exhibits clear discontinuities (Fermi 
surface) near ($\pi,0$) and ($\pi/2,\pi/2$). 
Near ($\pi,0$), the $n({\bf k})$ of both $d$-wave and AF states are 
considerably changed from that of the normal state, 
indicating that scattering actively takes place there. 
Near ($\pi/2,\pi/2$), the $n({\bf k})$ of the AF state is again noticeably 
modified depending largely 
on $t'/t$, but the $n({\bf k})$ 
of the $d$-wave state changes only slightly, and is almost independent 
of $t'/t$. 
Now, we are certain that, for $\Psi_Q^d$, the scattering of {\bf Q} 
is ineffective in the nodal direction, but relevant near the antinodal 
points.
\par

\subsection{$d$-wave pairing correlation\label{sec:Pair}}
%
\begin{figure*}[!t] 
\begin{center}
\includegraphics[width=16.0cm,clip]{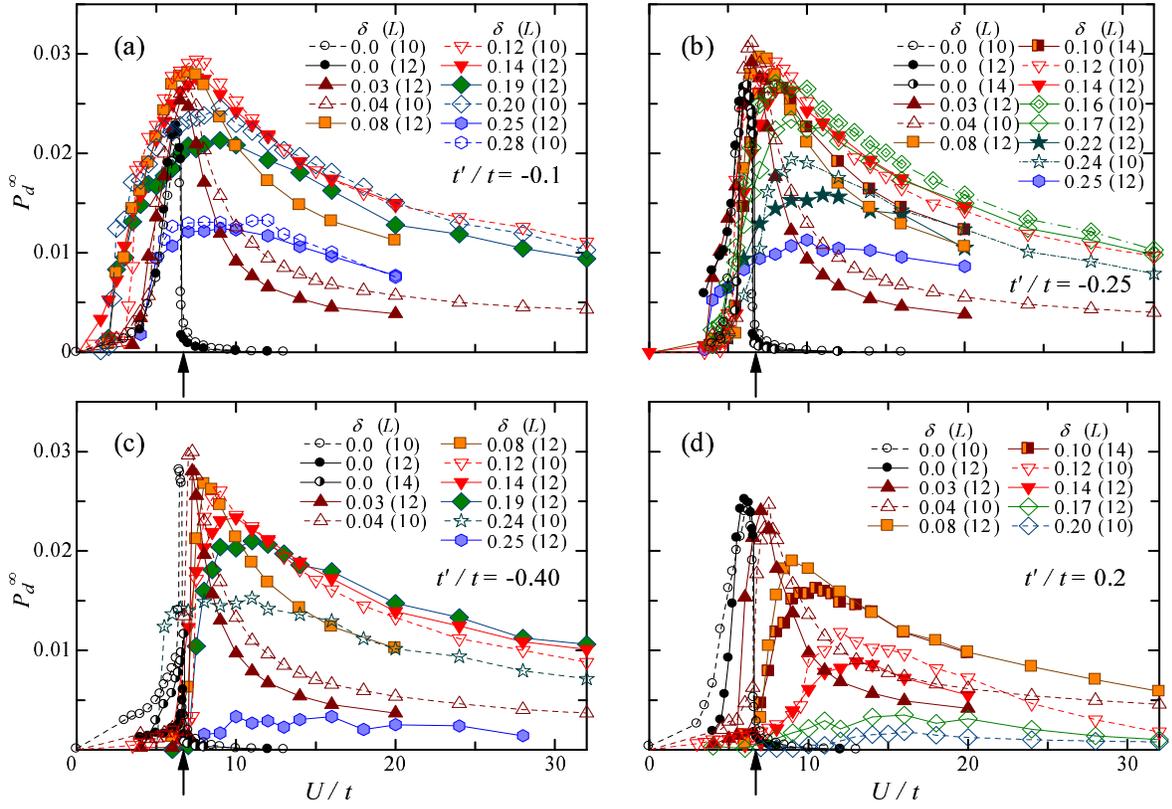}
\vskip -3mm
\caption{(Color online) 
$d$-wave pairing correlation function as function of $U/t$ for various 
values of doping rate $\delta$ (rounded off to two decimal places in 
legends).
The values of $t'/t$ are different among the four panels: (a)-(c) hole-doped 
and (d) electron-doped cases. 
Corresponding results for $t'/t=0$ are given in Fig.~\ref{fig:pdvsut2-0}.
The arrow in each panel indicates the Mott transition point at half filling 
(circles). 
}
\label{fig:pdvsuall} 
\end{center} 
\end{figure*}
\begin{figure*}[!t]
\begin{center}
\includegraphics[width=13.5cm,clip]{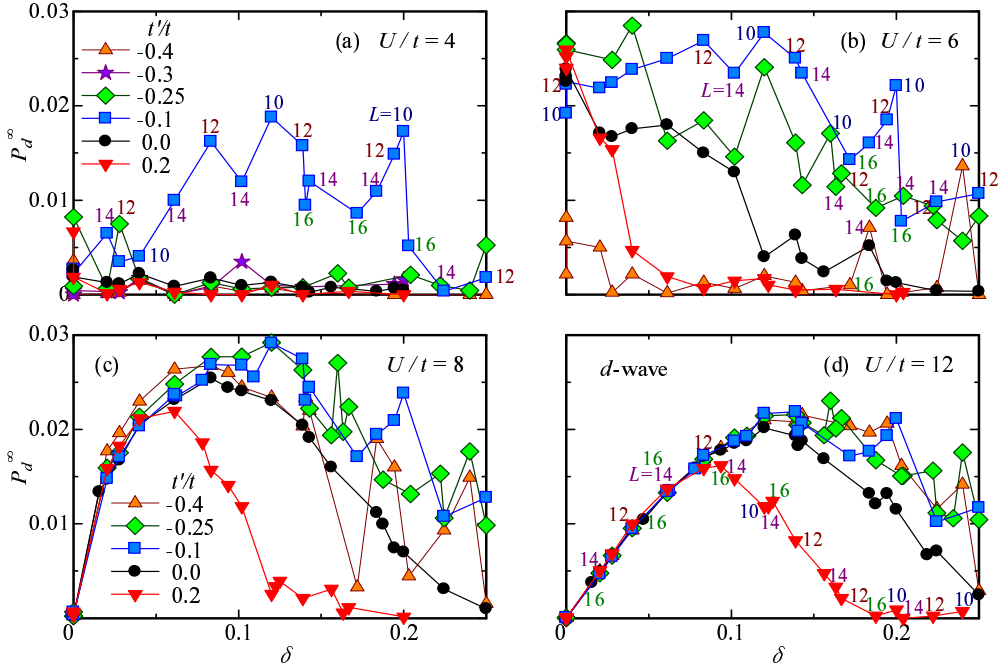}
\vskip -3mm
\caption{(Color online) 
$d$-wave pairing correlation function as function of doping rate $\delta$ 
for various values of $t'/t$. 
The values of $U/t$ are different among the four panels (a)-(d). 
The symbols indicating the values of $t'/t$ are common to the four panels. 
Small digits beside some data points indicate the lattice size 
$L$ ($=10$-16) used. 
}
\label{fig:pdvsnall} 
\end{center}
\end{figure*}

 
%
We turn to the $d$-wave SC correlation function in $\Psi_Q^d$. 
In Fig.~\ref{fig:pdvsuall}, the $U/t$ dependence of $P_d^\infty$, defined 
by eq.~(\ref{eq:pd}) and in Appendix\ref{sec:pairfunc}, is shown 
for four finite values of $t'/t$. 
In Fig.~\ref{fig:pdvsnall}, the $\delta$ dependence of $P_d^\infty$ is shown 
for four values of $U/t$. 
Although the overall behavior is similar to that in the case of $t'=0$ 
(Figs.~\ref{fig:pdvsut2-0} and \ref{fig:Pdvsn}), there are noteworthy 
differences owing to the effect of $t'$. 
(i) The magnitude of $P_d^\infty$ is somewhat enhanced for $t'/t=-0.1$ 
and $-0.25$, but is suppressed for $t'/t=0.2$, in accordance with 
the behavior of $\Delta E^d$ discussed in \S\ref{sec:dandAF}. 
(ii) For $U>U_{\rm c}$, the doping rate giving the maximum $P_d^\infty$ 
for a fixed $U/t$ shifts to the overdoped side as $t'/t$ decreases. 
(iii) For $t'/t=-0.1$, $P_d^\infty$ starts to increase at a smaller 
$U/t$. 
(iv) Although we do not show detailed data in Fig.~\ref{fig:pdvsuall}, 
for $t'/t\lsim -0.4$ and $\delta\gsim 0.2$, a state with a large $P_d^\infty$ 
is competitive with a state with a tiny $P_d^\infty$ ($\Delta_d/t\sim 0$), 
suggesting that the SC state collapses at approximately these parameters. 
\par

First, let us consider points (i) and (ii) in the strongly correlated 
regime. 
As in Fig.~\ref{fig:pdvsnall}(d), for $U (=12t)$ sufficiently larger 
than $U_{\rm c}$, $P_d^\infty$ forms a well-proportioned dome shape 
as a function of $\delta$. 
This $\delta$ dependence of $P_d^\infty$ is consistent with those of 
$T_{\rm c}$ and condensation energy observed in cuprates.  
We define $\delta_{\rm max}$ as the value of $\delta$ that gives 
the largest $P_d^\infty$ for given $U/t$ and $t'/t$. 
For $\delta<\delta_{\rm max}$, $P_d^\infty$ is proportional to 
$\delta$ and is almost independent of $t'/t$ and $L$. 
This means that SC in the underdoped regime is steady, and the magnitude 
of $P_d^\infty$ depends only on $U/t$, and not on minute band structures. 
On the other hand, for $\delta>\delta_{\rm max}$ in the hole-doped 
cases ($t'/t<0$), $P_d^\infty$ irregularly depends on $t'/t$ and $L$, 
and is scattered to some extent.\cite{note-n08} 
Thus, the strength of SC is sensitive to the shape of the Fermi surface 
and fragile in the overdoped regime. 
However, $P_d^\infty$ becomes larger than that for $t'=0$, probably 
because the $t'$ term bends the Fermi surface so that it may pass by 
the antinodal points. 
Similar behavior has been observed for the $t$-$J$ model.\cite{Shih,Watat-J} 
In the electron-doped case, $\delta_{\rm max}$ is smaller than those 
in the hole-doped cases, but the fluctuation with respect to $\delta$ 
and $L$ is small for $\delta>\delta_{\rm max}$, so that SC is steady.
In the transitional regime [Fig.~\ref{fig:pdvsnall}(c) 
for $U_{\rm c}<U\lsim U_{\rm co}$], 
although $P_d^\infty$ manages to preserve its dome shape, 
$\delta_{\rm max}$ is reduced to have a wide irregular range 
for $t'/t<0$.  
In the electron-doped case ($t'/t=0.2$), $P_d^\infty$ is suppressed 
compared with when $t'/t=0$ and almost vanishes in the overdoped regime. 
Some similar features were pointed out in a recent study by 
Tocchio \etal\cite{Tocchio2} 
\par

Next, we look at the case of intermediate correlation strength 
[$U=6t$, Fig.~\ref{fig:pdvsnall}(b)], which is slightly smaller 
than the Mott critical value $U_{\rm c}$. 
As discussed in ref.~\citen{YOT}, $P_d^\infty$ for $t'=0$ exhibits 
a sharp peak immediately below $U_{\rm c}$ $(\sim 7t)$ at half filling; 
such behavior is unchanged for $t'\ne 0$ (circle symbols 
in Fig.~\ref{fig:pdvsuall}). 
Because $U/t=6$ is close to this peak value, $P_d^\infty$ 
is practically the largest at half filling, and tends to decrease as 
$\delta$ increases. 
In particular, for $t'/t=-0.4$ and $0.2$, $P_d^\infty$ almost vanishes 
for $\delta>0.05$. 
Similarly to the case of $t'=0$, the behavior of $P_d^\infty$
here is contradictory to the dome shape of $T_{\rm c}$ and the 
condensation energy observed in cuprates, in contrast to the 
cases of $U>U_{\rm c}$.
\par 

Now we move to point (iii) regarding small values of $U/t$. 
As discussed in \S\ref{sec:Udep4} for $t'=0$, $P_d^\infty$ is negligible 
for $U/t$ as small as 4. 
As shown in Fig.~\ref{fig:pdvsuall}, this feature basically does not 
alter even if $t'$ is added, except for when $t'/t=-0.1$ 
[Fig.~\ref{fig:pdvsuall}(a)]. 
This exceptional enhancement of $P_d^\infty$ is evident 
in Fig.~\ref{fig:pdvsnall}(a), where appreciable magnitude appears 
only for $t'/t=-0.1$ among the six values of $t'/t$. 
We argue in the following that this exceptional increase in $P_d^\infty$ 
at $t'/t\sim -0.1$ for a small $U/t$ is useful to infer the origin of SC. 
\par

\begin{figure*}[!t]
\begin{center}
\includegraphics[width=14.0cm,clip]{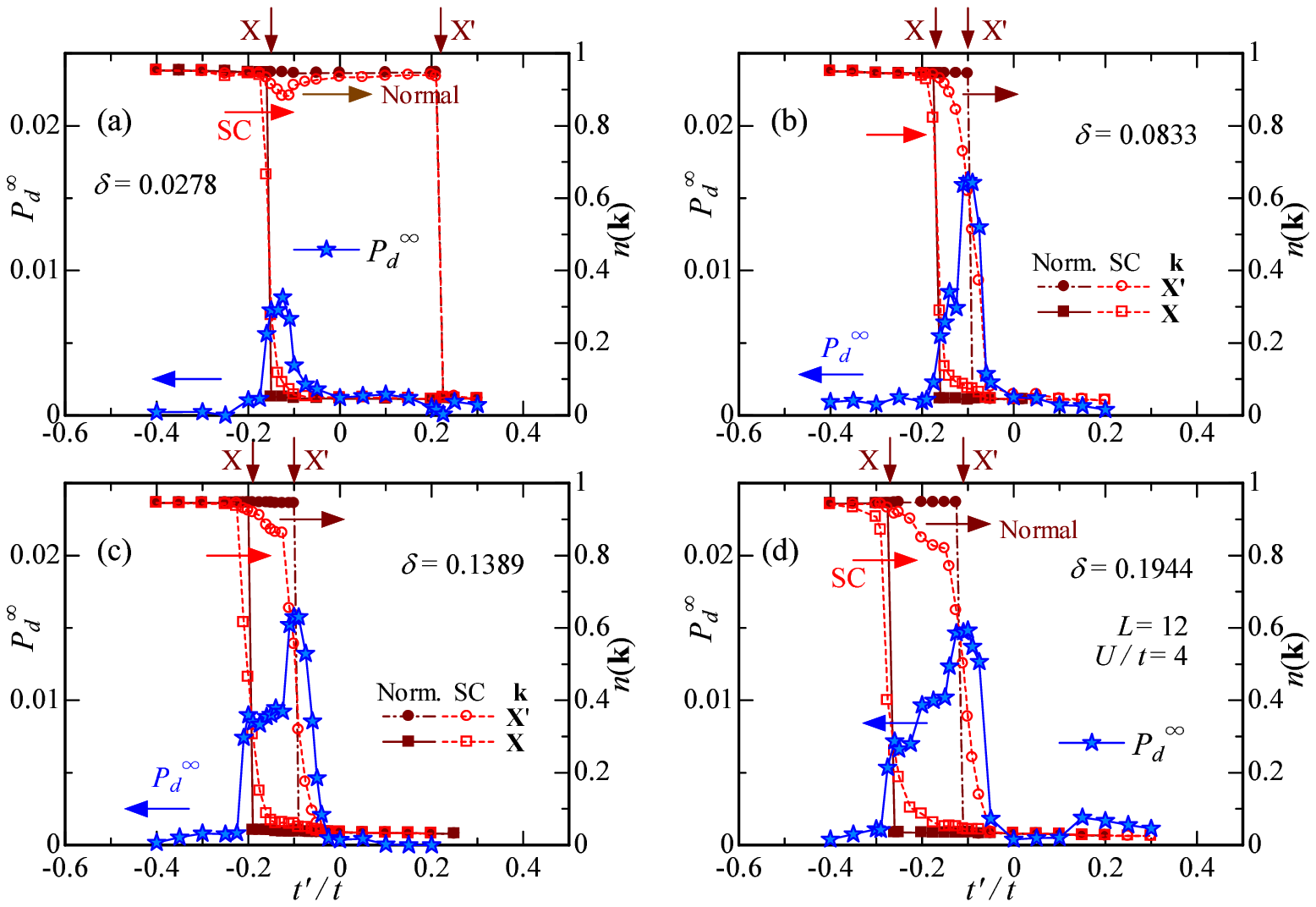} 
\vskip -2mm
\caption{(Color online) 
Correspondence between $d$-wave SC correlation function and jump in 
momentum distribution functions of $\Psi_Q^{\rm F}$ and $\Psi_Q^d$ at two 
{\bf k} points (${\bf k}={\bf X}$ and ${\bf X}'$) near $(\pi,0)$, 
as $t'/t$ varies, for a weak correlation ($U/t=4$). 
The cases of four doping rates are shown for $L=12$. 
The arrows on the upper axes indicate the values of $t'/t$ at which the Fermi 
surfaces of $\Psi_Q^{\rm F}$ pass ${\bf X}$ and ${\bf X}'$. 
}
\label{fig:diffX-comp} 
\end{center}
\vskip -3mm
\end{figure*}
%
In Fig.~\ref{fig:diffX-comp}, we show $P_d^\infty$ for $U/t=4$ with star 
symbols as a function of $t'/t$ for four $\delta$. 
The meaningful magnitude of $P_d^\infty$ is limited to a narrow range of 
$-0.16\lsim t'/t\lsim -0.075$ for underdoped densities 
[Figs.~\ref{fig:diffX-comp}(a) and \ref{fig:diffX-comp}(b)] and similarly 
a narrow range of $-0.275\lsim t'/t\lsim -0.075$ for an overdoped density, 
$\delta=0.1944$ 
[Fig.~\ref{fig:diffX-comp}(d)]. 
These results are consistent with those of a recent QMC 
study,\cite{Yanagisawa} 
which showed that the SC susceptibility vanishes for $t'=0$ 
but remains finite for $t'/t=-0.2$ for $U/t\le 5$ in the optimally 
doped regime. 
In Fig.~\ref{fig:diffX-comp}, we simultaneously plot the momentum 
distribution function $n({\bf k})$ calculated with $\Psi_Q^{\rm F}$ 
and $\Psi_Q^d$ at a couple of available ${\bf k}$ points near the antinodal 
point $\pi(1,0)$, i.e., 
${\bf X}=\pi(1,\frac{1}{L})$ and ${\bf X}'=\pi(1-\frac{2}{L},\frac{1}{L})$.
As $t'/t$ is increased, $n({\bf X})$ or $n({\bf X}')$ of $\Psi_Q^{\rm F}$ 
discontinuously drops from near unity to near zero when the Fermi 
surface passes through the ${\bf X}$ or ${\bf X}'$ point, as indicated 
by arrows on the upper axes. 
Note that, in each panel, the positions of the peak and shoulders of 
$P_d^\infty$ precisely coincide with those at which the Fermi surface 
overlaps with antinodal ${\bf k}$ points. 
As in Fig.~\ref{fig:nkcomp}, $(\pm\pi,0)$ and $(0,\pm\pi)$ are van Hove 
singularity points for $|t'/t|\le 0.5$, and connected to one another 
by the AF vectors ${\bf Q}=(\pi,\pm\pi)$, $(\pm\pi,\pi)$. 
It follows that the pair scattering with {\bf Q} and SC are 
enhanced at the values of $t'/t$ indicated by the arrows. 
This SC mechanism is consistent with the $d$-wave BCS theory.\cite{d-wave} 
\par

However, the mechanism for small values of $U/t$ has difficulties 
in explaining the features of high-$T_{\rm c}$ cuprates. 
First, the effective values of $t'/t$ of cuprates show a relatively 
wide range; single-layer La systems and double-layer Y and Bi systems 
have $t'/t\sim -0.1$ and $-0.3$, 
respectively.\cite{Fukuyama,Tohyama,Gooding,Raimondi,Pavarini}  
However, the above SC mechanism sensitively depends on $t'/t$ or the band 
structure, and the SC correlation is enhanced only in the limited range. 
It is unlikely that robust SC occurs at $t'/t=-0.3$ in the whole 
relevant range of $\delta$. 
Second, it is probable that the magnitude of $P_d^\infty$ discussed 
above is much weaker. 
As shown in Fig.~\ref{fig:pdvsnall}(a), $P_d^\infty$ has a large 
and relatively regular system-size dependence, and consequently seems 
to become negligible for $L\rightarrow\infty$. 
\par 

\begin{figure*}[!t]
\begin{center}
\includegraphics[width=14.0cm,clip]{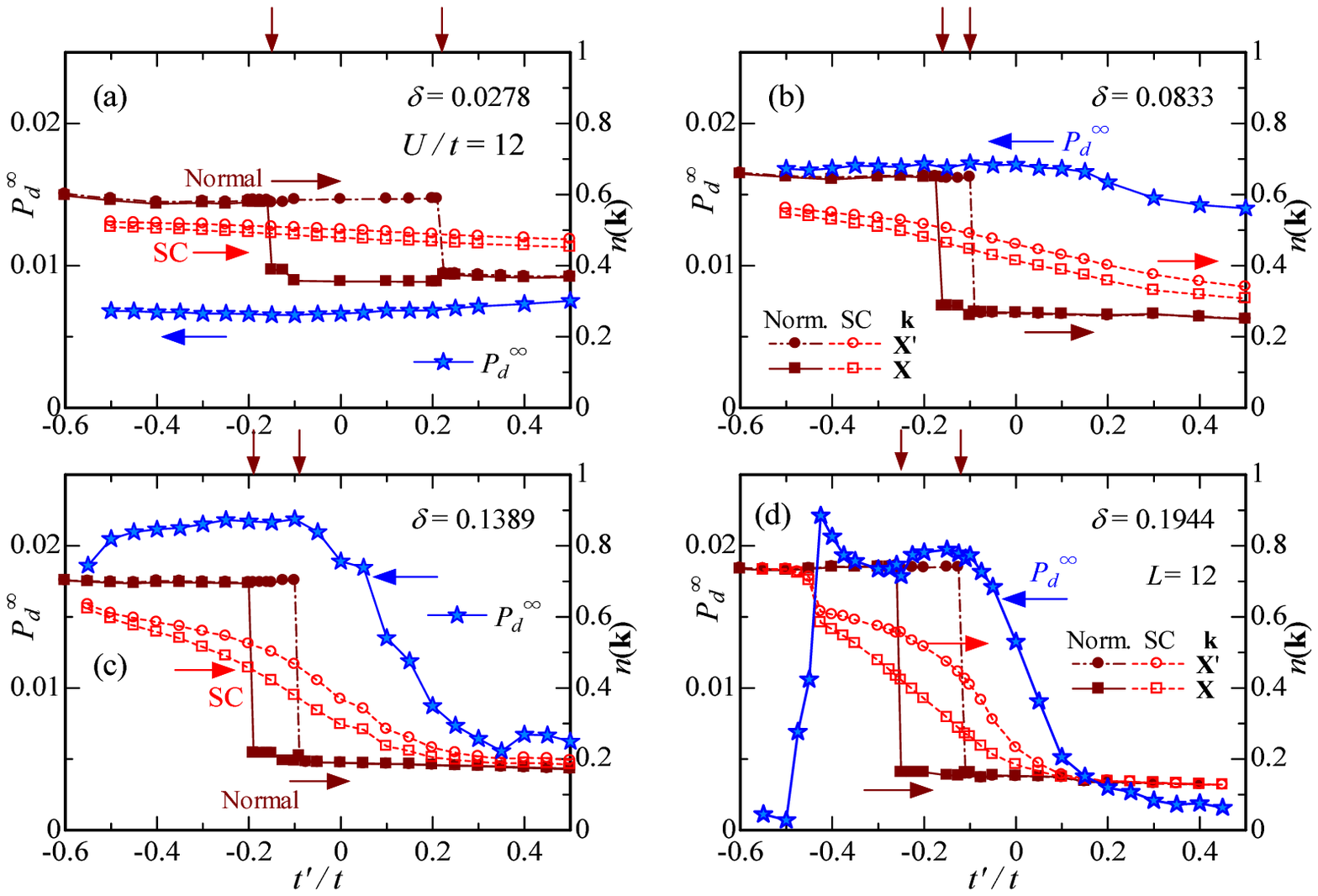} 
\vskip -3mm
\caption{(Color online) 
The same quantities as those in Fig.~\ref{fig:diffX-comp} are depicted 
in the same scales for a strong correlation ($U/t=12$). 
The band renormalization by correlations is not considered here, but 
its effect seems irrelevant except near half 
filling.\cite{Himeda-t',Kobayashi,Shih,Watat-J} 
}
\label{fig:diffX-comps} 
\end{center}
\vskip -3mm
\end{figure*}

\begin{figure*}[!t]
\begin{center}
\includegraphics[width=16.0cm,clip]{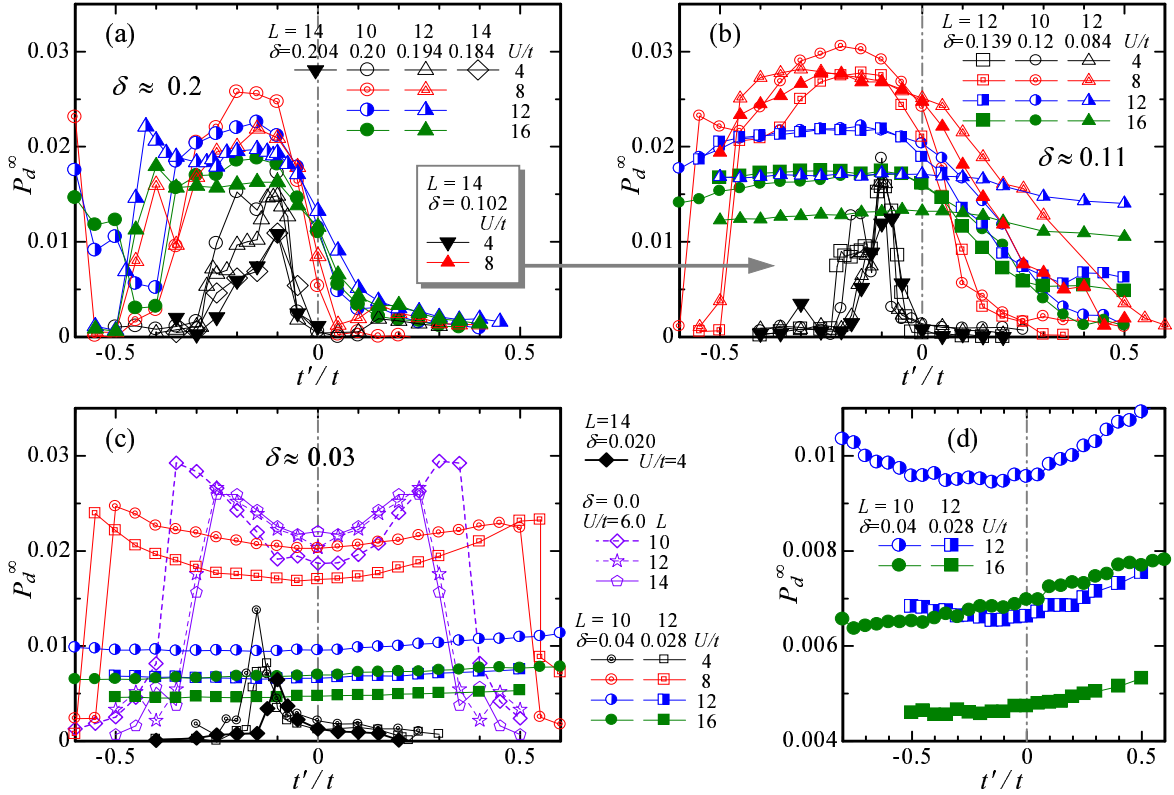}
\vskip -5mm
\caption{(Color online) 
$d$-wave pairing correlation function as function of $t'/t$ for several 
values of $U/t$. 
The doping rates $\delta$ are different among (a)-(c): 
(a) in the overdoped regime, $\delta\sim 0.20$, 
(b) in the optimally doped and underdoped regime, $\delta\sim 0.11$, and 
(c) nearly half filling $\delta\sim 0.03$. 
In (d), a magnification of (c) in the vertical axis is shown to emphasize 
the subtle variation in $P_d^\infty$. 
}
\label{fig:pdvsaall} 
\end{center} 
\end{figure*}

For comparison, we preform the same analysis in a strongly correlated 
case, $U/t=12$ (Fig.~\ref{fig:diffX-comps}). 
In the underdoped regime [panels (a) and (b)], $P_d^\infty$ has 
appreciable magnitudes in a wide range of $t'/t$, and the $t'/t$ 
dependence is very weak, even between electron-doped and hole-doped 
cases [see also Fig.~\ref{fig:pdvsaall}(c)].
At an optimum doping rate [Fig.~\ref{fig:diffX-comps}(c)], 
the electron-hole asymmetry stands out, and $P_d^\infty$ 
decreases as $t'/t$ increases in the electron-doped regime $t'/t>0$, 
but $P_d^\infty$ is steady and large throughout the range of $t'/t<0$. 
This tendency is preserved for an overdoped regime 
[Fig.~\ref{fig:diffX-comps}(d)], where $P_d^\infty$ is still large 
in the range of hole-doped cuprates ($-0.4\lsim t'/t\lsim -0.05$). 
On the other hand, the Fermi surface of $\Psi_Q^{\rm F}$ passes 
${\bf k}={\bf X}$ and ${\bf X}'$ at the same $t'/t$ as the weakly 
correlated cases, as indicated by arrows. 
For the underdoped and optimally doped cases, $P_d^\infty$ does not 
exhibit any special behavior at the positions of the arrows.  
Only in the overdoped case does $P_d^\infty$ display a slight tendency 
of forming a shoulder near an arrow. 
Anyway, in the strongly correlated regime, the strength of SC 
is almost independent of the locus of the Fermi surface of the 
underlying normal state, namely, the bare band structure, suggesting 
that $P_d^\infty$ is determined by a mechanism qualitatively different 
from conventional BCS-type theories, which start from the instability 
of the Fermi surface of the normal state. 
We will return to this subject in \S\ref{sec:discussions}. 
\par

%
To summarize the $t'/t$ dependence of $P_d^\infty$, we show $P_d^\infty$ 
for three different regimes of $\delta$ in Fig.~\ref{fig:pdvsaall}. 
For a weak correlation ($U/t=4$), the area of enhanced $P_d^\infty$ is 
limited ($-0.25\lsim t'/t\lsim-0.1$) regardless of $\delta$, and 
the magnitude tends to vanish as $L$ increases.
For strongly correlated cases ($U/t\ge 8$), $P_d^\infty$ is enhanced 
by moderate negative values of $t'/t$ when $\delta$ is in the overdoped 
and optimally doped regimes.  
On the other hand, near half filling, the magnitude of $P_d^\infty$ 
is steady, and tends to be independent of $t'/t$ as $U/t$ increases. 
\par

Finally, we point out that $P_d^\infty$ becomes a weakly 
{\it increasing} function of $t'/t$ near half filling and 
for very large $U/t$, as shown in Fig.~\ref{fig:pdvsaall}(d). 
Such unexpected behavior was discovered for very slightly doped $t$-$J$ 
models using the density matrix renormalization group\cite{White} 
and exact diagonalization\cite{Martins}, and seemed incompatible 
with the property of cuprates. 
Later, a VMC study\cite{Shih} based on the $t$-$J$ model revealed 
that this feature is restricted to the close vicinity of half filling. 
The present result supports this finding and adds a requirement 
that the interaction should be considerably strong (\eg, $U/t=16$).
\par 

\section{Antinodal Electrons and Superconductivity
\label{sec:discussions}}
In \S\ref{sec:origin}, we argued that electrons near antinodal 
points are crucial for $d$-wave SC.
In this section, we actually reveal a close relationship between 
the behavior of the momentum distribution function $n({\bf k})$ near 
${\bf k}=(\pi,0)$ and the SC correlation function $P_d^\infty$ 
in the strongly correlated regime. 
\par

\begin{figure*}[!t]
\begin{center}
\includegraphics[width=14.0cm,clip]{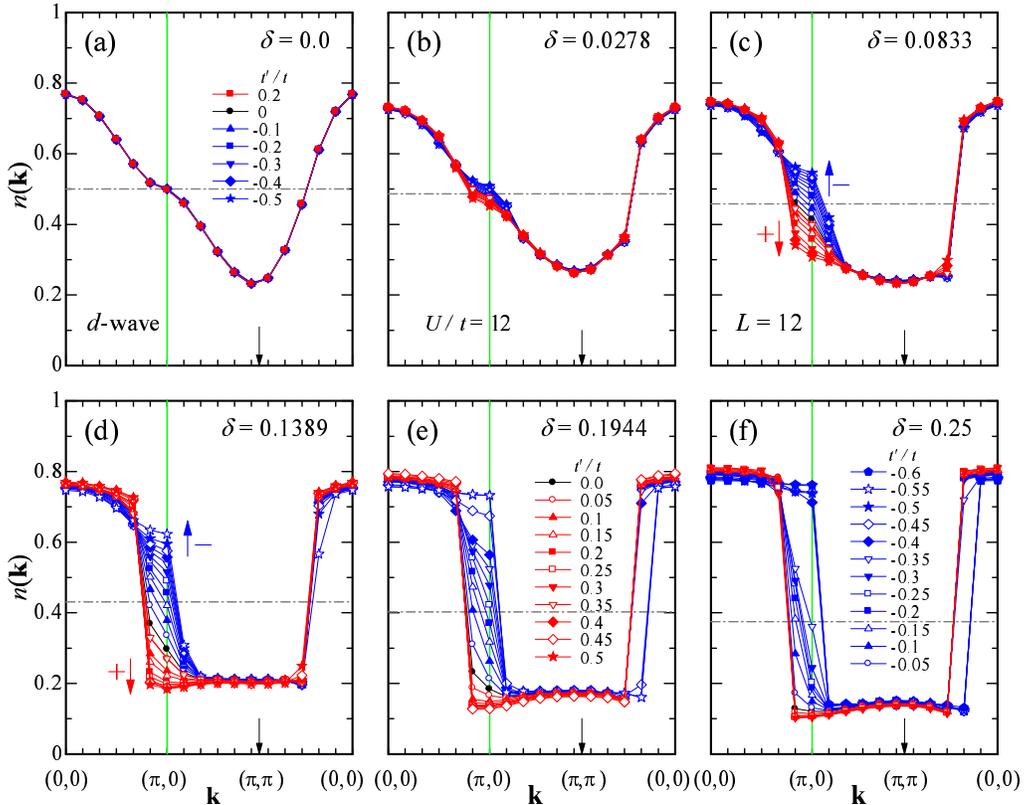} 
\vskip -4mm
\caption{(Color online) 
Momentum distribution function of $d$-wave state for various values 
of $t'/t$ for $U/t=12$ and $L=12$; the behavior basically does not change 
even if $U/t$ ($\gsim 10$) or $L$ varies. 
The doping rates are different among the panels. 
The symbols for $t'/t$ in (a), (e), and (f) are common to all 
panels. 
Owing to the boundary condition, $k_y$ for each {\bf k}-point shifts 
by $\pi/(2L)$. 
The arrows with $+$ [$-$] in (c) and (d) indicate the direction of 
variation for $t'/t>0$ [$t'/t<0$] when $|t'/t|$ increases; this $t'/t$ 
dependence does not alter for (b), (e), and (f). 
The $(\pi,0)$ point is marked with a vertical guide line, and the 
$(\pi,\pi)$ point with a small arrow on the abscissa. 
}
\label{fig:nkt2} 
\end{center}
\end{figure*} 
%
First, let us recall again the $t'/t$ dependence of $P_d^\infty$
for sufficiently large values of $U/t$ for comparison 
with that of $n({\bf k})$.
As discussed in ref.~\citen{YOT}, the properties of $\Psi_Q^d$ in the 
insulating 
phase ($U>U_{\rm c}$ at half filling) are almost independent of $t'/t$. 
For small $\delta$, the $t'/t$ dependence of $P_d^\infty$ is still weak 
as seen for $U/t=12$ and $16$ in Fig.~\ref{fig:pdvsaall}(c), but for 
$\delta\gsim 0.1$, $P_d^\infty$ comes to largely depend on $t'/t$, i.e., 
$P_d^\infty$ tends to increase for $t'/t<0$ and decrease for $t'/t>0$ 
as shown in Figs.~\ref{fig:pdvsaall}(b) and \ref{fig:pdvsaall}(a). 
\par 

With this behavior of $P_d^\infty$ in mind, we look at the $t'/t$ 
dependence of $n({\bf k})$ [eq.~(\ref{eq:nk})]. 
In Fig.~\ref{fig:nkt2}, $n({\bf k})$ is plotted for six $\delta$ 
for $U/t=12$. 
In each panel, data for various values of $t'/t$ are displayed together. 
At half filling [Fig.~\ref{fig:nkt2}(a)], $n({\bf k})$ is almost 
independent of $t'/t$ as mentioned, and does not have a discontinuity 
at any ${\bf k}$, because $\Psi_Q^d$ is Mott insulating 
($U_{\rm c}/t\sim 6.5$-7.2) with a charge density gap. 
Upon doping carriers, a discontinuity (Fermi surface) appears in the 
lattice-diagonal direction near $(\pi/2,\pi/2)$ [Figs.~\ref{fig:nkt2}(b)-
\ref{fig:nkt2}(f)],\cite{Randeria} because $\Psi_Q^d$ becomes SC 
with nodes of $\Delta_{\rm k}$ for $k_x=\pm k_y$ (and finally metallic).
Meanwhile, it is in the antinodal area near ($\pi,0$), where the behavior 
of $n({\bf k})$ markedly changes as $t'/t$ varies; the degree of change 
culminates around the optimum doping rate ($\delta\sim 0.15$). 
Compared with that in the antinodal area, the change in $n({\bf k})$ by $t'/t$ 
is insignificant in other areas of ${\bf k}$. 
Thus, we became aware of a close correspondence between the SC correlation 
($P_d^\infty$) and the electronic structure in the antinodal area. 
\par

In the overdoped regime ($\delta\gsim 0.2$), $n({\bf k})$ exhibits 
a Fermi-liquid-like discontinuity (Fermi surface) in the antinodal area 
for some values of $t'/t$, \ie, on the segment $(0,0)$-$(\pi,0)$ in the 
electron-doped cases ($t'/t>0$), and on $(\pi,0)$-$(\pi,\pi)$ for 
$t'/t\lsim-0.4$ [Figs.~\ref{fig:nkt2}(e) and \ref{fig:nkt2}(f)]. 
We find in Fig.~\ref{fig:pdvsaall}(a) that when such a discontinuity 
appears, robust SC does not occur. 
Inversely, when $n({\bf k})$ changes continuously and relatively slowly 
in the antinodal area [Figs.~\ref{fig:nkt2}(b)-\ref{fig:nkt2}(d)], 
$P_d^\infty$ has 
a large and steady magnitude [Figs.~\ref{fig:pdvsaall}(b) and 
\ref{fig:pdvsaall}(c)]. 
Such behavior appears in a broad range of $t'/t$, particularly 
in the underdoped regime. 
We will discuss this problem shortly. 
\par

\begin{figure}[hob]
\begin{center}
\includegraphics[width=8.5cm,clip]{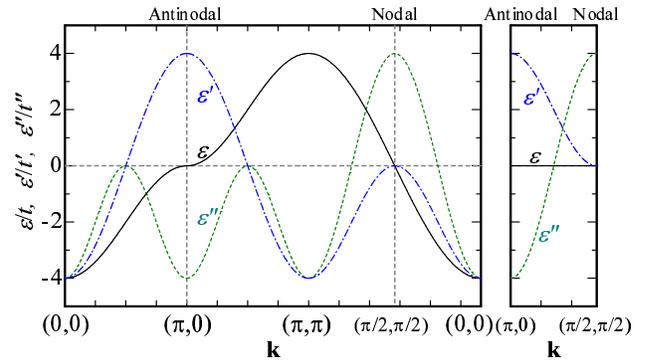}
\caption{(Color online) 
Elements of bare band dispersion relations along paths 
$(0,0)\rightarrow(\pi,0)\rightarrow(\pi,\pi)\rightarrow(0,0)$
(left panel) and $(\pi,0)\rightarrow(\pi/2,\pi/2)$ (right panel): 
$\varepsilon/t=-2(\cos k_x+\cos k_y)$, 
$\varepsilon'/t'=-4\cos k_x\cos k_y$, 
$\varepsilon''/t''=-2(\cos 2k_x+\cos 2k_y)$. 
$t''$ indicates the hopping integral to the third-neighbor sites 
($\pm 2,0$) and ($0,\pm 2$), which is disregarded in this paper. 
The nodal $(\pi/2,\pi/2)$ and antinodal $(\pi,0)$ points are marked 
by vertical dashed lines. 
}
\label{fig:Ek} 
\end{center} 
\end{figure}
%
The above feature of $n({\bf k}$) is more or less related to the band 
structure. 
In Fig.~\ref{fig:Ek}, the elements of the bare band dispersion relation 
$\varepsilon_{\bf k}\ [=\varepsilon+\varepsilon'(+\varepsilon'')]$ are 
depicted. 
The dispersion of the pure square lattice $\varepsilon$ ($t'=t''=0$) 
has a well-known flat part at $\varepsilon=0$ around $(\pi,0)$ (stationary 
in both $k_x$ and $k_y$ directions), where the Fermi surface passes 
for $\delta\sim 0$. 
The dispersion due to $t'$ ($\varepsilon'$) is also stationary but maximum 
at $(\pi,0)$. 
Consequently, $\varepsilon_{\bf k}$ and the locus of the Fermi surface 
around $(\pi,0)$ sensitively depend on $t'/t$. 
In contrast, at the $\varepsilon=0$ point in the nodal direction 
$(\pi/2,\pi/2)$, $\varepsilon'$ is zero and stationary, so that the change 
in $\varepsilon_{\bf k}$ by $t'/t$ is small near $(\pi/2,\pi/2)$. 
Incidentally, the third-neighbor hopping $t''$, which is omitted here 
but is often used as $t''=-t'/2$ in the literature, works similarly 
to that of $t'$ near $(\pi,0)$ because 
$\varepsilon''/t''\sim -\varepsilon'/t'$.\cite{note-renorm} 
\par 

To relate the above feature of $n({\bf k})$ to $P_d^\infty$ quantitatively, 
we need a quantity that appropriately expresses this feature. 
The magnitude of $n({\bf k})$ near ($\pi,0$) itself apparently does 
not represent the strength of SC. 
In fact, we found that the slope of $n({\bf k})$, 
\begin{equation}
|\nabla n({\bf k})|
=\sqrt{\left(\frac{\partial n({\bf k})}{\partial k_x}\right)^2+
       \left(\frac{\partial n({\bf k})}{\partial k_y}\right)^2}, 
\label{eq:nabla}
\end{equation}
around the antinodal point seems relevant.\cite{Kobayashi,YOK} 
The maximum of $|\nabla n({\bf k})|$ is often used as an index 
of a quasi-Fermi surface and is reduced to a pure Fermi surface 
for $|\nabla n({\bf k})|\rightarrow\infty$. 
Here, we estimate $|\nabla n({\bf k})|$ near ${\bf k}=(\pi,0)$ 
(abbreviated as $|\nabla n({\bf X})|$) from the finite-size VMC data 
as follows: 
$\partial n({\bf k})/\partial k_x$ and 
$\partial n({\bf k})/\partial k_y$ are obtained from the finite differences 
of $n({\bf k})$ between ${\bf X'}$ and ${\bf X}$ given 
in \S\ref{sec:Pair} and those between ${\bf X}$ and 
${\bf X''}=\pi(1,\frac{3}{L})$, respectively. 
\par

\begin{figure*}[!t] 
\begin{center}
\includegraphics[width=17.0cm,clip]{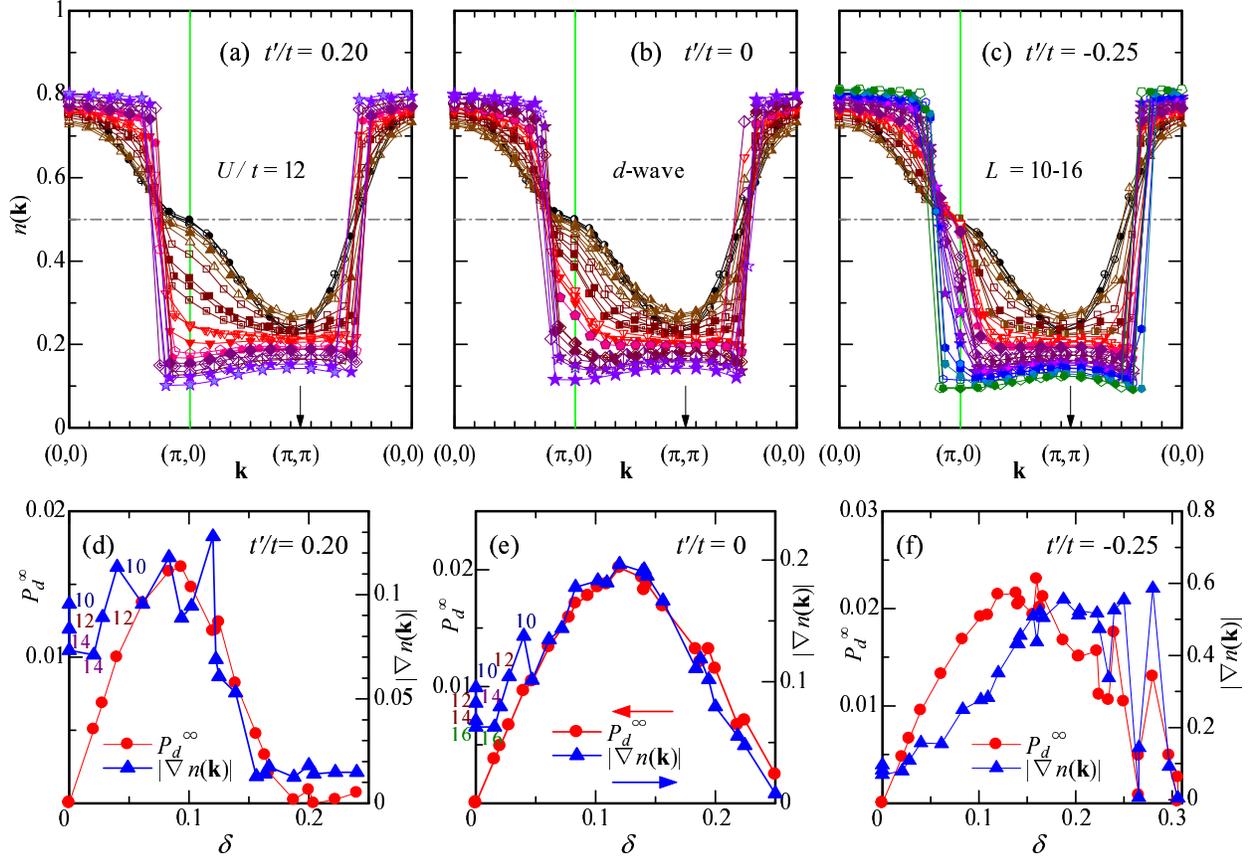} 
\vskip -5mm
\caption{(Color online) 
(a)-(c) Momentum distribution function of $\Psi_Q^d$ for various 
values of $\delta$ at $U/t=12$. 
The values of $t'/t$ are different among the three panels. 
The data for $L=10$-16 are simultaneously displayed. 
Detailed instructions for $\delta$ are omitted, but 
circles denote $\delta=0$, 
upward triangles $\delta=0.02$-0.04, 
squares $\delta=0.06$-0.11, 
downward triangles $\delta=0.12$-0.14, 
upward pentagons $\delta=0.15$-0.165, 
diamonds $\delta=0.165$-0.21,
stars $\delta=0.22$-0.25, 
hexagons $\delta=0.26$-0.29, and 
downward pentagons $\delta=0.29$-0.31. 
To adjust the scale in the section $(0,0)$-$(\pi,0)$-$(\pi,\pi)$, 
the locus of ${\bf k}$ points between $(\pi,\pi)$-$(0,0)$ slightly 
shifts depending on $L$.  
(d)-(f) Comparison between SC correlation function $P_d^\infty$ 
and $|\nabla n({\bf k})|$ near ${\bf k}=(\pi,0)$ estimated from the data 
in (a)-(c) for the three values of $t'/t$. 
We adjust the scales of the two quantities so as to roughly equalize 
the maximum magnitudes. 
The small digits for $\delta\sim 0$ in (d) and (e) indicate the system 
size $L$ used for the corresponding data point. 
}
\label{fig:nknall} 
\end{center}
\end{figure*} 
%
In Fig.~\ref{fig:nknall}, we plot $n({\bf k})$ for various $\delta$ 
($L=10$-16) for three values of $t'/t$ at $U/t=12$. 
In contrast to the $t'/t$ dependence (Fig.~\ref{fig:nkt2}), an appreciable 
variation exists at any ${\bf k}$, but the magnitude of 
variation in the antinodal area is still relatively large. 
We estimate $|\nabla n({\bf X})|$ for each $\delta$ from 
Figs.~\ref{fig:nknall}(a)-\ref{fig:nknall}(c), and plot it as a function 
of $\delta$ 
with triangles in Figs.~\ref{fig:nknall}(d)-\ref{fig:nknall}(f), 
respectively, where the 
$\delta$ dependence of $P_d^\infty$ is also displayed with circles. 
For $t'=0$, the behavior of $|\nabla n({\bf X})|$ is in close agreement 
with that of $P_d^\infty$ except near half filling, where, however, 
the disagreement obviously stems from the system-size 
dependence.\cite{notedisagree} 
Thus, $|\nabla n({\bf X})|$ is almost proportional to $P_d^\infty$ 
for $t'=0$.
For $t'/t=0.2$, the behavior of $|\nabla n({\bf X})|$ is also nearly 
proportional to that of $P_d^\infty$; $|\nabla n({\bf X})|$ exhibits a 
similar system-size dependence near half filling. 
For $t'/t=-0.25$, the overall tendency of $|\nabla n({\bf X})|$ 
coincides with that of $P_d^\infty$. 
\par

\begin{figure}[hob]
\begin{center}
\includegraphics[width=7.5cm,clip]{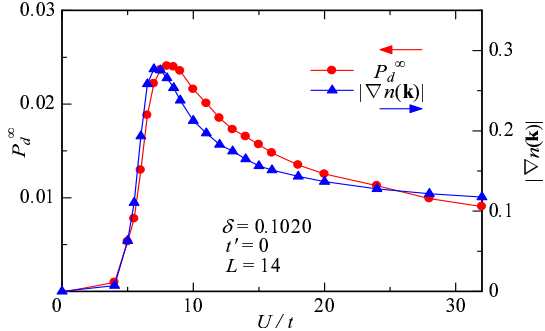}
\vskip -2mm 
\caption{(Color online) 
Comparison between SC correlation function $P_d^\infty$ and 
$|\nabla n({\bf k})|$ at ${\bf k}={\bf X}'\sim(\pi,0)$ as function of $U/t$ 
for $t'=0$ and $\delta=0.102$. 
The scales of the two quantities are adjusted so as to roughly equalize 
the maximum magnitudes. 
}
\label{fig:diffXn} 
\end{center}
\end{figure}
%
Finally, we touch on the $U/t$ dependence of $|\nabla n({\bf X})|$. 
If the simple $d$-wave BCS function without correlation factors
is used, we have $|\nabla n(\pi,0)|=0$ by differentiating 
\begin{equation}
n({\bf k})=|v_{\bf k}|^2=\frac{1}{2}
\left[1-\frac{\varepsilon_{\bf k}-\zeta}
             {\sqrt{\left(\varepsilon_{\bf k}-\zeta\right)^2+
                          \Delta_{\bf k}^2}}\right], 
\end{equation}
so that finite $|\nabla n({\bf X})|$ is considered as a correlation 
effect. 
Figure \ref{fig:diffXn} shows a comparison of $|\nabla n({\bf X})|$ and 
$P_d^\infty$ 
of $\Psi_Q^d$ for $t'=0$ as a function of $U/t$ in a slightly underdoped 
electron density.
The two quantities are both almost zero for $U/t\lsim 4$, and are in broad 
agreement for any $U/t$, suggesting that the electron distribution near 
($\pi,0$) is important for $d$-wave SC irrespective of the correlation 
strength. 
\par

In this section, we have demonstrated that the strength of SC, $P_d^\infty$, 
behaves similarly to $|\nabla n({\bf X})|$ near the antinodal point; 
this result confirms the discussions regarding the stability of the $d$-wave 
state in previous sections, and is consistent with recent 
experiments,\cite{Yang,ARPESanti} which indicate that the antinodal electrons 
contribute to the SC gap. 
On the other hand, this result is contradictory to the concept of the 
dichotomy of electronic roles in the wave-number space, in which electrons 
near the nodal direction contribute to SC, and electrons near the antinodal 
point exclusively contribute to the pseudogaps.\cite{twogap,JPSJ-ARPES} 
Similar results have recently been obtained by extended methods of 
DMFT.\cite{Aichhorn2,Civelli} 
\par

\section{Conclusions\label{sec:summary}}
%
With high-$T_{\rm c}$ cuprates in mind, we studied the $U/t$, $\delta$, and 
$t'/t$ dependences of the correlated $d_{x^2-y^2}$-wave singlet state 
$\Psi_Q^d$ and the AF state $\Psi_Q^{\rm AF}$ 
independently as doped Mott insulators, applying them to a 2D 
Hubbard ($t$-$t'$-$U$) model. 
The expectation values are calculated using a variational Monte Carlo 
method without additional approximations. 
The main results are recapitulated below. 
\par

(1) As $U/t$ increases, $\Psi_Q^d$ undergoes a sharp crossover of SC 
properties from a conventional BCS type to a kinetic-energy-driven type 
at $U=U_{\rm co}$ somewhat larger than the bandwidth ($8t$). 
As $\delta$ decreases, $U_{\rm co}$ is continuously connected to the Mott 
transition point $U_{\rm c}$ at half filling. 
For $U<U_{\rm co}$, SC, which is fragile, is 
enhanced by the pair scattering of the vector ${\bf Q}=(\pi,\pi)$ 
near the antinodes, according to the ordinary $d$-wave BCS theory. 
For $U/t\lsim 5$, steady SC corresponding to that of high-$T_{\rm c}$ cuprates 
is not found. 
The SC correlation function $P_d^\infty$ abruptly increases for $U/t\gsim 6$. 
Unconventional SC for $U>U_{\rm co}$ is robust for wide ranges of 
$t'/t$ and $\delta$, which coincide with the ranges where the doublon-holon 
binding correlation---the essence of Mott physics--- is effective, 
and the picture of the $t$-$J$ model is valid. 
This SC cannot be explained by the instability of the Fermi surface 
in the underlying normal state. 
\par

(2) By comparing the $\delta$ dependence of the $d$-wave SC correlation 
function $P_d^\infty$ with the dome shape of $T_{\rm c}$ and the 
condensation energy experimentally observed in cuprates, we showed that 
the effective value of $U$ for cuprates must be larger than at least 
the Mott critical value $U_{\rm c}\ (\sim 7t)$. 
Furthermore, we found from the analysis of kinetic energy that there are  
two kinds of holons for $U>U_{\rm co}$; a holon created during doublon 
formation does not contribute to conduction, but a holon introduced by 
doping becomes a carrier. 
Consequently, the number of doped holes (electrons) becomes equal to the 
number of carriers. 
This is a natural picture in the $t$-$J$ model and consistent with 
various experimental results of cuprates.\cite{Condprop,Uemura} 
On the other hand, for $U<U_{\rm co}$, all holons (electrons) work 
in the same manner as carriers, which make an ordinary large Fermi surface. 
This Fermi liquid feature is contradictory to that of cuprates. 
\par

(3) In view of the two-gap problem, the present result for $U>U_{\rm c}$ 
is interpreted as follows: 
A gap for the singlet pair formation $\Delta_d$, which possibly corresponds 
to a pseudogap, is a decreasing function of $\delta$. 
Meanwhile, the strength of SC represented by $P_d^\infty$ (related to 
the SC gap) has a dome-shaped $\delta$ dependence, and is written 
as the product of $\Delta_d$ and a factor indicating charge transportability 
such as the carrier density or the quasi-particle renormalization factor 
$Z$, which are increasing functions of $\delta$ for small $\delta$. 
This suppression of charge fluctuation is imposed by the Mott physics. 
This feature of the SC gap is consistent with various theories based on 
the $t$-$J$ model. 
\par

(4) A proper negative (positive) $t'/t$ term stabilizes the SC (AF) state 
and destabilizes the AF (SC) state. 
A case of a very small $\delta$ and a quite large $U/t$ is exceptional. 
Meanwhile, the AF state is intrinsically unstable in hole-doped cases 
($t'/t\lsim 0$), and the phase separates to the AF state with the local 
electron density $n=1$ and to the SC state with $n<0.85$. 
\par

(5) For the stability of AF state, the nesting of ${\bf Q}$ is primarily 
important on the whole Fermi surface. 
On the other hand, crucial for SC is the pair scattering of ${\bf Q}$ 
in the antinodal area, where the underlying band is flat. 
To check this aspect, we showed that the SC correlation $P_d^\infty$ is 
closely connected to the slope of the momentum distribution function 
$|\nabla n({\bf k})|$ near the antinodal point. 
This contradicts the often argued dichotomy of electronic roles in the 
{\bf k} space, i.e., electrons in the nodal (antinodal) area exclusively 
contribute to SC (pseudogap). 
\par

Finally, we briefly mention some future problems. 
(i) 
It is important to include long-range hopping terms in the band 
dispersion and band renormalization effects owing to the electron 
correlation. 
(ii) 
We should check the coexistence of SC and AF orders and their mutual 
exclusion. 
(iii) 
In connection with the pseudogap problem, normal states with some symmetry 
breaking should be studied. 
(iv) 
Making the best use of the recent progress in VMC techniques,\cite{VMCopt} 
we have to explore more precise trial wave 
functions.\cite{Tahara,Tocchio}  
(v) We should study the role of doublons and holons in the SC phase 
in more detail, for instance, to confirm a recent proposal.\cite{Phillips} 
\par

\begin{acknowledgments} 
The authors thank Tsutomu Watanabe, Shun Tamura, and Takafumi Sato 
for useful discussions. 
This work is partly supported by Grants-in-Aid from the Ministry of 
Education, Culture, Sports, Science and Technology. 
\end{acknowledgments}

\appendix
\section{Hubbard model for cuprates\label{sec:derivation}}
To consider CuO$_2$ planes, an appropriate starting model is 
the $d$-$p$ model composed of O-2$p$ and Cu-3$d_{x^2-y^2}$ orbitals. 
The $d$-$p$ model is reducible to simpler models that still capture 
the essence of CuO$_2$ planes. 
Since a doped electron enters a Cu-3$d_{x^2-y^2}$ orbital, the model 
is reduced to a Hubbard model on a square lattice by eliminating 
the degree of freedom for the O-2$p$ orbitals. 
In contrast, a doped hole enters an O-2$p$ orbital; a Zhang-Rice 
singlet is formed, through which the model for a low dopant density 
is reduced to the $t$-$J$ model on a square lattice 
with small $J/t$.\cite{Zhang-Rice} 
This $t$-$J$ model is connected to the Hubbard model with large 
$U/t$ $(=4t/J)$ through the strong-coupling expansion \cite{SCex} 
within the approximation of neglecting the pair-hopping (three-site) 
terms. 
Thus, the Hubbard model on a square lattice is derived as an effective
model for both electron-doped (ED) and hole-doped (HD) cuprates. 
Note, however, that the effective value of $U/t$ in the Hubbard model
may differ between ED and HD cases owing to the distinct derivation 
paths, even if the actual values of $U$ in the Cu-3$d_{x^2-y^2}$ 
orbitals are identical between the two. 
\par 

Another vital element of the model, eq.~(\ref{eq:Hamil}), is that 
one can map a more-than-half-filled case ($n>1$) to a 
less-than-half-filled case ($n<1$) by taking advantage of the 
antisymmetric level structure with respect to $n=1$. 
The band dispersion, eq.~(\ref{eq:band}), 
for $\tilde{\bf k}=(\pi-k_y,\pi-k_x)$, 
which is the symmetric point of $\bf k$ with respect to the Fermi surface 
at half filling for $t'=0$ (AF Brillouin zone boundary), 
is written as 
\begin{equation}
\tilde\varepsilon(\tilde{\bf k})=
2\tilde t(\cos k_x+\cos k_y)-4\tilde t'\cos k_x\cos k_y, 
\end{equation}
where tildes are the marks for $\tilde{\bf k}$. 
To satisfy $\varepsilon({\bf k})=-\tilde\varepsilon(\tilde{\bf k})$, 
we need the relations $\tilde t=t$ and $\tilde t'=-t'$, which are 
represented by a canonical (electron-hole) transformation,
\begin{equation}
c_{i\sigma}^\dag \rightarrow e^{i{\bf Q}\cdot{\bf r}_i}\ \tilde c_{i\sigma}, 
\label{eq:trans}
\end{equation} 
with ${\bf Q}=(\pi,\pi)$. 
Under this transformation, the model, eq.~(\ref{eq:Hamil}), is invariant 
except for the addition of a constant $U(N_{\rm s}-N)$. 
Consequently, calculations for an ED system with electron density 
$n=1+\delta$ ($\delta>0$) and $t'/t$ ($<0$) can be replaced by those 
for the less-than-half-filled case with $1-\delta$ and $|t'/t|$, 
if we consider that $t'/t$ is negative for both ED and HD cuprates. 
\par

\section{Details of doublon-holon factors\label{sec:dh}}
In this Appendix, we compare different forms of the D-H correlation 
factor ${\cal P}_Q$ [eq.~(\ref{eq:PQ})] to show that the difference 
in results among $Q_j^{\rm D}$ [eq.~(\ref{eq:DQ})] and other forms 
of $Q_j$ is insignificant. 
Here, ${\cal P}_Q$ between nearest-neighbor sites is mainly considered, 
because ${\cal P}_Q$ between further sites only makes a minor 
difference.\cite{Miyagawa} 
\par

\begin{table}
\caption{
Absolute values of total energy of the $d$-wave state, $|E/t|$, are 
compared among the three D-H correlation factors for $\delta>0$, 
$t'/t=0$, and $L=10$.
The last digits may have some errors. 
} 
\vspace{1mm}
\begin{tabular}{c|c|c|c|c|c|c}
\hline
$\delta$ & \multicolumn{3}{|c|}{$0.20$} & \multicolumn{3}{|c}{$0.12$} \\
\hline
$U/t$ & ${\cal P}_Q^{\rm D}$ & ${\cal P}_Q^{\rm S}$ & 
${\cal P}_Q^{\rm DH}$ & 
        ${\cal P}_Q^{\rm D}$ & ${\cal P}_Q^{\rm S}$ & 
${\cal P}_Q^{\rm DH}$ \\
\hline
 4    & 1.0754 & 1.0773 & 1.0773 & 0.9914 & 0.9935 & 0.9935 \\
 6    & 0.9237 & 0.9282 & 0.9284 & 0.7979 & 0.8035 & 0.8038 \\
 8    & 0.8216 & 0.8288 & 0.8291 & 0.6750 & 0.6841 & 0.6854 \\
10    & 0.7545 & 0.7626 & 0.7634 & 0.6010 & 0.6094 & 0.6121 \\
12    & 0.7084 & 0.7165 & 0.7177 & 0.5518 & 0.5583 & 0.5624 \\
16    & 0.6502 & 0.6576 & 0.6591 & 0.4903 & 0.4936 & 0.4988 \\
\hline
\end{tabular} 
\label{table:D-S-DH}
\end{table}
%
\begin{figure}
\begin{center}
\includegraphics[width=7cm,clip]{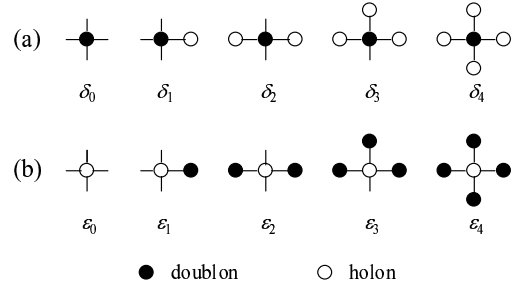}
\end{center}
\vskip -3mm
\caption{
(a) Local configurations around a doublon are classified according to 
the number of holons in the nearest-neighbor sites. 
(b) A similar classification around a holon. 
}
\label{fig:dh} 
\end{figure}
%
At half filling, symmetric projections such as ${\cal P}_Q$ with $Q_j^{\rm S}$ 
[eq.~(\ref{eq:SymQ})] (${\cal P}_Q^{\rm S}$) have been 
used\cite{Kaplan,Fazekas,YS3} on account of the particle-hole or 
doublon-holon symmetry. 
For $\delta>0$, however, this symmetry is broken, and the number of holons 
always exceeds that of doublons. 
Consequently, if ${\cal P}_j^{\rm S}$ is used, the configurations $\delta_2$, 
$\delta_3$, and $\delta_4$ illustrated in Fig.~\ref{fig:dh} seem to appear 
frequently as $\delta$ increases. 
In fact, a recent study\cite{Kobayashi} using a detailed parameterization 
for $Q_j$ has revealed that such configurations rarely appear for large 
$U/t$, because they increase the interaction energy $Ud$. 
Anyway, we consider another asymmetric form of $Q_j$, which is a natural 
extension of $Q_j^{\rm S}$: 
\begin{equation}
Q^{\rm DH}_j=\mu_{\rm d} d_j\prod_\tau(1-h_{j+\tau})+
             \mu_{\rm h} h_j\prod_\tau(1-d_{j+\tau}), 
\label{eq:DHQ} 
\end{equation} 
where binding parameters for holons $\mu_{\rm h}$ and doublons 
$\mu_{\rm d}$ are optimized independently. 
At half filling, $Q_j^{\rm DH}$ is reduced to $Q_j^{\rm S}$ 
($\mu_{\rm h}=\mu_{\rm d}$) to retrieve the symmetry. 
As $\delta$ increases, $\mu_{\rm h}$ is expected to decrease much faster
than $\mu_{\rm d}$ because $\varepsilon_2$ comes to appear less frequently 
than $\delta_2$; therefore $Q_j^{\rm DH}$ approaches $Q_j^{\rm D}$ 
($\mu_{\rm h}$ rapidly decreases). 
As shown in Table~\ref{table:D-S-DH},\cite{note-error} the variational 
energy of $Q_j^{\rm DH}$ is improved compared with that of $Q_j^{\rm S}$, 
especially 
for large $U/t$. 
In the following, we compare the asymmetric forms $Q_j^{\rm DH}$ and 
$Q_j^{\rm D}$. 
\par

\begin{figure}
\begin{center}
\includegraphics[width=7.5cm,clip]{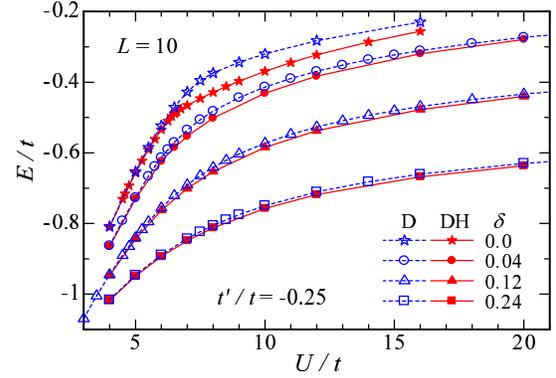}
\end{center}
\vskip -4mm
\caption{(Color online) 
The total energy of $\Psi_Q^d$ with ${\cal P}_j^{\rm DH}$ (solid symbols) 
is compared with that with ${\cal P}_j^{\rm D}$ (open symbols) 
as a function of $U/t$ for four $\delta$. 
At half filling, $Q_j^{\rm DH}$ is reduced to $Q_j^{\rm S}$ (solid 
stars).\cite{YOT} 
The tendency for other values of $t'/t$ is similar. 
}
\label{fig:edh} 
\end{figure}
%
\begin{table}
\caption{
Improvements in total energy by $\Psi_Q^{\rm DH}$ on $\Psi_Q^{\rm D}$ 
as percentages, $(E^{\rm D}-E^{\rm DH})/|E^{\rm D}|\times 100$,  
for various parameter values and $L=10$, which satisfy the closed-shell 
condition.
}
\vspace{1mm}
\begin{tabular}{c|c|c|c|r|c|c|c|r}
\hline 
$t'/t$      & \multicolumn{4}{|c|}{$-0.25$} & \multicolumn{4}{|c}{0} \\

\hline
$\delta$ & $0.24$ & $0.12$ & $0.04$ & $0.0$ & $0.20$ & $0.12$ & $0.04$ & $0.0$ \\
\hline
$U/t$ &  &  &  &  &  &  &  \\
 4    & 0.2  & 0.2 & 0.2 &  0.2 & 0.2 & 0.2 &  0.2 &  0.3 \\
 6    & 0.5  & 1.0 & 1.8 &  2.0 & 0.5 & 0.7 &  1.5 &  1.9 \\
 8    & 0.8  & 1.8 & 4.1 & 14.6 & 0.9 & 1.5 &  3.9 & 14.0 \\
10    & 1.0  & 2.0 & 4.2 & 15.3 & 1.2 & 1.8 &  4.0 & 15.0 \\
12    & 1.1  & 2.0 & 3.7 & 14.3 & 1.3 & 1.9 &  3.7 & 14.0 \\
16    & 1.2  & 1.8 & 2.8 & 11.4 & 1.4 & 1.7 &  2.7 & 11.3 \\
\hline
\end{tabular} 
\label{table:A1}
\end{table}

In Fig.~\ref{fig:edh}, the total energies of the $d$-wave state are 
compared between $Q_j^{\rm D}$ and $Q_j^{\rm DH}$ for $t'/t=-0.25$. 
In Table~\ref{table:A1}, we list the energy improvements by $Q_j^{\rm DH}$ 
compared with $Q_j^{\rm D}$ ($\Delta E=E^{\rm D}-E^{\rm DH}$) for some values 
of $\delta$ and $t'/t$ as functions of $U/t$. 
In every case, the energy is improved for $Q_j^{\rm DH}$. 
Comparing $\Delta E/t$ for $t'/t=-0.25$ and $0$, we find that $\Delta E$ 
only slightly depends on $t'/t$. 
For each $\delta$, $\Delta E/t$ is maximum near $U_{\rm co}/t=10$, 
especially at half filling, and $Q_j^{\rm DH}$ (=$Q_j^{\rm S}$) becomes highly 
advantageous for $U>U_{\rm c}$.
For a fixed $U/t$, $\Delta E/t$ decreases as $\delta$ increases. 
\par

\begin{figure}
\begin{center}
\includegraphics[width=7cm,clip]{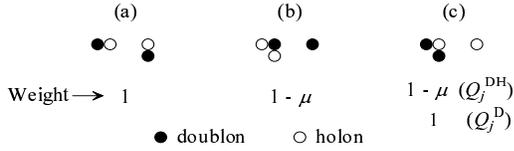} 
\end{center}
\vskip -3mm
\caption{
Weight assignment by $Q_j^{\rm DH}$ and by $Q_j^{\rm D}$ to local 
configurations with two doublons and two holons typical 
for $U\sim U_{\rm co}$ at half filling. 
All the sites in the background are singly occupied. 
In each case, only relative positions of doublons to holons are 
illustrated.
In (a), there are two $\delta_1$ and two $\varepsilon_1$; 
in (b), $\delta_2$, $\delta_0$, and two $\varepsilon_1$; 
and in (c), $\varepsilon_2$, $\varepsilon_0$, and two $\delta_1$. 
}
\label{fig:dhexmpl} 
\end{figure} 
%
\begin{figure}
\begin{center}
\includegraphics[width=6.5cm,clip]{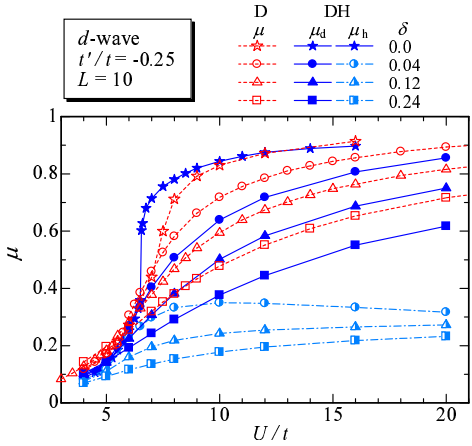} 
\end{center}
\vskip -5mm
\caption{(Color online) 
Comparison of optimized D-H correlation parameters between 
$Q_j^{\rm D}$ ($\mu$) and $Q_j^{\rm DH}$ ($\mu_{\rm d}$ and $\mu_{\rm h}$) 
for nearest-neighbor sites. 
}
\label{fig:paramuc} 
\end{figure} 
%
We briefly consider the reason why $Q_j^{\rm DH}$ is advantageous 
at half filling and for $U\sim U_{\rm co}$. 
Assume that only two doublons and two holons exist in a local electron 
configuration at half filling; typical ones frequently appearing 
for $U\sim U_{\rm co}$ are shown in Fig.~\ref{fig:dhexmpl}. 
The Gutzwiller factor assigns a common weight $g^2$ to each of these 
configurations. 
Regarding the D-H factors, ${\cal P}_Q^{\rm DH}$ and ${\cal P}_Q^{\rm D}$ 
give the same weights of 1 and $1-\mu$ for the configurations (a) and (b), 
respectively;\cite{Kobayashi} however, for an unfavorable configuration 
(c), ${\cal P}_Q^{\rm D}$ gives 1, but ${\cal P}_Q^{\rm DH}$ reduces 
the weight as $1-\mu$. 
Thus, $Q_j^{\rm DH}$ becomes better than $Q_j^{\rm D}$ at $\delta=0$ 
and for $U\sim U_{\rm co}$. 
When holes are doped, configurations such as $\varepsilon_2$ 
(also $\varepsilon_3$ and $\varepsilon_4$) appearing in 
Fig.~\ref{fig:dhexmpl}(c) are rapidly suppressed; accordingly, the 
difference between $Q_j^{\rm DH}$ and $Q_j^{\rm D}$ narrows. 
This aspect is clearly seen in the optimized D-H binding factors 
$\mu$ in $Q_j^{\rm D}$ and $\mu_d$ and $\mu_h$ in $Q_j^{\rm DH}$, 
as shown in Fig.~\ref{fig:paramuc}. 
As $\delta$ increases, $\mu_{\rm d}$ becomes slightly smaller, but 
$\mu_{\rm h}$ considerably and rapidly decreases compared with $\mu$ 
in $Q_j^{\rm D}$ denoted by open symbols. 
\par

\begin{figure}
\begin{center}
\includegraphics[width=7.0cm,clip]{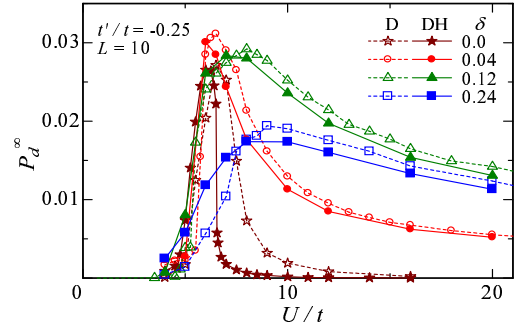}
\end{center}
\vskip -2mm
\caption{(Color online) 
Comparison of $d$-wave pair correlation function between two D-H 
correlation factors, ${\cal P}_Q^{\rm DH}$ and ${\cal P}_Q^{\rm D}$. 
The behavior of $P_d^\infty$ is similar for other values of $t'/t$.
} 
\label{fig:pdvsu} 
\end{figure} 

Because the optimized values of the other variational parameters are 
similar between the two cases of ${\cal P}_Q^{\rm DH}$ and 
${\cal P}_Q^{\rm D}$ (not shown), physical quantities, e.g., $n({\bf k})$, 
$S({\bf q})$, and $N({\bf q})$, thereof become similar (not shown). 
Among them, the $d$-wave pairing correlation function exhibits a relatively 
large difference, as shown in Fig.~\ref{fig:pdvsu}; the maximum 
$P_d^\infty$ is located at a slightly smaller $U/t$ for $Q_j^{\rm DH}$.
However, the difference remains quantitative. 
\par

In conclusion, the D-H correlation factor ${\cal P}_Q^{\rm D}$ 
possesses properties sufficiently close to those of an improved factor 
${\cal P}_Q^{\rm DH}$ at least for $\delta>0$. 
\par 

\section{Pairing correlation function\label{sec:pairfunc}}
We explain the details of how we determine long-distance values 
($P_d^{\infty}$) of the $d_{x^2-y^2}$-wave SC correlation function 
$P_d({\bf r})$ defined by eq.~(\ref{eq:pd}) with eq.~(\ref{eq:singlet}), 
to avoid misunderstanding. 
If $P_d({\bf r})$ remains finite for $|{\bf r}|\rightarrow\infty$, 
an off-diagonal long-range order exists. 
For finite systems, however, one must be careful in estimating 
$P_d({\bf r})$ for $|{\bf r}|=\infty$. 
\par 

\begin{figure}
\begin{center}
\includegraphics[width=6.0cm,clip]{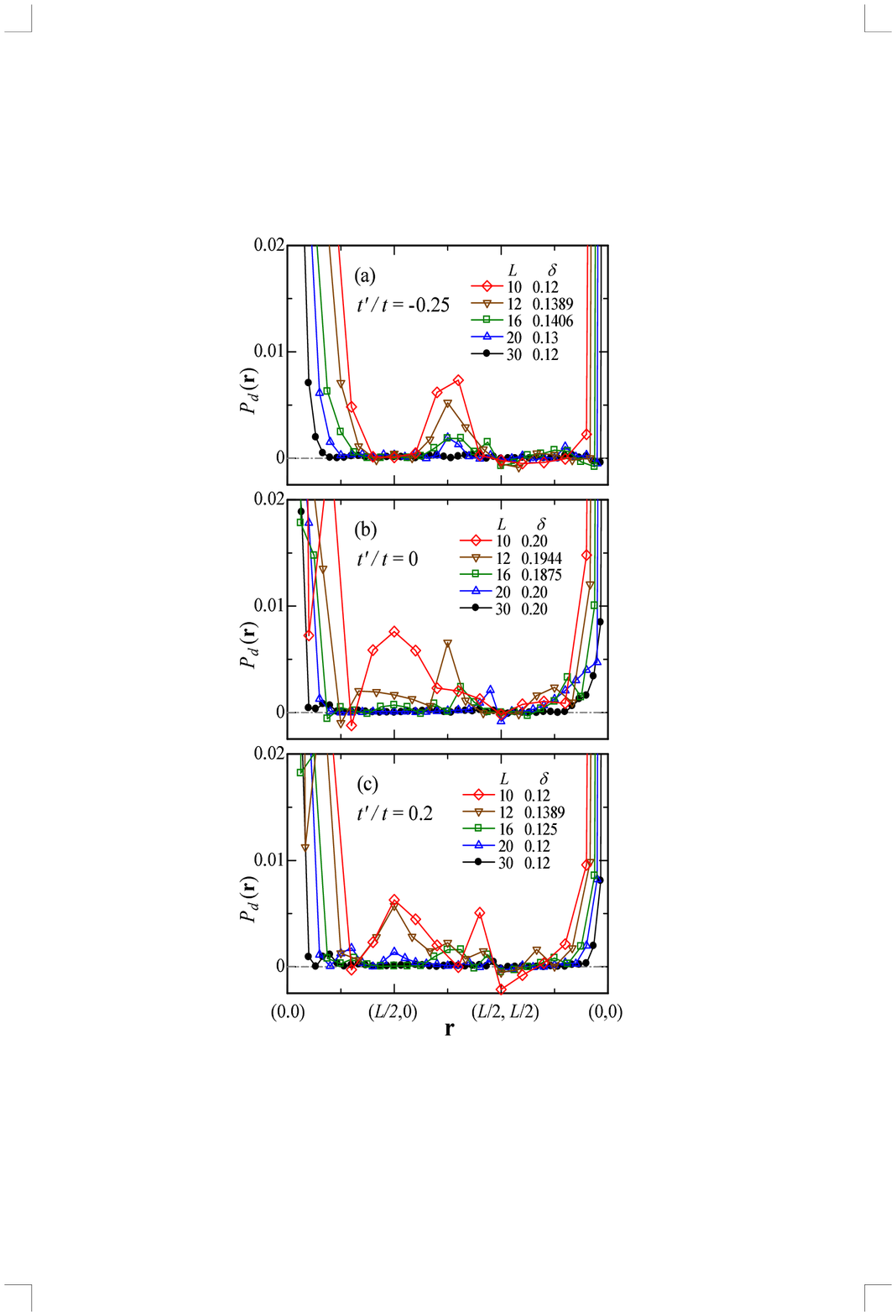}
\end{center}
\vskip -2mm
\caption{(Color online)
System-size dependence of $d$-wave SC correlation function $P_d({\bf r})$ 
for noninteracting systems ($U/t=0$). 
The path of ${\bf r}$ is shown in Fig.~\ref{fig:pdralpha}(c). 
The values of $t'/t$ and $\delta$ are 
(a) $-0.25$ and $\sim 0.13$, 
(b) $0$ and $\sim 0.20$, and 
(c) $0.2$ and $\sim 0.13$, respectively. 
$\delta$ is chosen so as to satisfy the closed-shell condition under 
the same boundary condition as that in the VMC calculations. 
The data are obtained using the analytic formula. 
}
\label{fig:pdrnonint} 
\end{figure}
\begin{figure}
\begin{center}
\includegraphics[width=6.0cm,clip]{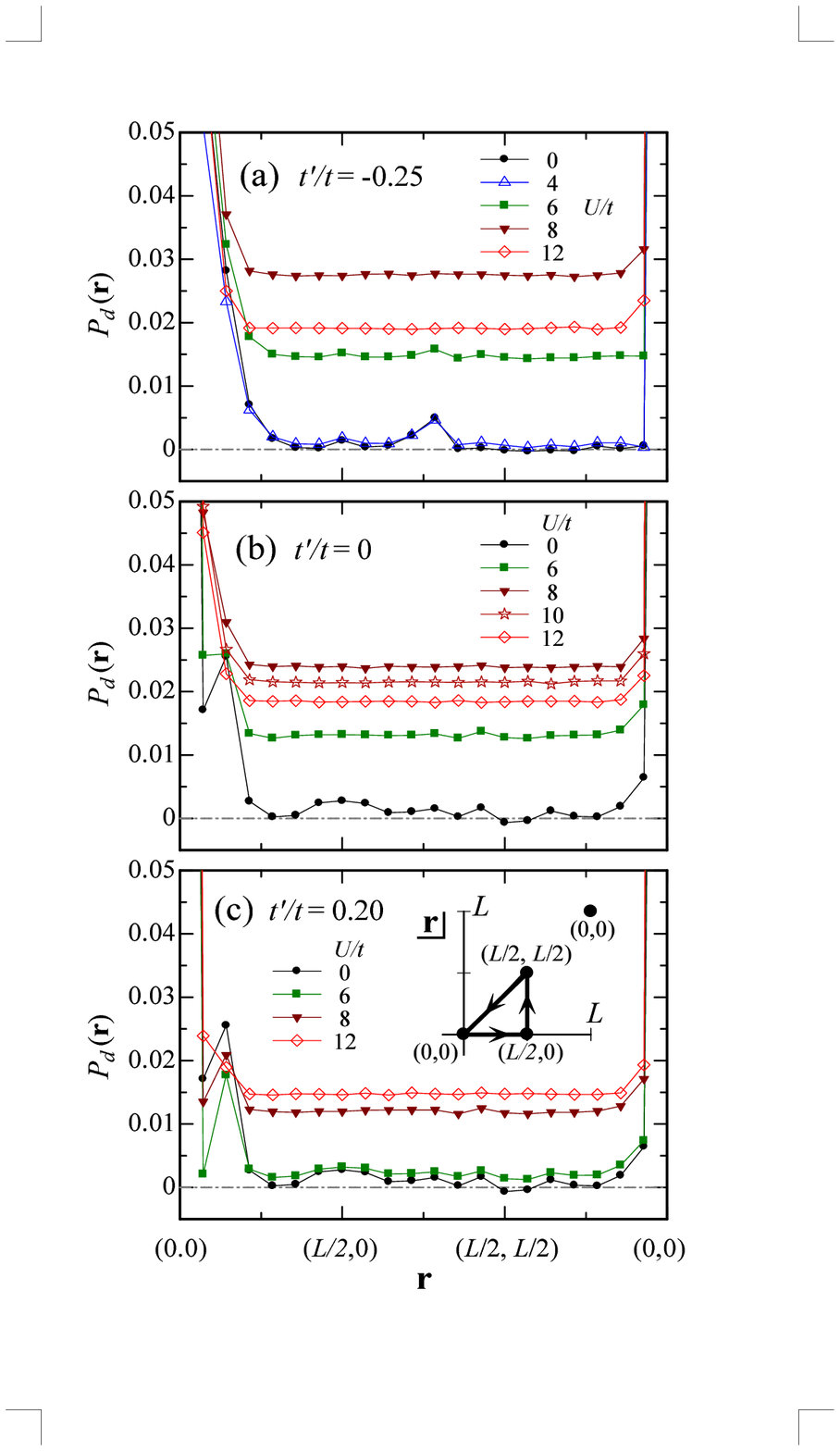} 
\end{center}
\vskip -2mm
\caption{(Color online)
$U/t$ dependence of $d$-wave SC correlation function $P_d({\bf r})$ 
for $L=14$ and $\delta=0.102$. 
The path of ${\bf r}$ is depicted in (c); ${\bf r}_\infty$ is not 
on this path. 
The values of $t'/t$ are (a) $-0.25$ (hole doped), (b) $0$, and
(c) $0.2$ (electron doped). 
In these systems, $U_{\rm max}/t=8.0$, 8.1, and 10.5 for $t'/t=-0.25$, 
$0$, and $0.2$, respectively.
In each case, the closed-shell condition is satisfied. 
Note that the scale of the vertical axis is larger than that in 
Fig.~\ref{fig:pdrnonint}. 
The data are obtained using the analytic formula for $U/t=0$ and VMC 
otherwise.
}
\label{fig:pdralpha} 
\end{figure}

%
In Fig.~\ref{fig:pdrnonint}, the behavior of $P_d({\bf r})$ for the 
noninteracting case ($U/t=0$) is compared among different system sizes. 
Here, the path of ${\bf r}$ is chosen as shown in Fig.~\ref{fig:pdralpha}(c). 
Because SC does not occur for $U/t=0$, $P_d({\bf r})$ must vanish 
for large $|{\bf r}|$. 
However, for small systems such as $L=10$ and $12$, $P_d({\bf r})$ still 
has an appreciable magnitude at distant points, e.g., ${\bf r}=(L/2,0)$ and 
${\bf r}=(L/2,L/2)$. 
Furthermore, the location of a large magnitude of $P_d({\bf r})$ depends 
on $t'/t$ and $\delta$. 
If we require that $P_d({\bf r})$ should vanish for every distant point 
in this scale, we need to use a system with $L\gsim 30$. 
Such a tendency remains for fairly large values of $U/t$, as shown 
in Fig.~\ref{fig:pdralpha}. 
Fortunately, we found that $|P_d({\bf r})|$ for 
${\bf r}=(L/2-1,L/2)\equiv{\bf r}_\infty$, which is next to the furthermost 
point, is substantially zero (less than $10^{-4}$) at $U=0$ 
regardless of $t'/t$, $\delta$, and $L$, and remains smaller 
than those for ${\bf r}\ne{\bf r}_\infty$ for small $U/t$. 
Hence, we employ $P_d({\bf r_\infty})$ as $P_d^\infty$ for small $U/t$ 
($0\le U\le U_{\rm max}$), where $U_{\rm max}$ indicates the 
$U$ of the largest $P_d({\bf r})$ for a fixed $t'/t$.  
Thereby, the artificial increase in $P_d({\bf r})$ owing to the finite 
sizes can be eliminated. 
\par

%
For large $U/t$ ($U>U_{\rm max}$), $P_d^\infty$ can be determined more 
confidently. 
The $U/t$ dependence of $P_d({\bf r})$ for fixed $L$ and $\delta$ is shown 
in Fig.~\ref{fig:pdralpha}. 
$P_d({\bf r})$ exhibits finite-size fluctuations for $U\lsim U_{\rm max}$, 
but becomes almost constant for $|{\bf r}|\ge 3$\cite{note-metric} 
regardless of $t'/t$ and $L$ for $U\gsim U_{\rm max}$. 
This behavior is the same as that found for the $t$-$J$ model,\cite{Shih} 
and reflects the short-range nature of strong correlation. 
Thus, we use the average $P_d({\bf r})$ for $|{\bf r}|\ge 3$ 
as $P_d^\infty$ for $U>U_{\rm max}$. 
\par

In summary, as a suitable indicator of the $d$-wave SC order, we safely 
define the long-distance value of $P_d({\bf r})$ for finite systems by 
\begin{equation}
P_d^\infty=\left\{
\begin{array}{ll}
P_d({\bf r_\infty}) & (U<U_{\rm max}) \\
\displaystyle
\frac{1}{M}\sum_{|{\bf r}_j|\ge 3}^M P_d({\bf r}_j) & (U\ge U_{\rm max}) 
\end{array}
\right., 
\end{equation}
where ${\bf r}_\infty=(L/2-1,L/2)$ and $M$ is the number of vectors 
${\bf r}_j$ satisfying $|{\bf r}_j|\ge 3$. 
\par




\begin{thebibliography}{99}
\bibitem{note-errmu2} 
In ref.~\citen{YOT}, we displayed the data of $\mu'$ at half filling 
as being twice as large as the correct values in error. 
The ranges of ordinates in Figs.~4(d) and 15(c) in ref.~\citen{YOT} 
should be $[-0.05,0.05]$ and $[0, 0.15]$, respectively. 
This does not affect the interpretation of the physics.
\bibitem{note-Kondo}
This function form does not correspond to that of a perturbation 
study of the BCS function for $U\rightarrow 0$:\cite{Kondo}
$\Delta\propto\exp(-t^2/U^2)$. 
Such a singular term should exist, but probably is dominant 
only for extremely small $U$.
\bibitem{note-strong}
As mentioned in ref.~\citen{YTOT}, total energies for $U>U_{\rm co}$ 
at half filling fit very well with $-\gamma t^2/U$ ($\gamma$: const.), 
which corresponds to the result of the strong-coupling ($t/U$) expansion.
\bibitem{note-Eh}
In a strict sense, the hopping of a holon (or doublon) of a D-H pair 
is counted in $E_{\rm h}$ if the holon is separated by more than 
one lattice spacing from the partner doublon (in a process higher 
than the second order). 
However, such a process occurs very rarely for large values
of $U/t$, 
and thus may be neglected.\cite{Miyagawa} 
Incidentally, the contribution of three-site (pair-hopping) terms, 
if we take them into account, is included in $E_{\rm d}$.
\bibitem{note-n08}
Actually, it is not easy to optimize cases of $t'/t=-0.4$, $n=0.80$, 
$L=10$, and $10\lsim U/t\lsim 20$, because the variational energies become 
almost the same for $0.05\lsim\Delta_d/t\lsim 0.16$, which yields 
$0.002\lsim P_d^\infty\lsim 0.015$, suggesting that this parameter set 
is situated near the transition point. 
We refrain from using the data of this parameter set in the figures.
\bibitem{note-renorm} Although we use a bare band here, it is 
more appropriate to consider a renormalized band owing to the effect 
of $U$, which is a future problem.
\bibitem{note-error}
In Table~\ref{table:D-S-DH}, the energy of $Q_j^{\rm S}$ is unexpectedly 
close to that of an improved asymmetrized factor $Q_j^{\rm DH}$ even 
for $n=0.8$. 
If we disregard the argument of D-H symmetry, $Q_j^{\rm S}$ may be used 
for $n<1$. 
In this connection, we erroneously described in ref.~\citen{YTOT} that 
$Q_j^{\rm S}$ does not work for $n\ne 0$ and $U>0$.
\end{thebibliography}
\end{document}